\documentclass{aa}
\usepackage{graphicx}
\usepackage[]{times}
\newcommand{\teff}{T_{\rm eff}}

\newcommand{\rstar}{{\rm R}_{\star}}

\newcommand{\ha}{H$\alpha$}
\newcommand{\Ha}{H$\alpha$}
\newcommand{\hb}{H$\beta$}
\newcommand{\hg}{H$\gamma$}
\newcommand{\hd}{H$\delta$}

\newcommand{\HeI}{He\,{\sc i}}

\newcommand{\FeI}{Fe\,{\sc i}}
\newcommand{\FeII}{Fe\,{\sc ii}}

\newcommand{\MgI}{Mg\,{\sc i}}
\newcommand{\MgII}{Mg\,{\sc ii}}
\newcommand{\NaI}{Na\,{\sc i}}

\newcommand{\OI}{O\,{\sc i}}

\newcommand{\OIII}{O\,{\sc iii}}

\newcommand{\CrII}{Cr\,{\sc ii}}
\newcommand{\CII}{C\,{\sc ii}}

\newcommand{\SII}{S\,{\sc ii}}
\newcommand{\SIII}{S\,{\sc iii}}

\newcommand{\NII}{N\,{\sc ii}}

\newcommand{\kms}{km\,s$^{-1}$}

\newcommand{\be}{B[e]}
\newcommand{\bephen}{B[e]~phenomenon}
\newcommand{\cpdsi}{{CPD\,$-$57$\degr$2874}}
\newcommand{\cpdzw}{{CPD\,$-$52$\degr$9243}}
\newcommand{\cdvi}{{CD\,$-$24$\degr$5721}}

\begin{document}
 %
   \title{Kinematical structure of the circumstellar environments of galactic \be -type stars   
\thanks{Based on observations collected at the European Southern Observatory, 
Chile, and at the German-Spanish Astronomical Centre, Calar Alto, 
operated by the Max-Planck-Institut f\"ur Astronomie, Heidelberg,
jointly with the Spanish National Commission for Astronomy.}
}

\author{F.-J.~Zickgraf}
  
\institute{Hamburger Sternwarte, Gojenbergsweg 112, 21029 Hamburg, Germany}
\offprints{F.-J. Zickgraf}
\date{Received date; accepted date}
\abstract{
High resolution line profiles  are presented for selected forbidden and  
permitted emission lines of a sample of galactic  \be -type stars. The 
spectral resolution corresponds to 5-7\,\kms\, with the exception of some   
line profiles which were observed with a 
resolution of 9-13 \,\kms .
All \ha\ profiles are characterized by a narrow split or single emission component 
with a width of  $\sim150-250$\,\kms\ (FWHM)
and broad wings with a full width  of 
$\sim1000-2000$\,\kms . The \ha\ profiles  can be classified into three groups:
double-peaked  profiles representing the majority, single-peaked emission-line profiles, 
and normal P\,Cygni-type profiles. Likewise, the forbidden lines  
exhibit in most cases double-peaked profiles. In particular, the majority of stars shows
split [\OI ]$\lambda$6300\AA . Double-peaked profiles  
are also found in several stars for [\NII ]$\lambda$6583\AA\ and  
[\FeII ]$\lambda$7155\AA\ although these lines in many stars exhibit single-peaked emission 
profiles. The split forbidden line profiles have peak separations of as little as $\sim10$\,\kms , 
and were therefore only discernible for the first time in the high-resolution spectra.
The ratio of violet to red emission peak intensities, $V/R$, is predominantly smaller 
or equal to 1. Theoretical  profiles were calculated for the 
optically thin case.  A latitude-dependent stellar wind 
with a radial expansion and a velocity decreasing from the pole to the equator was adopted. 
This configuration can produce split line profiles if viewed under some angle with respect 
to the line of sight.
In addition an equatorial dust ring with various optical depths was assumed. 
It can explain line asymmetries observed in some stars.
Moreover, the $V/R$ ratios can be understood in terms of this model. 
The comparison of the observed line profiles with the models thus
confirms the assumption of disk-like line-formation regions as commonly adopted 
for \be -type stars. 
\keywords{Stars: circumstellar matter -- Stars: early-type -- Stars:
emission-line, Be -- Stars: mass-loss}
}

\titlerunning{Circumstellar environments of galactic \be -type stars}
\authorrunning{F.-J. Zickgraf}
\maketitle

\section{Introduction}
\label{intro}
The class of \be -type stars is characterized by the {\em \bephen}
(Lamers et al. \cite{Lamersetal98}). 
This term summarizes the presence of strong Balmer 
emission lines, narrow permitted and forbidden low-excitation emission lines
of  \FeII, [\FeII] and [\OI], and in particular a strong 
near to mid-IR excess. 
It is attributed to hot circumstellar dust ($T_{\rm dust} \sim 
1000$\,K) and is a distinguishing characteristic with respect to
other classes of peculiar emission-line stars.  The presence of dust requires 
regions of high density and a temperature low enough to allow dust
condensation. A recent review of the properties of this
still enigmatic class of emission-line stars was given by Zickgraf 
(\cite{Zickgraf98}) during the first workshop dedicated entirely to this
type of stars (Hubert \& Jaschek \cite{HubertJaschek98}). 
It was shown that although \be -type stars share the mentioned properties,
indicating very similar physical conditions in their circumstellar environments 
with regard to temperature, density, and velocity, they form by no means 
a homogeneous group. 
Rather, they comprise a variety of
object classes with vastly differing evolutionary stages of low, medium and
high mass stars. To account for the diversity of 
intrinsic classes Lamers et al.
(\cite{Lamersetal98}) suggested a new classification scheme for \be -type stars
including \be\ supergiants (sg\be ), Herbig-type \be\ stars (HAe\be ), 
compact planetary nebulae (cPN\be ), and certain symbiotic objects (sym\be ). 
A large number of \be\ stars are, however, not yet classified. Lamers et al. 
summarized them in the group of unclassified \be -type stars (uncl\be ).  
The relation of \be\ stars to classical Be stars, often a matter of
confusion, was discussed recently by Zickgraf (\cite{Zickgraf00}).

The common property of \be -type stars of all sub-types seems to be the presence 
of non-spherical circumstellar environments. 
Polarimetry and spectro\-polarimetry of galactic as well as of Magellanic Cloud 
\be\ stars clearly demonstrated that independent of the \be\ subgroup the 
scattering particles in the circumstellar envelopes are  distributed 
non-spherically (e.g. Barbier \& Swings \cite{BarbierSwings82}, Zickgraf 
\& Schulte-Ladbeck \cite{ZS89}, Magalhaes \cite{Magalhaes92}, Schulte-Ladbeck et al.
\cite{SchulteLadbecketal94}, Oudmaijer et al. \cite{Oudmaijeretal98}, 
Oudmaijer \& Drew \cite{OudmaijerDrew99}). The most likely configuration is 
disk-like as suggested e.g. by Zickgraf et al. (\cite{Zickgrafetal85}, 
\cite{Zickgrafetal86},\cite{Zickgrafetal89})  based on spectroscopic observations 
in the 
optical wavelength region 
and in the satellite UV of \be\ supergiants in
the Magellanic Clouds (MCs). These observations strongly suggested that 
the stellar winds can be described by a two-component model. 
In this picture a cool and dense equatorial wind emerging from a single star is
responsible for the  formation of the narrow low-excitation emission lines. 
It is also supposed to be the site of dust formation. 
The polar region is dominated by a hot and fast expanding OB star wind with the high wind
velocities observed normally for stars of this type.  
A similar model had been proposed earlier by Swings 
(\cite{Swings73a}) for the galactic \be -type 
star HD\,45677 also based on spectroscopic observations.  
In contrast to the post-main sequence MC sg\be s it seems 
to be a (near) main-sequence object. Likewise, the pre-main sequence
Herbig Ae/Be stars are supposed to
possess circumstellar disks.

Disk-like circumstellar environments could also be caused by binarity. 
Apart from objects belonging to the subclass of sym\be\ several \be\ stars
have in fact been shown to be components of a binary system. 
In the SMC two \be\ supergiants, \object{Hen\,S18} and \object{R\,4}, were found 
to possess lower mass companions (Zickgraf et al. \cite{Zickgrafetal89}, 
\cite{Zickgrafetal96}). Likewise, in the Milky Way a couple of \be\ stars were found 
to be binaries, e.g. 
MWC\,623 (Zickgraf \& Stahl \cite{ZickgrafStahl89}), \object{AS\,381} 
(Miroshnichenko et al. 
\cite{Miro02a}), and CI Cam (= MWC\,84). Further instances are possibly MWC\,349A  
(Hofmann et al. \cite{Hofmannetal02}) and MWC\,342 (Miroshnichenko \& Corporon 
\cite{MP99}). It is, however, not clear whether in these objects the \bephen\ itself 
is actually caused by their binary nature. For some objects this seems not to be the 
case. In Hen\,S18, R\,4, and MWC\,623 the \bephen\ can be 
ascribed  to the B star component in the binary systems. These \be\ stars behave 
like  single stars (Zickgraf et al. \cite{Zickgrafetal89}, 
\cite{Zickgrafetal96}, Zickgraf \cite{Zickgraf01}). AS\,381 on the other hand shows signs of mass transfer  
suggesting that interaction could play a role in the occurence of the \bephen\ in 
this object (Miroshnichenko et al. 
\cite{Miro02a}). At this time the role of binarity is thus controversial. 

Spectroscopic studies showed that the low-excitation lines attributed to the
disks are narrow and thus indicative for low wind velocities in the line forming
region. Typically, line widths (FWHM) of the order of less than 
$\sim$100\,\kms\ to 300\,\kms\ are
observed (e.g. Swings \& Andrillat \cite{SwingsAndrillat81}, 
Zickgraf et al. \cite{Zickgrafetal86}). 
Given the early spectral types of the underlying stars such small wind
velocities are unusual.

In the case of stars
viewed edge-on the direct investigation of the velocity structure of the 
disk winds is possible by studying
absorption lines formed in the disk. This method was used by Zickgraf et al. 
(\cite{Zickgrafetal96}) to study three \be\ supergiants in
the MCs using satellite UV spectroscopy. The observations 
of UV resonance lines showed that the disk winds are in
fact very slow, at least in the case of massive supergiants. The expansion
velocities measured were of the order of 70-100\,\kms , i.e. typically a
factor of 10 less than usually observed for stars of similar spectral type. 
This may also hold for members of other \be\ star classes.

For viewing angles
deviating from edge-on one can make use of the low-excitation emission lines 
to study the kinematics of the disk winds.
Of particular interest are lines from forbidden 
transitions because they are optically thin. Therefore radiation transfer
does not complicate the interpretation of the line intensities and profiles. 
Furthermore, the forbidden lines should form at a large distance from the
central star. Hence, in the case of a radially 
accelerated outflow (as e.g. the usually adopted $\beta$-type velocity 
law) the radial velocity component in the line forming region 
should have reached the terminal wind speed. 
Because of the small velocities involved the investigation of the 
emission-line profiles requires high spectral resolution. If one aims at 
a resolution of about 1/10 of the terminal velocity a spectral resolution of about 
$\sim5-10$\,\kms\ is necessary for the wind velocities 
measured e.g. by Zickgraf et al. (\cite{Zickgrafetal96}) for \be\ supergiants. 

In order to study the disk winds using emission-line profiles 
a sample of galactic \be -type stars listed in  Table \ref{spectype} was observed with high
spectral resolution. In Sect. \ref{obs} the observations are described. 
The observed line profiles are described in Sect. \ref{res}. The density conditions in the
line formation region of the forbidden lines are discussed in Sect. \ref{condition}. The
role  of rotation and expansion is investigated in Sect. \ref{rotation}. In Sect. 
\ref{thinprofile} model calculations of optically thin line profiles are presented 
and compared with the observed lines. Finally, conclusions are given in Sect. 
\ref{sum}. The Appendix contains the observational data in Sects. \ref{parameter} 
and \ref{helio}, and  remarks on individual stars in Sect. \ref{remarks}.
An atlas of the high-resolution spectra is presented in Sect. 
\ref{plots}\footnote{Figs. \ref{prooi}  to \ref{prohe66} are
available  only electronically.}. 

\begin{table}
\caption[]{Observed sample of \be -type stars.
References for spectral types are: 
WS85 = Wolf \& Stahl (\cite{WolfStahl85}), 
McG88 = McGregor et al. (\cite{McGregoretal88}), 
WW89 = Winkler \& Wolf (\cite{WinklerWolf89}),
LeB89 = Le Bertre et al. (\cite{LeBertreetal89}),
Th\'e94 = Th\'e et al. (\cite{Theetal94}),
Sw73 = Swings (\cite{Swings73a}),
C99 = Clark et al. (\cite{Clarketal99}),
Lei77 = Leibowitz \cite{Leibowitz77}),
L98 = Lamers et al. (\cite{Lamersetal98}),
Isr96 = Israelian et al. (\cite{Israelianetal96}),
Drew97 = Drew et al. (\cite{Drewetal97}).
}
\begin{tabular}{lll}
\noalign{\smallskip}    
\hline
\hline
\noalign{\smallskip}    
star             &  spec. class.  & references \\
\noalign{\smallskip}    
\hline
\noalign{\smallskip}    
\object{MWC\,17} & uncl\be  & L98        \\
        & (sym\be , cPN\be ?) &(L98, Lei77)\\
\object{MWC\,84} (CI Cam) & sgB[e], X-ray binary  & C99       \\
        &(uncl\be )&(L98)  \\
\object{MWC\,137} & HAE\be & Th\'e94        \\
\object{MWC\,297} & HAE\be , B1.5Ve&  Drew97       \\
\object{MWC\,300} & sg\be &  WS85       \\
\object{MWC\,342} & uncl\be  & L98       \\
\object{MWC\,349A} & uncl \be &  L98      \\
\object{MWC\,645} & uncl\be  & L98       \\
\object{MWC\,939} & uncl\be &  L98       \\
\object{MWC\,1055} &uncl\be &         \\
\object{HD\,45677} &HAE\be, B2V[e] & Sw73, Isr96, L98     \\
\object{HD\,87643} &sg\be &  McG88     \\
\object{Hen\,230} & uncl\be &       \\
\object{Hen\,485} &  uncl\be &     \\
\object{Hen\,1191} & cPN\be & LeB89     \\
\object{CD$-24\degr5721$} &uncl\be &   \\
\object{CPD$-57\degr2874$} &sg\be & McG88 \\
\object{CPD$-52\degr9243$} &sg\be & Sw81, WW89 \\
\noalign{\smallskip}    
\hline
\noalign{\smallskip}    
\end{tabular}
\label{spectype}
\end{table}

\section{Observations and data reduction}
\label{obs}

\begin{table} 
\caption[]{Journal of observations.} 
\begin{tabular}{lccl} 
\noalign{\smallskip}     
\hline 
\hline 
\noalign{\smallskip}    
date & wavel. range & spectral res. & instrument \\
     & [\AA ]           &  $R = \lambda / \Delta\lambda$ &            \\ 
\hline 
\noalign{\smallskip}    
Dec. 7, 1986  & 4545 - 4581 & 55\,000 & CES \\
Dec. 8, 1986  & 6541 - 6598 & "& CES \\
Dec. 9, 1986  & 5862 - 5913 & "& CES \\
              & 6275 - 6325 & "& CES \\
Dec. 10, 1986 & 4273 - 4307 & "& CES \\
Sep. 9, 1987  & 6535 - 6600 & 23\,000 & CA coud\'e \\
Sep. 10, 1987 & 6650 - 6720 & " & CA coud\'e \\
Sep. 11, 1987 & 6280 - 6350 & " & CA coud\'e \\
Sep. 11, 1987 & 6574 - 6605 & 45\,000 & CA coud\'e \\
Sep. 12, 1987 & 6290 - 6325 & "  & CA coud\'e \\
Sep. 13, 1987 & 5868 - 5903 & " & CA coud\'e \\
Sep. 14, 1987 & 7147 - 7182 & " & CA coud\'e \\
Mar. 27, 1988 & 6275 - 6325 & 55\,000& CES \\
Mar. 28, 1988 & 7134 - 7178 & "& CES \\
Mar. 29, 1988 & 6543 - 6600 & "& CES \\
Mar. 30, 1988 & 6538 - 6595 & "& CES \\
              & 6431 - 6484 & "& CES \\
Mar. 31, 1988 & 6538 - 6595 & "& CES \\
              & 5861 - 5912 & "& CES \\
              & 6275 - 6325 & "& CES \\
              & 7134 - 7178 & "& CES \\
Jun. 19, 2000 & 3950 - 7550 & 34\,000& FOCES \\
Feb. 19\&22, 2002 & 3950 - 7550 & 34\,000& FOCES \\
\hline 
\noalign{\smallskip}    
\label{journal}
\end{tabular}
\end{table}

\begin{table*}[tbh]
\caption[]{Lines observed with CES in 1986 ($+$) and in 1988 ($\times$).} 
\begin{tabular}{lccccccc} 
\noalign{\smallskip}    
\hline
\hline 
\noalign{\smallskip}    
star             &\ha\ +[\NII ]  & {[\OI ]}      &{[\FeII ]} &{[\FeII ]}&\FeII &\FeII       &\HeI +\NaI\,D\\
                 &              & $\lambda$6300\AA\  &$\lambda$7155\AA\ 
		 &$\lambda$4287\AA\ &$\lambda$6456\AA\ &$\lambda$4549/56\AA\ & $\lambda$5876\AA\  \\
\noalign{\smallskip}    
\hline
\noalign{\smallskip}    
MWC\,939          &$\times$     &$\times$        &$\times$   &          &$\times$ &          &            \\
Hen\,230          &$\times$     &$\times$        &$\times$   &          &$\times$ &          &           \\
Hen\,485          &$+\times$&$+\times$&$\times$&$+$ &$\times$ &$+$&$\times$ \\
Hen\,1191         &$\times$&$\times$&$\times$&         &$\times$ &        &          \\
CD$-24\degr$5721  &$+$     &$+$        &           &$+$    &         &$+$&$+$ \\
CPD$-52\degr$9243 &$\times$&$\times$&$\times$&         &$\times$ &        &$\times$ \\
HD\,45677         &$\times$&$\times$&$\times$&         &$\times$ &        &        \\
HD\,87643         &$+\times$&$\times$&$\times$&         &$\times$ &$+$&$\times$        \\
CPD$-57\degr$2874 &$\times$&$\times$&$\times$&         &$\times$ &        &$\times$  \\
\noalign{\smallskip}    
\hline
\noalign{\smallskip}    
\end{tabular}
\label{cesobs}
\end{table*}

\begin{table*}[tbh] 
\caption[]{Lines observed  at Calar Alto Observatory.
Coud\'e observations with a resolution of 45\,000 are indicated by the letter "h",
coud\'e observations obtained with the lower resolution of 23\,000 are indicated 
by "m". Supplementary observations with FOCES are denoted by the letter ``F''.} 
\begin{tabular}{lcccccccccccc} 
\noalign{\smallskip}    
\hline
\hline 
\noalign{\smallskip}    
star     &\ha\ +[\NII ] &[\NII ]    & {[\OI ]}       &{[\FeII ]}  & \HeI +\NaI\,D & \HeI \\
         &              & $\lambda$6583\AA\ & $\lambda$6300\AA\  &$\lambda$7155\AA\ 
		 &$\lambda$5876\AA\& $\lambda$6678\AA\ \\
\noalign{\smallskip}    
\hline
\noalign{\smallskip}    
MWC\,17  &m             &h          & h              &h           & h& m\\
MWC\,84  &m              &h          & m              &h           & h& m\\
MWC\,137 &m              &  F       & h              &F           & h& m \\
MWC\,297 &m              &          & h             &           & h& \\
MWC\,300 &m             &h          & h              &h           & h&  \\
MWC\,342 &m              &  F         & h              &h           & h& \\
MWC\,349A&m             &h          & h              &h           & h&  \\
MWC\,645 &m             &           & h             &h           & & m \\
MWC\,939 &m             & h         &               &h           & F &  F\\
MWC\,1055&m             & F          &  h              &     F      & F & m \\
\noalign{\smallskip}    
\hline
\noalign{\smallskip}    
\end{tabular}
\label{caobs}
\end{table*}

The spectroscopic observations were carried out in 1986 and 1988 with the Coud\'e
Echelle Spectrometer (CES) at the 1.4\,m CAT at ESO, La Silla, and in 1987 with 
the coud\'e  spectrograph at the 2.2\,m telescope at the Centro Astronomico 
Hispano Aleman (CAHA) on Calar Alto, Spain. 
For a few stars with incomplete coud\'e data  the observations were supplemented 
by echelle spectra obtained with FOCES at Calar Alto Observatory in June 2000 and
February 2002.
The journal of observations is given in Table \ref{journal}.

Due to the small spectral coverage of about $\simeq30-60$\,\AA\ provided by the  
coud\'e spectrographs strong emission lines characteristic for \be -type stars 
were selected and the observed wavelength ranges adjusted around these lines. In Tabs. \ref{cesobs} and 
\ref{caobs} the observed lines are listed for each studied object. 
During the 1987 observing run on Calar Alto the northern \be -type star MWC\,623 
was included in the sample. The results on this star have been presented 
already by Zickgraf \& Stahl (\cite{ZickgrafStahl89}) and Zickgraf
(\cite{Zickgraf01}) and are therefore omitted here.

The CES spectra  were collected during two 
campaigns in November 1986 and March 1988. The short camera of the 
spectrograph was equipped with 
a RCA CCD (ESO CCD \#8, 640$\times$1024 pixels, 15\,$\mu$m pixel
size). For details on the instrumentation see Dekker et al. (\cite{Dekker86}). 
The resulting (measured) spectral resolution was $R = 55\,000$, corresponding 
to a velocity resolution of $\Delta v = 5.5$\,\kms .   

The coud\'e  observations on Calar Alto were obtained with the $f/12$ camera 
of the coud\'e spectrograph equipped with a RCA CCD chip (1024$\times$640 pixels, 
15\,$\mu$m pixel size). Most spectra were observed with a 
linear dispersion of 2.2\,\AA\,mm$^{-1}$. A few were obtained with  4.5\,\AA\,mm$^{-1}$. The lower dispersion was
used during nights with reduced meteorological quality mainly for the observation 
of \ha . With a slit width of 0.5\arcsec\ on the sky the  projected slit on the 
chip had a width of 4 pixels. In order to improved the S/N ratio  two 
pixels could therefore be binned in the direction of the dispersion without loss
of resolution. The resulting measured
spectral resolution for the two linear dispersions used was about 45\,000 
and 23\,000, respectively, corresponding to a velocity resolution of 
7\,\kms\ and 13\,\kms , respectively. Another two pixels were binned perpendicular 
to the direction of dispersion in order to increase the $S/N$ ratio. 

Supplementary observations were obtained with the echelle 
spectrograph FOCES (cf. Pfeiffer et al. \cite{Pfeifferetal98}) at the 2.2\,m 
telescope of Calar Alto Observatory in June 2000, and in February 2002. 
The spectrograph was coupled to the telescope with the
red fibre. The detector was a 1024$\times$1024 pixel Tektronix 
CCD chip with 24\,$\mu$m pixel size. With a diaphragm diameter of 
200\,$\mu$m and an entrance slit width of 180\,$\mu$m a spectral resolution of
34\,000 was achieved, i.e. 9\,\kms . A full discussion of the FOCES spectra 
will be given elsewhere 
(Zickgraf 2003, in preparation). 
Here only the lines 
observed also with the coud\'e spectrographs will be considered.

During all observing campaigns wavelength calibration was obtained with  
Th-Ar lamps. For flat fielding built-in lamps were used. The coud\'e spectra were 
reduced by application of standard procedures (bias subtraction, flat-fielding, 
wavelength calibration, normalization) of the ESO-MIDAS image processing 
software package, context {\em longslit}. For the FOCES data the ESO-MIDAS 
context {\em echelle} was used. All spectra were finally rebinned to heliocentric 
wavelengths.

The spectra in the red spectral region are strongly affected by narrow
telluric absorption features. To correct for these lines, the normalized 
spectra were divided by the normalized spectrum of a hot comparison star with a 
line free continuum or with possible photospheric lines removed during the normalization 
procedure. For the \ha\ lines the correction spectrum was created from the object
spectra themselves. 
First each  spectrum was smoothed. Then the original spectrum was
divided by the smoothed spectrum. The final correction spectrum was then
created by averaging several of these individual spectra observed during 
the same night as the spectrum to be corrected. 

\begin{figure*}
\includegraphics[angle=-0,width=17cm,height=23.0cm,bbllx=40pt,bblly=47pt,bburx=527pt,bbury=713pt,clip=]{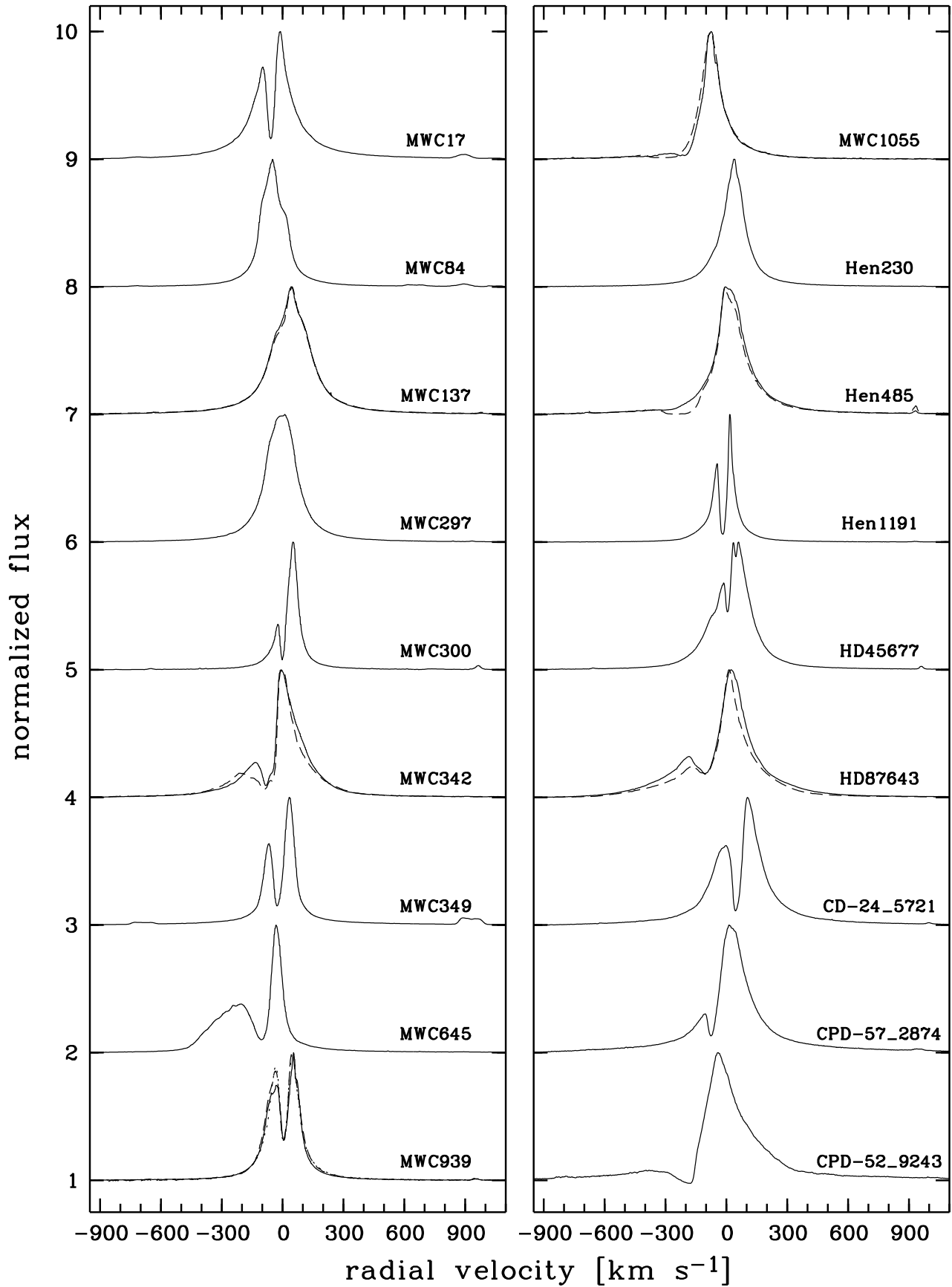}
\caption[]{
Line intensity profiles of \Ha\ as a function of heliocentric radial velocity. 
All lines were normalized
to the peak flux. A few stars were observed more than once. Profile variability was
found in MWC342,  MWC\,939, MWC1055, HD\,87643, and Hen\,485 (see text). 
}
\label{havel}
\end{figure*}

The observed spectral sections  are displayed in the Appendix in 
Figs. \ref{prooi}  to \ref{prohe66} together with remarks on the individual 
objects in Sect. \ref{remarks}. For \Ha\ see Fig. \ref{havel}. 

\begin{figure*}
\includegraphics[angle=-0,width=17.0cm,height=23.4cm,bbllx=34pt,bblly=47pt,bburx=550pt,bbury=740pt,clip=]{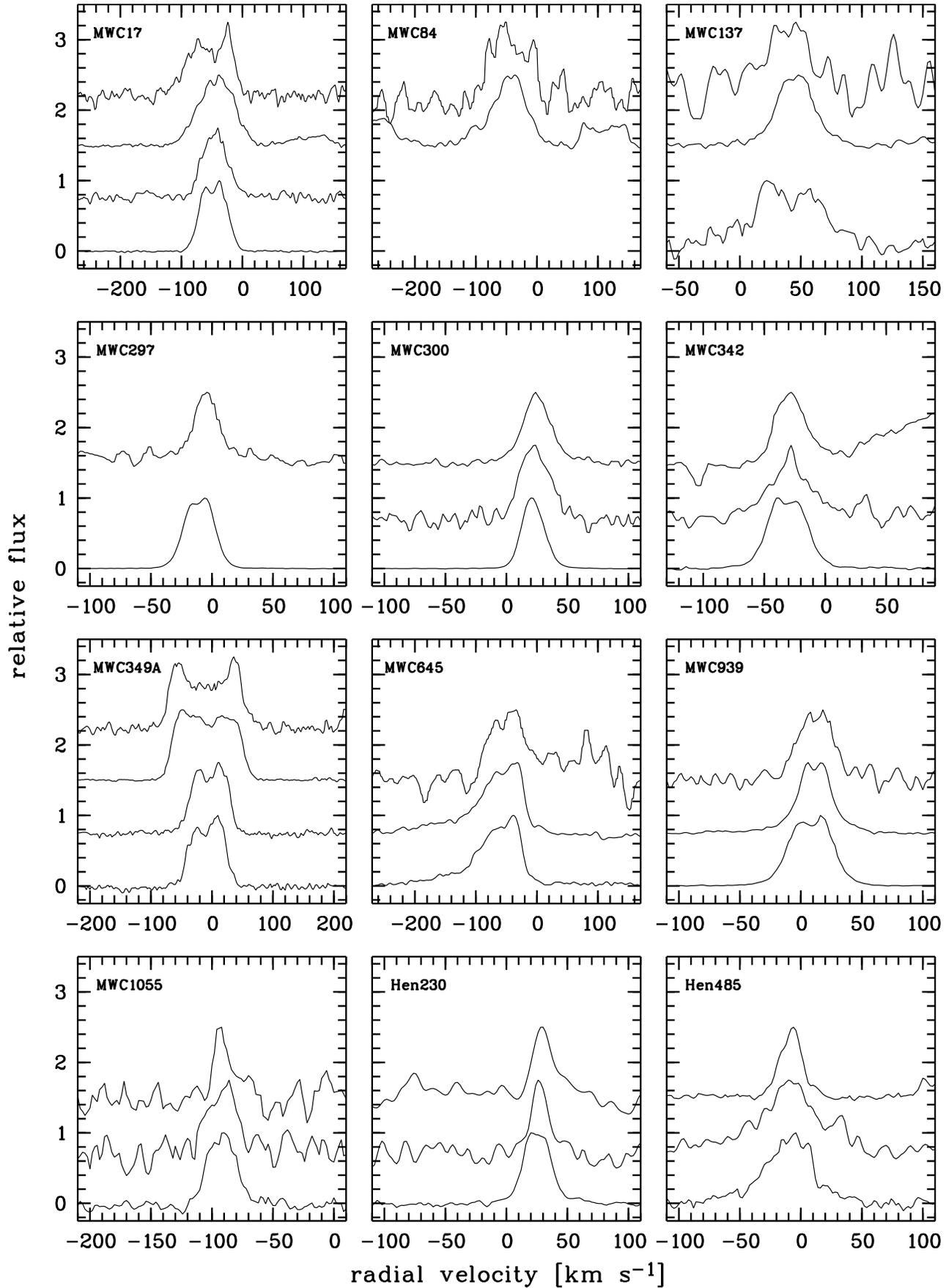}
\caption[]{{\bf a}. 
Line intensity profiles profiles of the forbidden lines as a function of heliocentric 
radial velocity. 
All lines were normalized
to the peak flux. From bottom to top the profiles of [\OI ]$\lambda$6300\AA , 
[\FeII ]$\lambda$7155\AA, [\NII ]$\lambda$6583\AA, and [\SIII ]$\lambda$6312\AA\
are plotted with shifts in relative intensity of 0, 0.75, 1.5, and 2.25, respectively. 
}
\label{forbidden}
\end{figure*}

\begin{figure*}
\addtocounter{figure}{-1}
\includegraphics[angle=-0,width=17.0cm,bbllx=34pt,bblly=390pt,bburx=550pt,bbury=745pt,clip=]{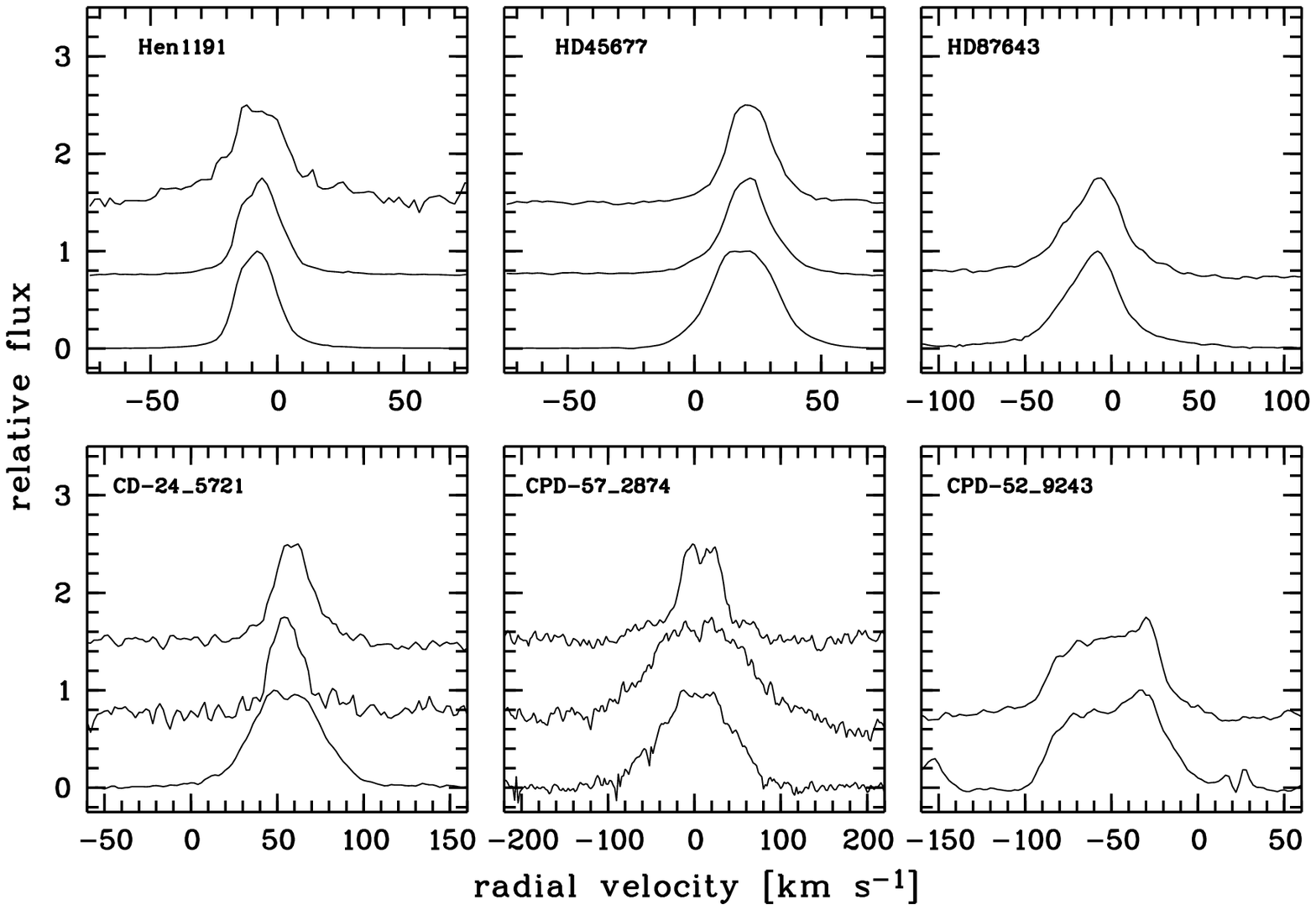}
\caption[]{{\bf b}. 
Line intensity profiles of the forbidden lines as a function of heliocentric radial velocity,
continued.  

}
\end{figure*}

\begin{figure*}[tbh]
\includegraphics[angle=-0,width=17.0cm,bbllx=35pt,bblly=45pt,bburx=534pt,bbury=344pt,clip=]{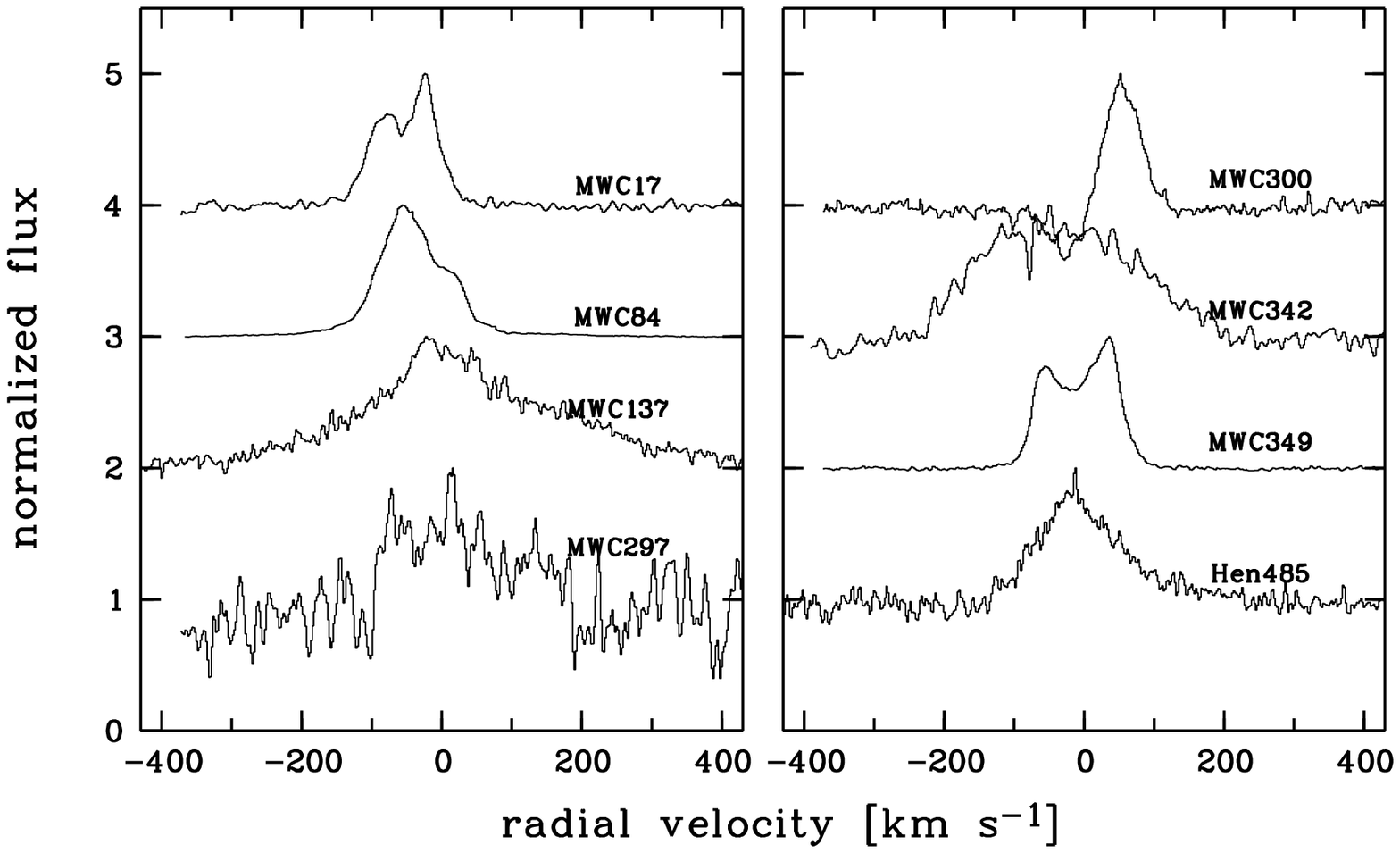}
\caption[]{Line intensity profiles of \HeI$\lambda$5876\AA\ emission line profiles as a 
function of heliocentric radial velocity.  For MWC\,137 the line observed
in 2002 is shown.
}
\label{heivel}
\end{figure*}

\section{Observed line profiles}
\label{res}
In the following the observed characteristics of the line profiles are summarized.
Table \ref{para1} in the Appendix lists relevant line parameters. 
Measurements of heliocentric radial velocities are also listed in the Appendix in  
Table \ref{rad2}.  The  profiles of \Ha\ , of \HeI\ emission lines, and of the 
forbidden lines rebinned on a velocity scale are displayed in Figs. \ref{havel} 
\ref{heivel} and \ref{forbidden}a, b.

The line profiles
can be categorized into four groups:
\begin{itemize}
\item[--] group 1: normal P~Cygni-type line profiles with an absorption 
component reaching below the continuum on the violet side of the profile and 
an emission component on the red side,  
\item[--] group 2: single-peaked pure emission lines without absorption 
components, 
\item[--] group 3: double-peaked emission lines with a central or almost 
central absorption component, or at least an intensity dip on one of the line 
flanks, 
\item[--] group 4: absorption lines. 
\end{itemize}
These profile groups correspond  to Beals types I, V,  III, 
and VII-VIII, respectively,  defined by  Beals (\cite{Beals55}). The profile
types of the observed lines are summarized in Table \ref{proftype}.

\subsection{\ha\ profiles}

\begin{table*}[tbh]
\tabcolsep=7pt
\caption[]{Classification of the line profile types of the programme 
stars into 
groups 1 to 4: 1=  normal P~Cygni profile,  2= single-peaked emission
line, 3= double-peaked emission line, 4 =  absorption line. Additional 
classification codes are: ? = weak line, no further
classification possible, 0 = no line visible, -- = not observed.
For class 3 a minus or plus sign denotes objects with $V/R \le 1.0$ or $V/R > 1.0$,
respectively.
For \NaI\,D no group is listed because of the confusion due to interstellar
absorption components. For this doublet only the presence of emission (``em'') or 
pure absorption (``abs'') is indicated.
} 
\begin{tabular}{lllllllll} 
\noalign{\smallskip}    
\hline
\hline
\noalign{\smallskip}    
star                 &\ha    &[\OI ]&[\FeII ] &[\NII ]   &[\SIII ]&\FeII & \HeI\ &\NaI\,D\\
\noalign{\smallskip}
\hline
\noalign{\smallskip}
MWC\,17              & $3-$  & $3-$ & 2       & $3-$	 & $3-$      & --    &  $3-$	 & abs  \\
MWC\,84              & 2     & ?    & ?       &$3:-$	 & 2:	  & --  &  2	 & em \\
MWC\,137             & 2     & $3+$ &  ?$^f$  &$ 3-^f$   & ?	  & --    &  4$^c$, 2$^f$ & abs   \\
MWC\,297             & 2     & $3-$ & --      & 2$^e$	 & 0	  & --    &  2     & abs  \\
MWC\,300             & $3-$  & 2    & 2       & 2	 & 0	  & --    &  1     & em  \\
MWC\,342             & $3-$  & $3+$ & 2       & 2$^f$	 & 0	  & --    &  2$^d$ & em 	  \\
MWC\,349A            & $3-$  & $3-$ & $3-$    & $3+$	 & $3-$  & --	 &  $3-$     & em  \\
MWC\,645             & $3-$  & $3-$ & $3-$    & 2:$^e$   & 0	  & --    &  -- & --  \\
MWC\,939             & $3-$  & $3-$ & $3-$    & $3-$	 & 0	  & $3+$	 &  4$^f$ & em$^f$  \\
MWC\,1055            & 1:    & $3:-$& 2$^f$   & 2$^f$	 & 0	  & 2$^f$ &  4$^f$ & em$^f$   \\
Hen\,230             & 2     & 2    & 2       & 2	 & 0	  &  2    &  --    & --   \\
Hen\,485             & 1:-2  &2     & 2       & 2	 & 0	  &  2:   &    2   & em  \\
Hen\,1191            & $3-$  & 2    & 2       & $3:+$	 & 0	  &  2    &   --   & --    \\
HD\,45677            & $3-$  & $3-$ & 2       & 2	 & 0	  & $3+$      &  --    & --  \\
HD\,87643            & $3-$  & 2    & 2       & 0	 & 0	  & 2	  &  4     & em  \\
CD$-24\degr$5721     & $3-$  & $3+$ & 2       & $3:-$	 & 0	  & 4	  &  4     & em:  \\
CPD$-57\degr$2874    & $3-$  & $3:-$& 3:$-$   & $3+$	 & 0	  & $3-$     &  1:    & em   \\
CPD$-52\degr$9243    & 1     & $3:-$& $3:-$   & 0	 & 0	  & 1	  &  4     & em \\
\noalign{\smallskip}
\hline     
\noalign{\smallskip}    
\multicolumn{1}{l}{$a$: \FeII$\lambda$6586}&
\multicolumn{4}{l}{$c$: \HeI$\lambda$6678 in emission}\\
\multicolumn{8}{l}{$d$: \HeI$\lambda$6678 possible blue shifted absorption
component}\\
\multicolumn{3}{l}{$e$: observed with $R = 23\,000$}&\multicolumn{4}{l}
{$f$: observed with FOCES}\\
\end{tabular}
\label{proftype}
\end{table*}

A general characteristic of all  \ha\ profiles displayed in Fig. \ref{havel} is that 
they exhibit a narrow  single or split emission component with 
a full width at half maximum (FWHM) of 
about 3-5\,\AA , i.e. $\sim150-250$\,\kms , and broad wings on both sides of the 
emission component extending up to typically $\sim20-25$\,\AA , i.e. 
$\sim1000$\,\kms . These wings are generally ascribed to electron scattering (e.g.
Zickgraf et al. \cite{Zickgrafetal86}).

Only one star, \object{CPD$-52\degr$9243}, shows a P~Cygni profile  which resembles the ``normal''
(group 1) profile type.  \object{Hen\,485} in 1988 and \object{MWC\,1055}
may also be classed with group 1, although the absorption components do not reach 
below the continuum level. 

The \ha\ profiles of four stars, \object{MWC\,84}, \object{MWC\,137}, 
\object{MWC\,297}, and \object{Hen\,230}, fall into group 2, which 
exhibits pure emission line profiles. 
The FWHM is 
of the order of 3-5\,\AA . Note, however, 
that the lines are not symmetric. The asymmetry is particularly 
pronounced  in the case of \object{MWC\,84}. 

Most of the investigated stars belong to group 3 exhibiting double-peaked 
\ha\ emission lines. In all of the eleven cases of this group the blue 
emission peak is weaker than the red peak. In no case the central absorption 
components reaches below the continuum level. For the peculiar line profiles 
of \object{HD\,45677} and  \object{MWC\,645} see Sect. \ref{remarks}. 

For several stars \ha\ was observed more than once. The profiles are plotted in Fig. 
\ref{havel}: MWC\,137 in 1987 (solid line)
and in 2002 (dashed line); MWC\,342 in 1987 (solid line)
and in 2000 (dashed line); MWC\,939  in 1987 (solid line)
and  1988 (dashed line), the profile observed in 2000 is indistinguishable that of 1988; 
MWC1055 in 1987 (solid line) and 2000
(dashed line); HD\,87643  in 1986 (solid line) and 1988 
(dashed line); Hen\,485  in 1986 (solid line) and 1988 (dashed line). 
For MWC\,137 the profiles of 1987 and 2002 are nearly indistinguishable.

\subsection{Metal lines}
\subsubsection{[\OI ] lines}
The spectral section with the [\OI ]$\lambda$6300\AA\ line is displayed in Fig. 
\ref{prooi}. It also contains the line
[\SIII ]$\lambda$6312\AA\ (s. Sect. \ref{sulfur3}) and a line of neutral 
magnesium, \MgI$\lambda$6318\AA . This line is present in all objects.

The line strength of [\OI ]$\lambda$6300\AA\ differs widely from object to object, the two extremes
being \object{MWC\,84} and \object{Hen\,1191}. Whereas in
\object{MWC\,84} the line peak is at a 5\% level above the continuum
in \object{Hen\,1191} the [\OI ] emission line is 
extremely strong reaching as much as 140 times the continuum flux. 

\begin{figure}[tbh]
\resizebox{\hsize}{!}{\includegraphics[angle=0,bbllx=20pt,bblly=255pt,bburx=370pt,bbury=600pt,clip=]{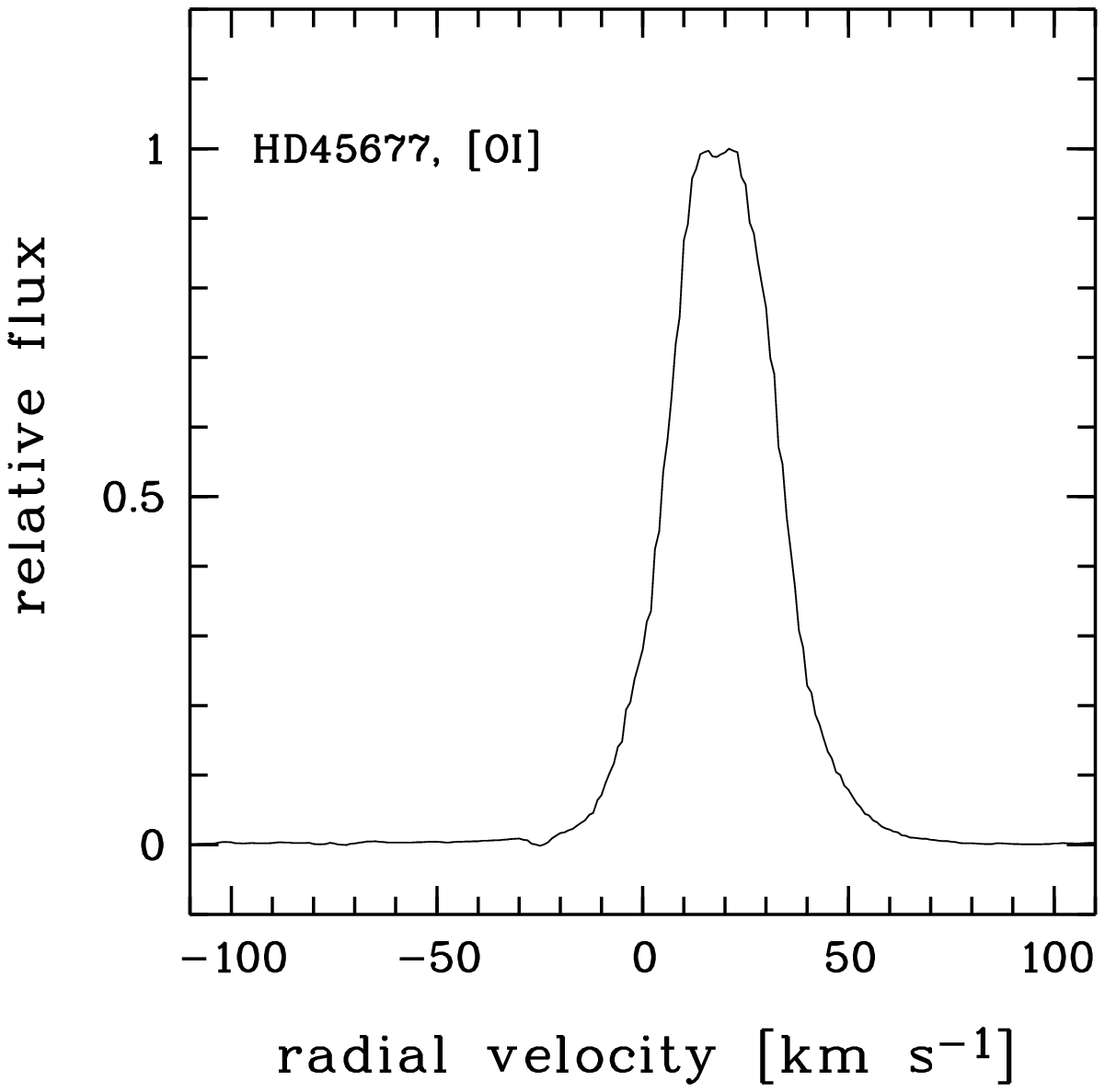}}
\caption[]{[\OI ] $\lambda$6300\AA\ profile of HD\,45677. The double peak is just
resolved with a peak separation of 6\,\kms .
}
\label{hd45oi}
\end{figure}

Thirteen of the 18 objects exhibit double-peaked [\OI ] profiles. In some cases
the line splitting is weak yet detected, as e.g. in HD\,45677 (cf. Fig.\ref{hd45oi}) and \object{MWC\,297}, 
or at least indicated as in MWC\,1055 and CPD$-52\degr$9243. 
In 8 cases the flux of the blue peak is weaker than that of the red peak. 
In 3 stars the blue peak is stronger, i.e. \object{MWC\,137}, \object{MWC\,342},
and \object{CD$-24\degr$5721}. In HD\,45677 the two peaks are equally strong. 
\object{CPD$-57\degr$2874} exhibits a nearly
flat-topped profile which was classified type 3 due to the weak flux increase at the
blue and red side of the nearly flat top. The [\OI ] profile of \object{MWC\,645} 
is strongly asymmetric. The remaining objects have single-peaked profiles.
Note that in the [\OI ] profile of Hen\,230 the line top is sloping to the red side. 

\subsubsection{[\NII ] lines} 
The wavelength region around  [\NII ]$\lambda$6583\AA\ is displayed in Fig. \ref{pronii}. 
In two stars, \object{CPD$-52\degr$9243} and \object{HD\,87643}, 
the [\NII ] line is  absent. 
The lines visible in the spectra of these stars 
around $\lambda$6585\,\AA\  are probably due to \FeII\,$\lambda$6586.69\AA .
Heliocentric radial velocities are $v_{\rm rad} = -72$\,\kms 
for \object{CPD$-52\degr$9243} 
and $v_{\rm rad} = -18$\,\kms\ for \object{HD\,87643}. 
In \object{MWC\,1055} [\NII ] is only weakly discernible.  
The majority of stars, however, exhibits clearly visible, in many
cases strong, [\NII ]$\lambda$6583\AA\ emission. Eight stars show double-peak
profiles. Hen\,1191 shows a sloping line top inclined towards the red side 
similar to the [\OI ] line of Hen\,230, however with a weak peak on the blue edge.
Due to this feature the line was classified type 3. Eight stars exhibit 
single-peaked emission lines. However,
in two of these cases, respectively, the profiles were observed with the 
lower resolution of 23\,000 and 34\,000. 
They are labelled ``$e$'' and "f" in Table \ref{proftype}. 
Note that each of these stars shows a double-peaked (type 3) 
[\OI ] profile. 

\subsubsection{[\FeII ]}
\label{fe71}
The spectral section with [\FeII ]$\lambda$7155\AA\ is shown in Fig. \ref{profe71}. 
Note for that \object{CD$-24\degr$5721} the forbiddden lines 
[\FeII ]$\lambda\lambda$4287, 4276\AA\ were observed instead of the red line 
(Fig. \ref{profe42}). In 
five 
cases the [\FeII ] profiles are 
double-peaked similar to [\OI ]. However, contrary to [\OI ] the majority of 
objects, 
i.e. 10,  
exhibits single-peaked profiles, 2 of them on a  resolution
level of 9\,\kms .  

\subsubsection{[\SIII ] lines} 
\label{sulfur3}
The [\SIII ]$\lambda$6312\AA\ lines are displayed in Fig. \ref{prooi}. 
Only four stars exhibit this higher-excitation emission line, i.e. 
\object{MWC\,17}, \object{MWC\,84}, \object{MWC\,137}, and \object{MWC\,349A}. 
In \object{MWC\,137} it is very weak and not much can be said about its
profile. The [\SIII ] line of \object{MWC\,84} is also weak. 
The strongest [\SIII ] was found in \object{MWC\,349A}. 

\subsubsection{\FeII\ lines}
\label{fe64}
For only half of the sample permitted \FeII\ lines were observed, mostly  
\FeII$\lambda$6456\AA , but also \FeII\ lines around 4550\AA\ for a few stars instead.
The wavelength region around \FeII$\lambda$6456\AA\ is displayed in Fig. \ref{profe64}. 
Three of the observed stars exhibit single-peak
emission lines. Four stars show  double-peaked profiles. For Hen\,485 the
double-peak structure is only weakly indicated.
CPD$-52\degr 9243$ is the only star showing a P\,Cygni
profile of group 1. CD$-24\degr 5721$ is exceptional. This star shows narrow
absorption lines (Fig. \ref{profe45}). 
The lines identified in the observed spectral sections of this star are 
listed in Table \ref{cd24feii}.

\subsubsection{\NaI\,D lines}
\label{nad}
The lines of the \NaI\ D doublet are shown in Fig. \ref{prohe58}. The spectral section
shown in this figure also contains the line of  \HeI$\lambda$5876\AA\ (see below).
Most stars clearly show circumstellar \NaI\ emission. Only 4 of
the 14 observed stars do not show an emission component of the doublet.
In most cases the absorption components are blends of multiple narrow absorption lines
which are very likely mainly due to interstellar absorption. This makes it difficult 
to detect circumstellar absorption features.  Because of this problem only
the overall appearance of emission or absorption is listed in  Table \ref{proftype}. 
Exceptions are  CPD$-52\degr 9243$ and 
possibly Hen\,485, cf. Sect. \ref{cpd52} and \ref{hen485}. 
Heliocentric radial velocities of the absorption components are listed in Table 
\ref{natrium}.

\subsection{\HeI\  lines} 
The \HeI$\lambda$5876\AA\ lines are displayed in Fig. \ref{prohe58}. They 
appear in all four varieties of profile types. However, only one star exhibits a 
clear P\,Cyg profile of type 1, namely \object{MWC\,300}. In 
\object{CPD$-57\degr 2874$} an emission component seems to partly 
fill in the absorption component. Two stars show split type 3 profiles and five
stars single-peaked emission profiles. Six stars show an absorption line.
For four stars no observation of \HeI$\lambda$5876\AA\ were obtained.
In MWC\,137 strong variability was found between 1987 and 2002. The \HeI\ line
changed from absorption to emission (cf. Sect. \ref{mwc137}).

\subsection{Summary}
An important result of the observations presented here is the detection of one or more 
double-peaked emission lines in many objects (cf. Table~\ref{proftype}). 
Actually, 15, possibly 16, of the 18 objects show at least one line with a 
double-peaked profile. This profile type is found for both, 
permitted and forbidden lines, but not necessarily for each line of a 
particular star. Eleven of the 18 objects have  split \ha\ profiles. Split 
forbidden lines are found in 13 objects. Twelve stars  exhibit split [\OI ] 
lines. The fraction of split lines  of [\NII ] and [\FeII ] is smaller. 
Only 8 of 18 stars exhibit split [\NII ] lines, and  
5 
of 17 stars have split 
[\FeII ] lines. There are only 2 cases where \ha\ is 
double-peaked, 
but all 
forbidden lines are single-peaked emission lines. These are HD\,87643 and MWC\,300. 
According to Oudmaijer et al. (\cite{Oudmaijeretal98}),
and  Wolf \& Stahl  (\cite{WolfStahl85} and Winkler \& Wolf (\cite{WinklerWolf89}), 
respectively, they belong to the \be\ supergiants and are 
most likely viewed under intermediate to pole-on inclination angles. Note, 
however, that the nature of these stars still is controversially discussed 
(see also Sect. \ref{mwc300}).

A remarkable feature of the double-peaked profiles is that most, i.e., $\sim$85\%, of 
the observed lines have an intensity ratio of the violet to red component of $V/R \le 1$.
Of the 
43 
detected type 3 lines only 8 show a $V/R$ ratio larger than 1. 
These are  6 of 26  forbidden, and 2 of 16  permitted lines (cf. Table \ref{para1}). 
The latter are all \FeII\ lines.

\section{Density conditions in the forbidden-line forming zone}
\label{condition}
The interpretation of the observed line profiles might be complicated by the fact that
the sample of \be -type stars is not homogeneous with respect to the intrinsic object
characteristics. The discussion by Lamers et al. (\cite{Lamersetal98}) showed
that the connection between the different classes of \be -type stars is 
the uniformity of the \bephen\ which calls for invoking a common cause for its 
occurence in different environments. In the following we will therefore take the view 
of looking primarily at the  \bephen\ itself rather than at specific object classes. 

The forbidden-lines in the spectra of \be -type stars are dominated by lines of 
low-excitation ions of neutral or
singly ionized metals. Higher excitation lines like [\SIII ] are rare.
This indicates that the temperature in the line emitting region is about $10^4$\,K
(Lamers et al. \cite{Lamersetal98}). 
The forbidden lines probe the outer low-density zone of the line formation region.
A measure for the maximum density  in this region is the critical
density, $N_{\rm cr}$, for which downward collisional and radiative rates are equal. 
In the approximation of a 2-level ion with  upper level $u$ and lower level $l$ 
it is given by
\begin{equation}
N_{\rm cr} = \frac{A_{ul}}{q_{ul}}
\label{ncrq}
\end{equation}
with  the radiative transition probability $A_{ul}$, and the rate coefficient for 
collisional de-excitation $q_{ul}$ (Osterbrock \cite{Osterbrock89}). 

At $T_{\rm e} = 10^4$\,K the critical density of the neutral line 
[\OI ]$\lambda$6300\AA\  is 3\,10$^6$\,cm$^{-3}$ (e.g. B\"ohm \& Catala 
\cite{BoehmCatala94}). [\NII ]$\lambda$6583\AA\ has a lower critical 
density than [\OI ]$\lambda$6300\AA , $N_{\rm cr} = 8.6\,10^4$\,cm$^{-3}$ (Osterbrock 
\cite{Osterbrock89}). 
For [\SIII ]$\lambda$6312 the critical density is 1.4\,10$^7$\,cm$^{-3}$ 
This value was obtained with the IRAF task {\em ionic} by Shaw \& Dufour 
(\cite{ShawDufour94}). 

For the metastable levels of singly ionized iron giving rise to the observed 
forbidden transitions the critical density can be estimated from Eq. (\ref{ncrq}).
Following Beck et al. (\cite{Becketal90}) this relation can be rewritten as
\begin{equation}
N_{\rm cr} = 3.7\,10^6 \:\frac{g_u\,A_{ul}}{\Omega (u,l)}\:\left(\frac{T}{1000}\right )^{\frac{1}{2}}
\label{ncAul}
\end{equation}
with the collision strength $\Omega (u,l)$, the statistical weight $g_u$  and the 
electron temperature $T$. According to  Viotti (\cite{Viotti76})  $\Omega (u,l)$ is 
given by 
\begin{equation}
\Omega (u,l) \approx 0.2\,\lambda ^4\,g_u\,A_{ul}
\label{omegaul}
\end{equation}
with the wavelength $\lambda$ in microns. For [\FeII ](14F)$\lambda 7155$\,\AA\ 
this leads to $N_{\rm cr} \simeq 2.2\,10^8$\,cm$^{-3}$ for $T_{\rm e} = 10^4$\,K. 
The critical densities are summarized in Table \ref{ncr}.

\begin{table}[tbh]
\caption[]{Critical densities, $N_{\rm cr}$, of the observed forbidden lines 
for an electron temperature of
$T_{\rm e} = 10^4$\,K, and total ionization energies, $\chi$, for the production of the
respective ion (Cox \cite{Cox00}). 
The last line gives $D_0 = N_0/N_{\rm cr}$ for a density $N_0 = 10^{11}$\,cm$^{-3}$ 
at $r = 1\,\rstar $.
½¾
}
\begin{tabular}{lllll}
\noalign{\smallskip}    
\hline
\hline
\noalign{\smallskip}    
ion                   &[\OI ]& [\FeII ]& [\NII ]& [\SIII ]\\
line                   &$\lambda 6300$\,\AA & $\lambda 7155$\,\AA& $\lambda 6583$\,\AA &
		   $\lambda 6312$\,\AA \\
\noalign{\smallskip}    
\hline
\noalign{\smallskip}    
$N_{\rm cr}\,$[cm$^{-3}$] & 3\,10$^6$ & $2.2\,10^8$ & $8.6\,10^4$& $1.4\,10^7$\\
$\chi$\,[eV]              & 0.0 & 7.90 & 14.53 & 33.70\\
$D_0$              & $3.3\,10^4$ & $4.5\,10^2$ & $1.2\,10^6$ & $7.1\,10^3$\\
\noalign{\smallskip}    
\hline
\noalign{\smallskip}    
\end{tabular}
\label{ncr}
\end{table}

The forbidden lines not only differ with respect to the critical density but also have
different ionisation potentials. The ionisation
energy necessary to form  \FeII\ is 7.9\,eV. For \NII\  an energy of 14.53 \,eV is
required. With $\chi = 33.70\,$eV  \SIII\ has the highest ionisation
energy of the observed forbidden lines.
Hence, the forbidden lines probe a density interval of about three 
orders of magnitude, $\sim10^5$-$10^8$\,cm$^{-3}$,
and a range of ionisation  from neutral, [\OI ], to [\SIII ] with an ionisation
potential of $\sim34$\,eV. 

\section{Disk wind: radial expansion vs. rotation}
\label{rotation}
A spherically symmetric and radially expanding wind is expected to form flat-topped 
profiles if the  lines are optically thin. This has already been shown by Beals 
(\cite{Beals31}). The forbidden lines in particular form at large distances from the
central star where the wind  has reached the terminal velocity (see below). A constant velocity
wind is expected to form  box-shaped lines if the emissivity is constant throughout the
emitting volume. 

In the observed sample of \be -type stars there is just one 
case, \cpdsi , where a line profile comes close to flat-topped, however not 
box-shaped. This is the line [\OI ]$\lambda6300$ of this object. The vast majority 
of the observed profiles are clearly different from flat-topped and 
therefore are inconsistent with a spherically symmetric and optically thin 
line formation region. 
Deviations from flat-topped profiles could be produced in the case of
spherical symmetry by additional 
extinction due to dust
distributed evenly throughout the line formation region. 
This has been discussed e.g by
Appenzeller et al. (\cite{AJO84}) for T\,Tauri stars.
However, the profile shape expected for this configuration is not observed in any
of the \be -type stars of the sample  presented here.
The polarimetric 
observations and the forbidden line profiles therefore strongly indicate that the  
\bephen\ is correlated with an anisotropic  distribution of the 
circumstellar matter. 

Split profiles of \ha\ similar to those shown in Fig. \ref{havel} are frequently found
in classical Be stars, although the \ha\ equivalent widths in these stars are
usually much smaller than in \be -type stars  and the underlying photospheric absorption
component is often discernible. The double-peaked Be star profiles are generally 
assigned to a disk-like geometry of the line forming region in connection with
rotation.

Mihalas \& Conti (\cite{MihalasConti80}) discussed the formation of
Beals type III, i.e.  type 3, line profiles in the context of the combination of
rotation and expansion in a disk-like circumstellar environment. Adding expansion 
could in particular explain the  blueshifted absorption components of \ha\ and 
the $V/R$ ratios smaller than 1. It would introduce an asymmetry of the line 
profiles by shifting the central reversals towards shorter wavelengths as 
observed for most \be -type stars. For the forbidden lines, however, this mechanism 
would not work because
the lines are optically thin and therefore absorption does not contribute.
Nevertheless, the combination of expansion and rotation could at least 
explain the observed double-peaked profiles of \ha . 

The double-peaked profiles of the optically thin lines could quite naturally  
be produced in rotating disks as shown e.g. by P\"ollitsch (\cite{Poellitsch81}).
Keplerian disks could for example exist around binary \be\ stars (see Sect. \ref{intro}). 
The profiles calculated by P\"ollitsch display, however, two emission peaks with 
$V/R=  1$ due to the axial symmetry. Profiles  of this type are found only in a 
few cases, e.g. [\SIII ] of MWC\,349A, [\NII ] of \cpdsi , [\NII ] and [\FeII ] of 
MWC\,939, and [\OI ] of \cdvi\ and \cpdsi . 
However, the majority of double-peak lines has $V/R < 1$
including other lines of the mentioned stars.
It is therefore not obvious that rotation
is the likely explanation for the double-peaked profiles.
Rather, the line profiles seem to be determined by radial outflow.

Let us assume as an example a disk-like configuration with a rotational 
velocity $v_0$ at 1\,$R_{\star}$  
of $v_0 = 300$\,\kms\ and a constant
radial expansion velocity of $v_{\rm exp}$. Angular
momentum conservation requires $v_{\rm rot}(r) = v_0\,R_{\star}/r$. 
Hence at a distance of 10 $R_{\star}$ the rotation velocity would have dropped 
to 30\,\kms . At this point an expansion velocity of $v_{\rm exp} \approx$ 
50-100\,\kms\ as
observed for the disks of \be\ supergiants 
would  dominate 
the velocity field. Further out in the disk, at 
$r \ga 100\,R_{\star}$, 
rotation would not play a role anymore. 
In classical Be stars and sg\be s densities  of
about $10^{12\ldots 13}$\,cm$^{-3}$ have been  observed 
for wind zones near the star (e.g. Waters  \cite{Waters86}, 
Zickgraf et al. \cite{Zickgrafetal89}). The density thus
would have dropped to $\la 10^{8\ldots 9}$\,cm$^{-3}$ at $r \ga 100\,R_{\star}$. 
At this density and distance the forbidden lines are formed. 
Therefore the forbidden line zone should be dominated by expansion rather than 
rotation  under the assumptions made above. 
Whereas the forbidden lines are formed in tenuous regions at large distances from the
central star \ha\ is formed in the inner regions of the disk. Here at distances
of $\la 10\,\,R_{\star}$ the velocity field could still be dominated by rotation.

\begin{figure}[tbh]
\resizebox{\hsize}{!}{\includegraphics[angle=-90,bbllx=190pt,bblly=55pt,bburx=545pt,bbury=715pt,clip=]{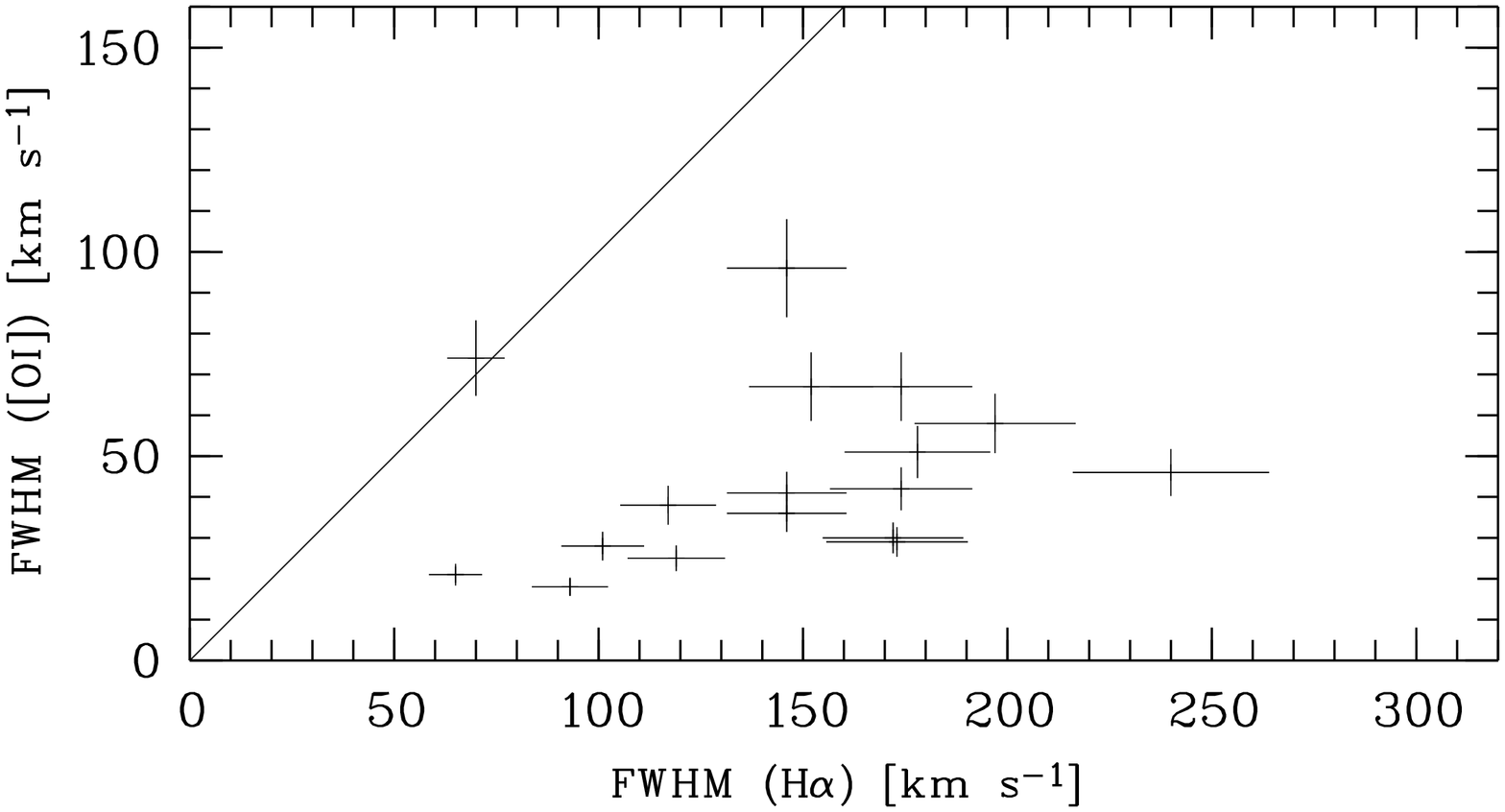}}
\caption[]{ FWHM of [\OI ] vs. \ha .  
Here and in the following figures the  solid line designates a ratio of 
line widths  of 1.
}
\label{fwhm_oiha}
\end{figure}

\begin{figure}[tbh]
\resizebox{\hsize}{!}{\includegraphics[angle=-90,width=8.7cm,bbllx=190pt,bblly=55pt,bburx=545pt,bbury=715pt,clip=]{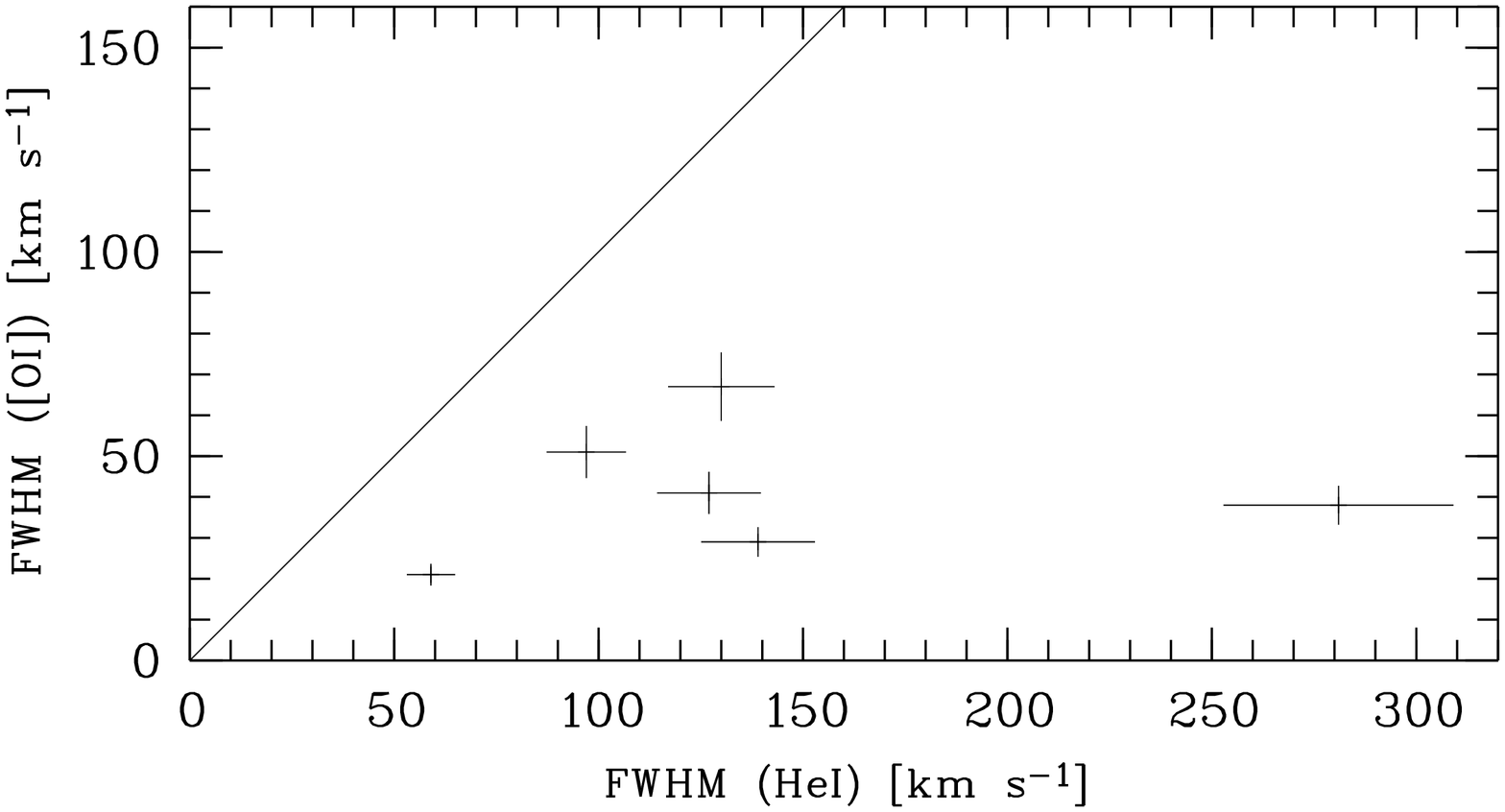}}
\caption[]{ FWHM of [\OI ] vs. \HeI .    
}
\label{fwhm_oihei}

\end{figure}

\begin{figure}[tbh]
\resizebox{0.47\hsize}{!}{\includegraphics[width=8.7cm,angle=-90,bbllx=50pt,bblly=30pt,bburx=555pt,bbury=575pt,clip=]{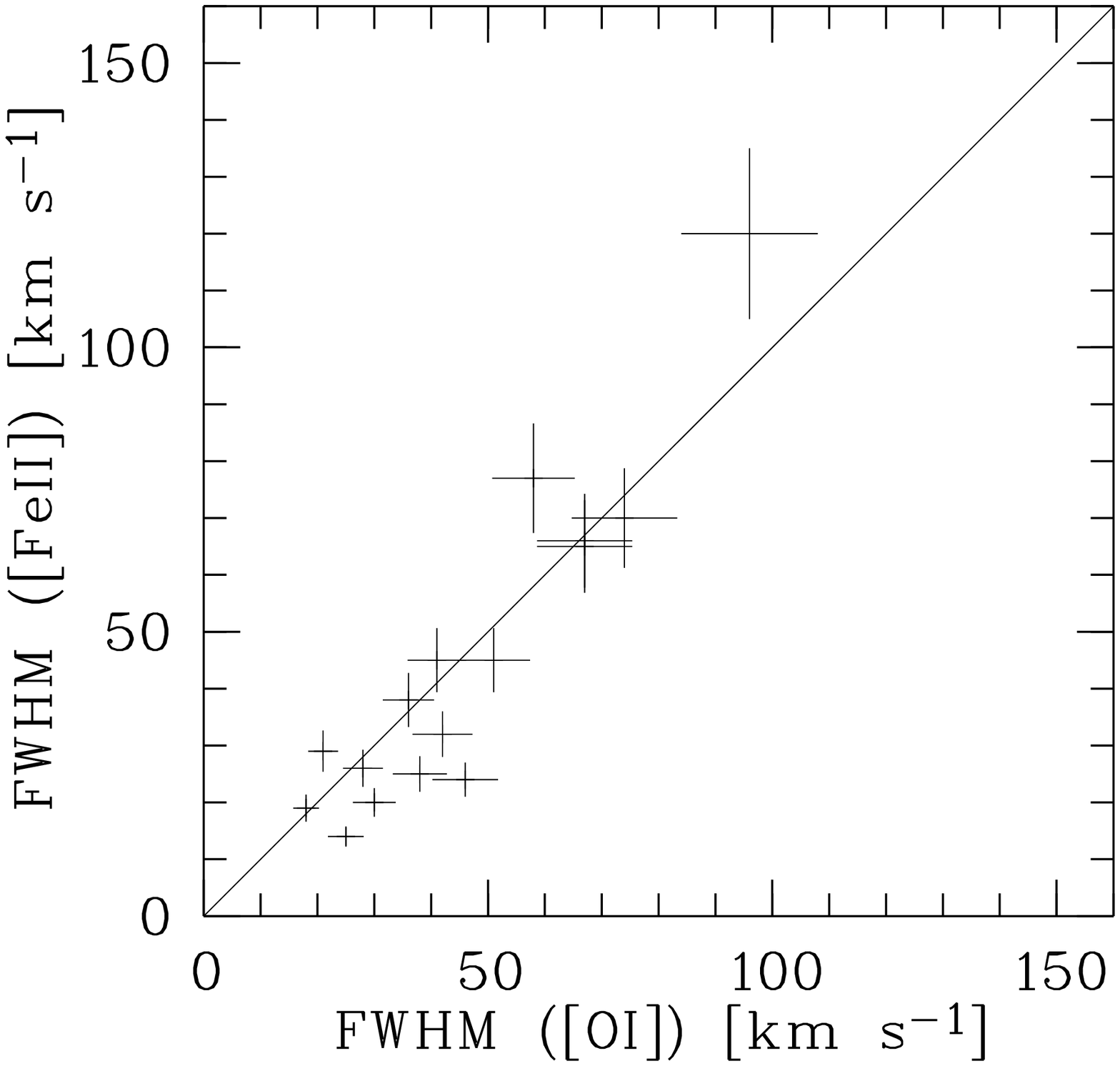}}\hspace{0.02\textwidth}\hfill\resizebox{0.47\hsize}{!}{\includegraphics[width=8.7cm,angle=-90,bbllx=50pt,bblly=45pt,bburx=555pt,bbury=575pt,clip=]{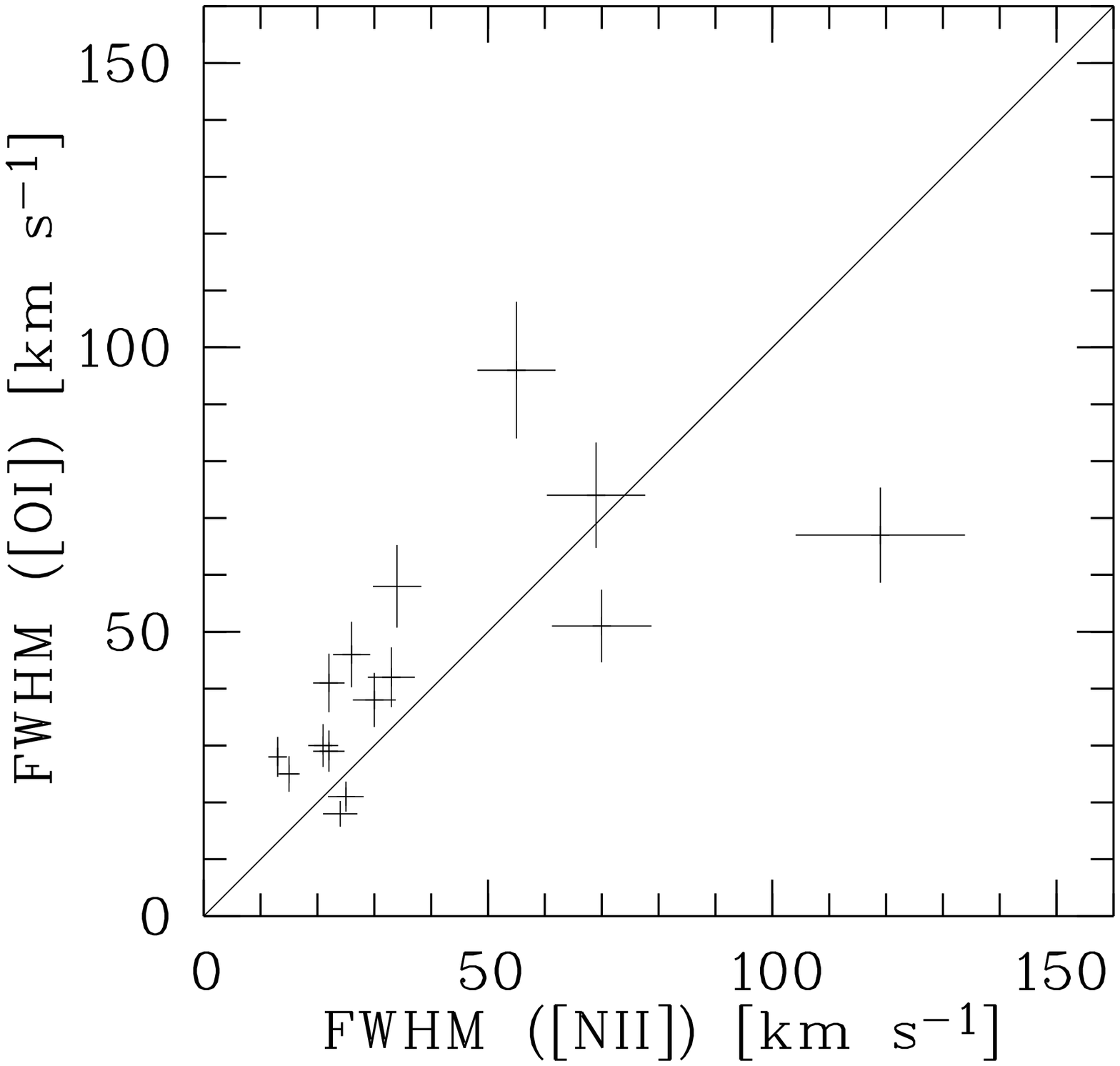}}
\resizebox{0.47\hsize}{!}{\includegraphics[width=8.7cm,angle=-90,bbllx=50pt,bblly=30pt,bburx=555pt,bbury=575pt,clip=]{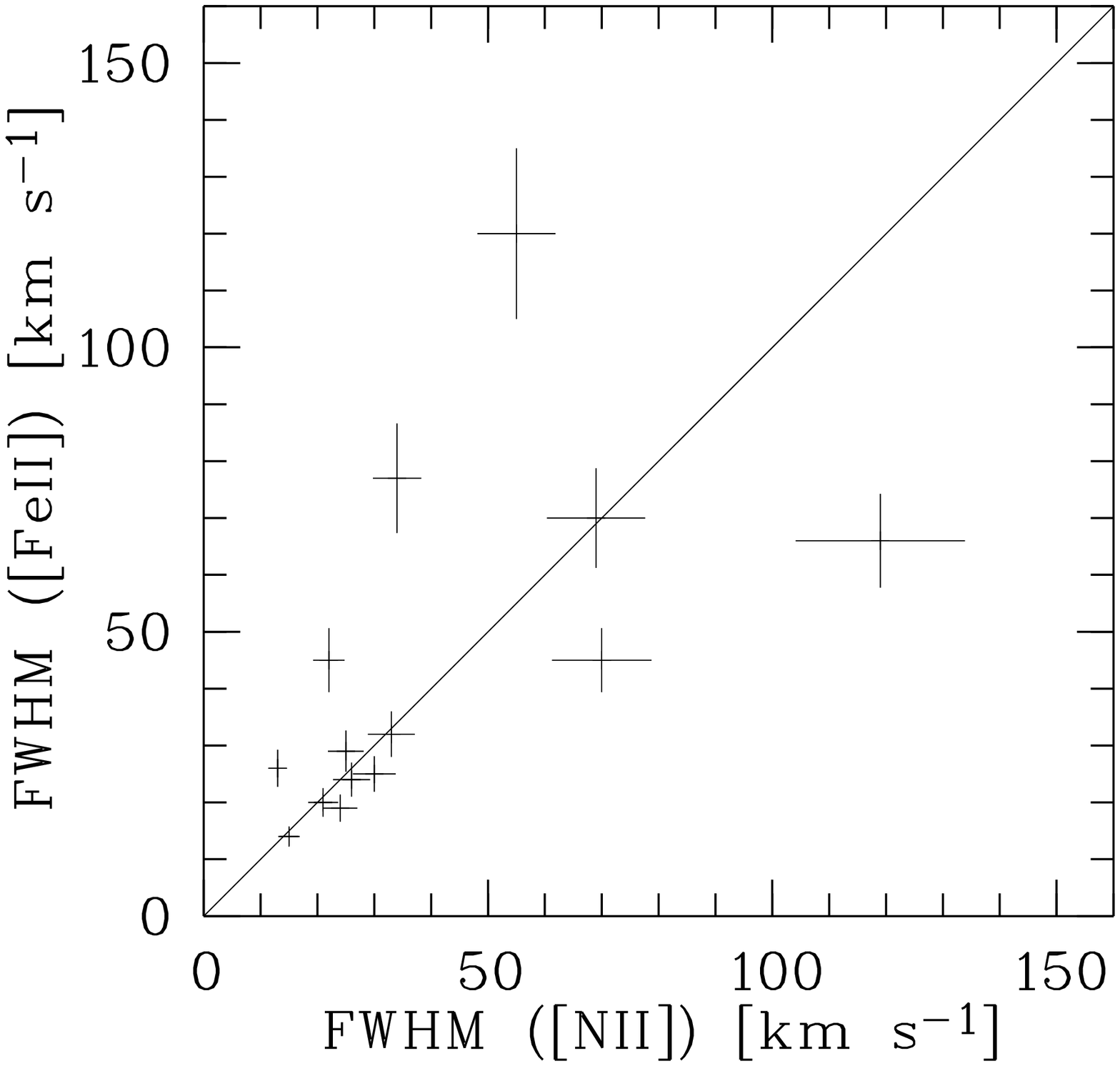}}
\caption[]{ FWHM of [\FeII ] vs. [\OI ] (upper left panel),  [\OI ] vs. 
[\NII ] (upper right panel), and [\FeII ] vs.  [\NII ] (lower panel).   
}
\label{fwhm_oinii}
\end{figure}
The observed line widths may help to better understand the possible role of 
expansion and rotation. 
In a disk-like circumstellar environment 
in which rotation dominates over expansion the forbidden lines 
are expected to be narrower than the permitted lines because the rotational velocity decreases outwards. 
If rotation is negligible compared to the expansion velo\-city of a wind  accelerated outwards
the forbidden lines should have a larger width 
than the permitted lines. The latter are formed in the accelerating inner wind zone. The forbidden lines
originate at large distance from the star where the wind has reached the terminal velocity. 

In Figs. \ref{fwhm_oiha} and \ref{fwhm_oihei} the FWHM   
of  [\OI ]$\lambda$6300\AA\ is plotted versus the  FWHM of \ha\  and \HeI , respectively.  
They clearly show that the
low-excitation forbidden line is on the average significantly narrower 
than the permitted lines. MWC\,645 is exceptional because of the
narrow red peak of \ha . The line widths are thus consistent with the first assumption,
i.e. the velocity field in the inner wind zone could be dominated by rotation.

The comparison of the line widths of the forbidden lines shown in Fig. 
\ref{fwhm_oinii}  reveals a correlation which is  
consistent with the  assumption that the velocity in the line forming region is
constant, and hence that rotation is 
not  
important far from the central star. With few exceptions the [\OI ] line is approximately as broad as the  [\FeII ] line. 
For [\NII ] the result is similar, however, with larger scatter. 
The comparison of [\NII ] and [\FeII ] displayed in the lower left panel of Fig. 
\ref{fwhm_oinii} shows four stars with broader [\FeII ] than [\NII ]. Note that 
in two of these cases, MWC\,137 and MWC\,1055,  the lines are very weak and the 
line widths are uncertain. Only Hen\,485 and \cpdsi\ show significantly broader [\FeII ] 
than [\NII ]. The general trend is towards equal widths or smaller widths for [\FeII ].
The comparison with 
[\SIII ] is not meaningful and therefore not shown because only three stars exhibit 
this line, i.e. MWC\,17, MWC\,137, and MWC\,349A. Inspecting the line widths 
listed in Table \ref{para1} for [\SIII ] no clear trend emerges for the few lines. 

\begin{figure}
\resizebox{\hsize}{!}{\includegraphics[angle=-90,bbllx=180pt,bblly=55pt,bburx=570pt,bbury=720pt,clip=]{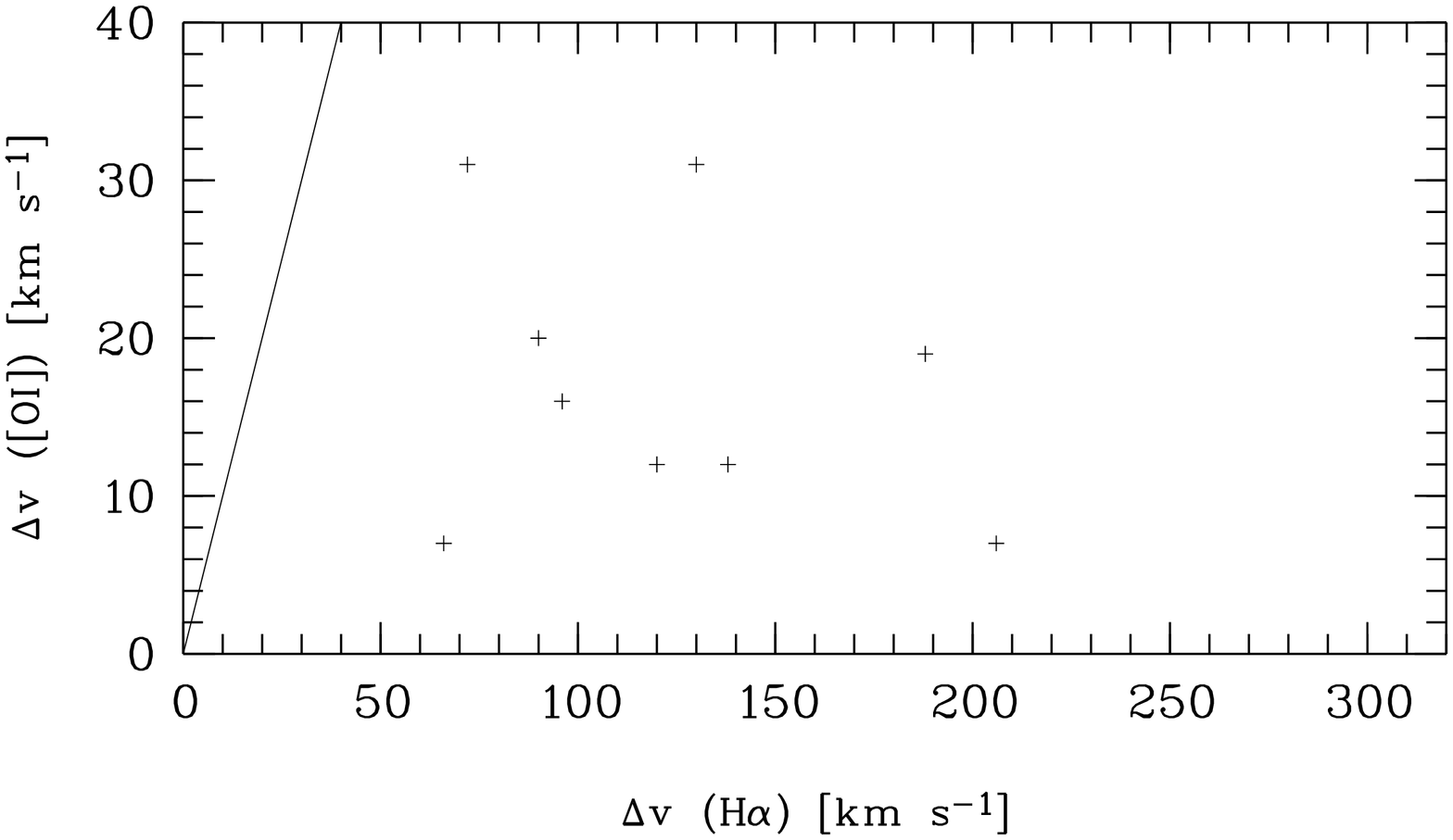}}
\caption[]{Line splitting of \ha\ vs. [\OI ]. 
The solid line represents the locus of equal splitting for both lines.
}
\label{delta_oiha}
\end{figure}

\begin{figure}
\resizebox{0.47\hsize}{!}{\includegraphics[width=8.7cm,angle=-90,bbllx=190pt,bblly=60pt,bburx=565pt,bbury=430pt,clip=]{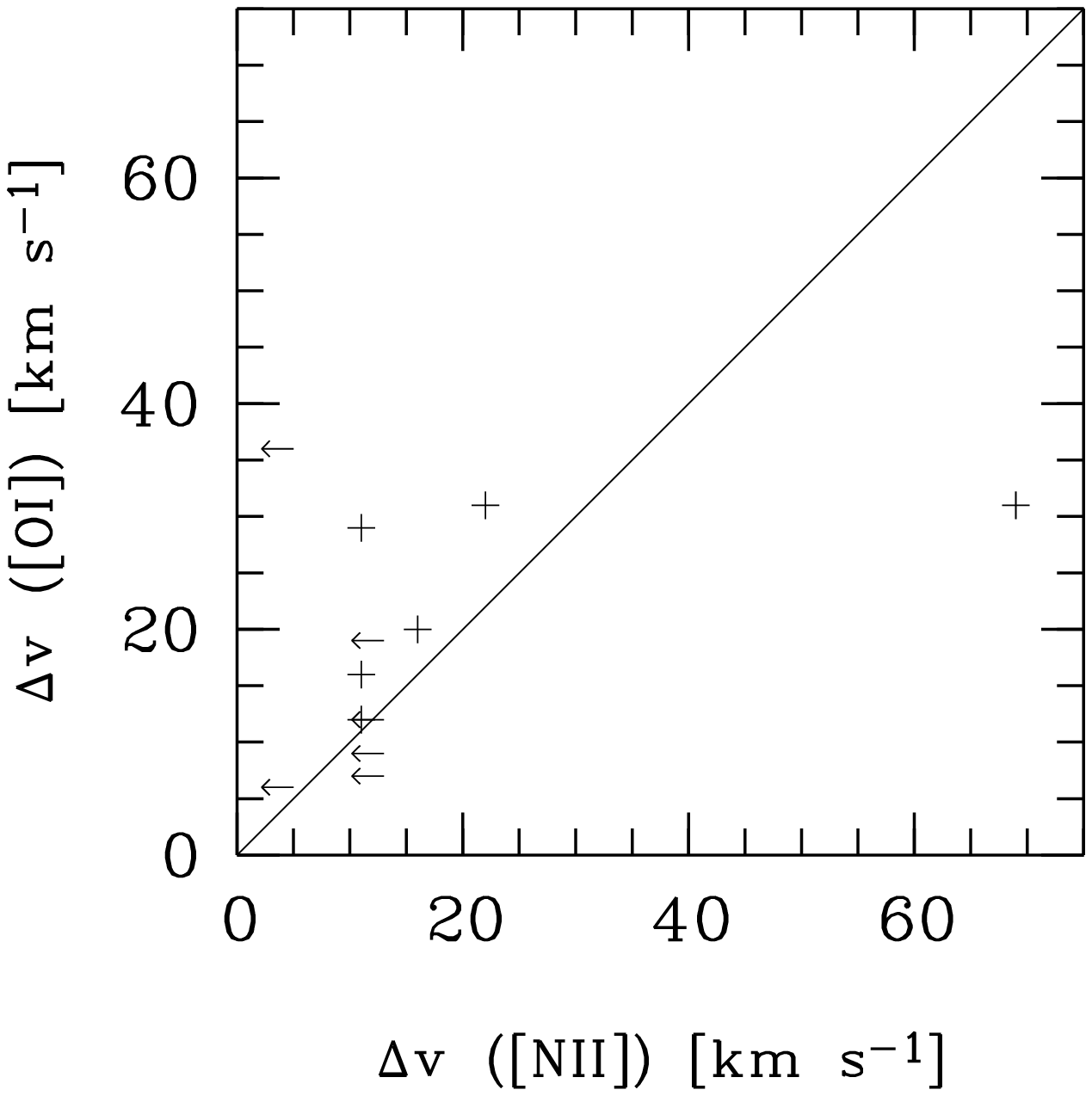}}
\hspace{0.02\textwidth}\hfill\resizebox{0.47\hsize}{!}{\includegraphics[width=8.7cm,angle=-90,bbllx=190pt,bblly=60pt,bburx=565pt,bbury=430pt,clip=]{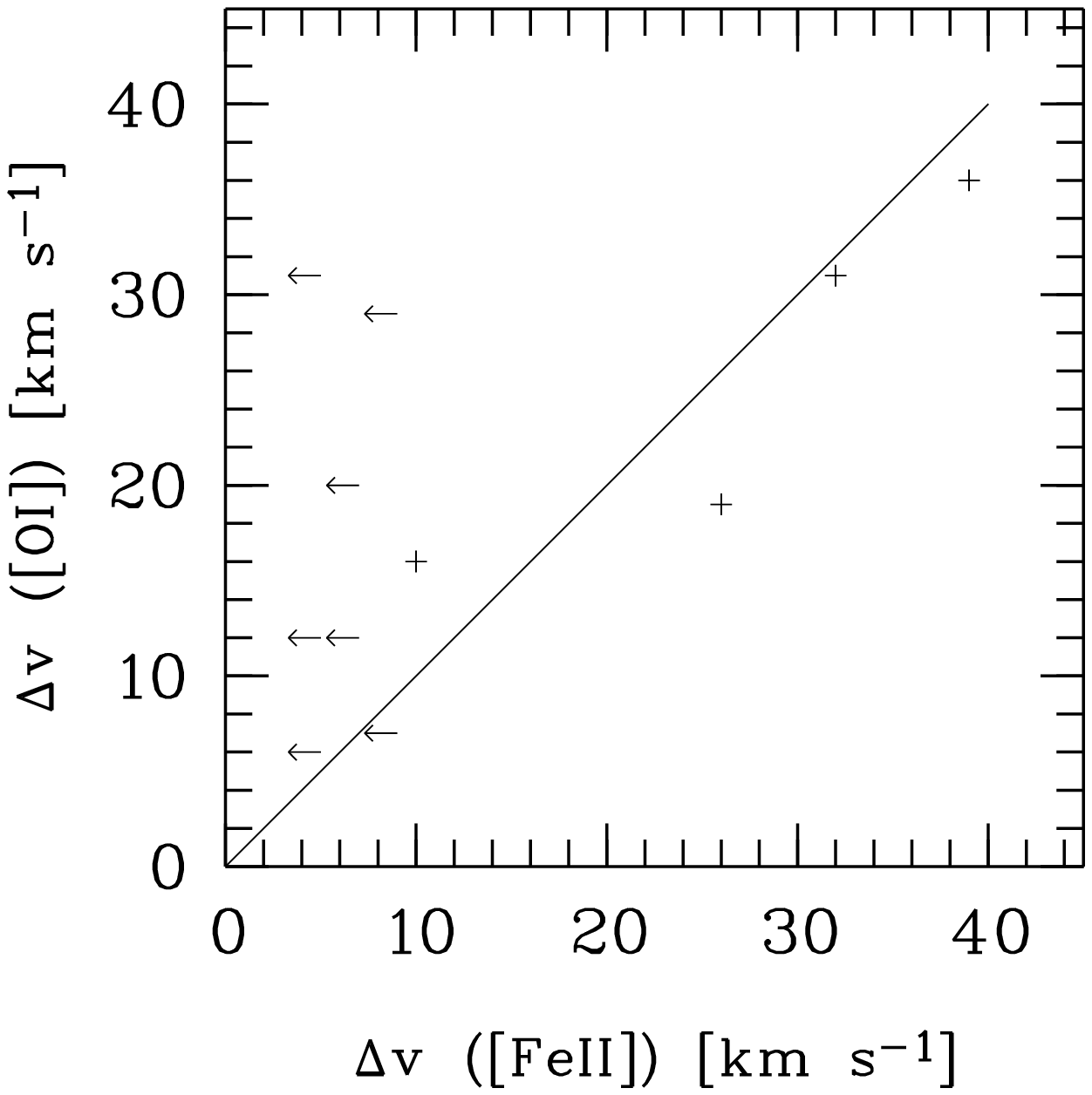}}
\caption[]{Line splitting of [\OI ] vs. [\NII ] (left panel) and of [\OI ] vs.
[\FeII ] (right panel). Upper limits are plotted as arrows.   
Equal peak separation is represented by the diagonal lines in each panel.
}
\label{delta_oinii}
\end{figure}

The line splitting is depicted in Figs. \ref{delta_oiha} and \ref{delta_oinii}. 
For single lines the velocity corresponding to the spectral resolution was adopted as 
upper limit of the line splitting.
For the line splitting no clear correlation between different lines exists. However, 
\ha\ exhibits a much larger splitting than the forbidden metal lines, which is again 
indicative of a higher velocity in the \ha\ forming region. A few stars with both, split 
[\FeII ] and [\NII ] are found close to the 
diagonal 
line of equal splitting
shown in Fig. 9. 
For [\OI] and [\NII ] the
scatter is large.  Most upper limits of the line splitting of [\NII ]$\lambda6583$\AA\  are 
close to the line of equal peak separation.  

The line widths are thus consistent with the assumption that in the inner wind zone
rotation could play a role. In the outer regions where 
the forbidden lines are formed a constant velocity wind seems to prevail.

Alternative to rotation, split optically thin  emission-line profiles can result 
from a radial outflow  with a hollow-cone 
geo\-metry as discussed e.g. by Appenzeller et al. (\cite{AJO84}). Such a 
configuration may be considered an approximation of the radially outflowing 
equatorial disk wind adopted by Zickgraf et al. (\cite{Zickgrafetal85},\cite{Zickgrafetal86}) 
for the \be\ supergiants. However, as before the axial symmetry of this configuration  
entails the problem  of understanding the $V/R$ ratios.
In the case of T\,Tauri stars asymmetric line profiles were explained e.g by
Appenzeller et al. (\cite{AJO84}) and Edwards et al. (\cite{Edwardsetal87}) by
assuming an opaque dust disk. Because the \bephen\ is characterized by the existence of
circumstellar dust, 
a dust disk is also a possible explanation for line asymmetries in \be-
type stars. As will be shown in Sect. \ref{thinprofile} it is not required that 
this disk is opaque.  
The opaque disk configuration 
with a constant velocity law 
leads to $V/R > 1$ in contrast to what is observed for the majority of \be -type stars, namely 
$V/R \le 1$. However, modifying the model parameters somewhat profiles similar to the
observed ones could be produced. This will be shown in the following section.

\section{Theoretical profiles of optically thin lines}
\label{thinprofile}

\begin{figure*}
\includegraphics[angle=-90,width=18cm,bbllx=50pt,bblly=60pt,bburx=550pt,bbury=770pt,clip=]{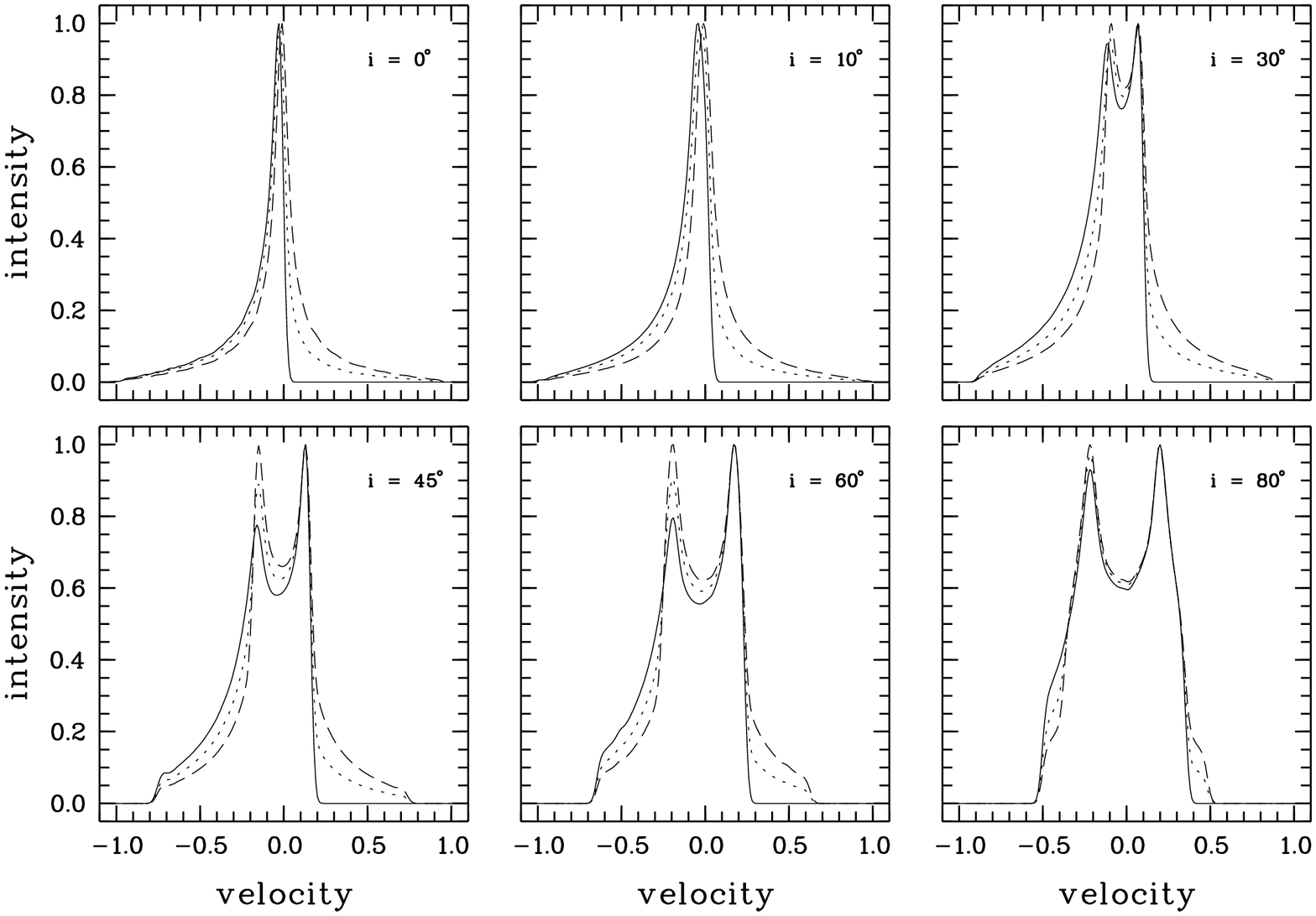}
\caption[]{Profiles of optically thin lines for different inclination angles $i$ and
a dust disk in the equatorial plane with various optical depths, $\tau = 1000$ (solid
line), $\tau = 1$ (short dashed line), and $\tau = 0.1$
(long dashed line). 
$D_0$ and $R_{\rm d}$ were  set to 1.0 and 1\,$\rstar$, respectively. 
The ordinate is the
flux normalized to the maximum. The abscissa is the radial velocity normalized to the
maximum wind speed.
}
\label{profil_all}
\end{figure*}
\subsection{Line profiles for a latitude-dependent wind model}
\label{model}
A detailed discussion of optically thin emission-line profiles  was given by 
Edwards et al. (\cite{Edwardsetal87}) (E87 hereafter) for the case of forbidden 
lines of T\,Tauri stars. They calculated  
line profiles for axially symmetric and radially expanding winds with an 
equatorial opaque dust disk. 

In order to study the observed forbidden lines 
of the \be- type stars profiles of optically thin lines  were calculated following 
the method of E87 which  is shortly summarized here. The adopted geometry 
is depicted in Fig. 5 of E87. 
The requirement of an opaque dust disk was relaxed by allowing arbitrary optical depths
for an equatorial dust ring (see below). The inclination angle $i$ is measured with 
respect to the polar axis. Polar coordinates $r$,
$\theta$, $\phi$ of the vector $\vec{r}$ are defined in the stellar reference 
frame with $r$ being the distance from the
star, $\theta$ the angle of the vector $\vec{r}$ with the polar axis and $\phi$ 
the rotation angle around the polar axis. 

For the velocity law the latitude-dependent
model of E87 was chosen for two reasons. Firstly, it represents a disk-like 
structure resembling the two-component wind model for \be\ supergiants suggested by 
Zickgraf et al. (\cite{Zickgrafetal85}) with a fast polar and a slow equatorial wind.
Secondly, the profiles for the latitude-independent
wind shown in E87 do not resemble the observed line profiles of the \be-type stars neither
for optically thick nor optically thin dust absorption.

In the latitude-dependent model the wind was chosen to have a 
constant velocity $v({\theta})$ in  radial direction depending on the 
latitude $\theta$ according to 
\begin{equation}
v({\theta}) = v_{\rm max} f({\theta})
\label{vlat}
\end{equation}
with $v_{\rm max}$ being the velocity in polar direction.
The latitude-dependent factor $f({\theta})$ describes the variation of 
the wind velocity from the pole towards the equator.  
Assuming e.g. a B supergiant wind
in the direction towards the pole with a maximum expansion velocity of  
$\sim500-1000$\,\kms\  an equatorial wind velocity of 
$\sim50-200$\,\kms\  would correspond to 
$f(\pi/2) \sim 0.05 - 0.25$. This assumption is 
consistent with the equatorial velocities of $\sim100$\,\kms\  measured 
for \be\ supergiants by Zickgraf et al. (\cite{Zickgrafetal96}). 

For the model 
calculations presented below  in Figs. \ref{profil_all} to \ref{t1000konus} a linear 
decrease of $f({\theta})$ from 1 to 0.2 between $\theta = 0$ and $\pi/2$ was  
adopted. Note that the minimum value of $f(\pi/2)$ specifies the line width
at FWHM. 
The profile shape, however, is independent of this value. 
The wind velocity 
was normalized to 1 relative to the velocity in the direction
of the polar axis. Other functional dependencies of $f({\theta})$ 
are  possible. 
For example,
in the bi-stable wind model suggested 
by Lamers and Pauldrach (\cite{LamersPauldrach91}) for outflowing disk winds of 
early-type stars $f({\theta})$ could be described by a step or ramp-like function which
takes constant values within certain ranges around the equator and the pole with some
transition region in between. However, the general characteristics of the profiles remain
similar unless the constant velocity part of the wind prevails. Then the models approach 
the latitude-independent case.

The electron density $N_{\rm e}$ at point $\vec{r}$ normalized to the critical 
density $N_{\rm cr}$  is given by
\begin{equation}
D_{\rm e}(r,\theta,\phi) = N_{\rm e}(r,\theta,\phi)/N_{\rm cr} = D_0 
(R_{\star}/r)^2
  v({\theta})^{-1},
\label{density}
\end{equation}
where $D_0$ is the density ratio at $r = R_{\star}$ and $\theta = 0$.  
For a given density distribution $N_{\rm e}(r)$ 
the parameter $D_0$ thus depends on the critical density of the line considered. 
Lower critical density lines correspond to larger $D_0$ values. 
A line with $D_0 = 1, 1000,$ or 10\,000 has reached the critical density at $r = 1\,\rstar , 
33\,\rstar ,$ or $100\,\rstar ,$ respectively, in the polar direction.
Note that due to the $v({\theta})$ dependence the density distribution is disk-like 
with the density increasing towards the equatorial plane.
According to E87 the emissivity $j(r,\theta,\phi)$ for a 2-level atom is 
\begin{equation}
j(r,\theta,\phi) \propto D_{\rm e}^2 [1 + D_{\rm e}(\delta +1)]^{-1}
\label{emissivity}
\end{equation}
where $\delta$ is the ratio of the collisional rate coefficients of the lower and upper
level. The line profiles turned out 
to be insensitive  to $\delta$ 
which could therefore be set to $\delta = 0$. The emissivity was set to zero for 
$r < r_{\rm min}$ with  $r_{\rm min}$ given by the
condition of $D_{\rm e}(r_{\rm min},\theta,\phi) = 1$, i.e. no line emission for 
$N_{\rm e} > N_{\rm cr}$. For each volume element 
$\Delta V$ the radial velocity along the direction of the line of sight was calculated from
\begin{equation}
v(r,\theta,\phi) = -v({\theta})\,\cos \beta
\label{vradial}
\end{equation}
with  
\begin{equation}
\cos \beta = \cos \theta \cos i + \sin \theta \sin i \sin \phi,
\label{cosb}
\end{equation}
$\beta$ being the angle between $\vec{r}$ and the line of sight.
Finally, the flux contributions $j(\vec{r})\Delta V$ of all volume elements 
within $r \le r_{\rm max}$ were summed up in the 
suitable velocity intervals evenly distributed between $-1$ and $+1$. 

Edwards et al. assumed the presence of an opaque disk blocking entirely the receding
part of the wind. For the \be\ stars an equatorial ring structure was adopted instead
with an inner ring radius $R_{\rm d} \ge R_{\star}$ and an optical depth of 
$\tau_{\rm d} \ge 0$. The case $R_{\rm d} = R_{\star}$ 
and $\tau_{\rm d} = \infty$ corresponds to the configuration of E87.  
The ring structure with $R_{\rm d} > R_{\star}$ takes into account that the inner rim of
the dust disk should depend on the dust condensation radius.  
According to Lamers \& Cassinelli (\cite{LamersCassinelli98}) the equilibrium temperature of the
dust varies as 
$T_{\rm d} \simeq \teff (2 R_{\rm d}/R_{\star})^{-2/5}$. 
For the \be -type stars it 
can therefore be estimated 
to be  \mbox{$\sim10^2\,R_{\star}$} 
to \mbox{$\sim2\,10^3\,R_{\star}$} for $\teff$ between
10$^4$\,K and 
2.5\,10$^4$\,K 
and $T_{\rm d} \sim10^3$\,K.
Volume elements below the equatorial plane, i.e. at $\theta > \pi/2$, 
contribute 
$F_{\rm d} = j(\vec{r})\Delta V \exp(-\tau_{\rm d})$ to the observed flux unless  
the line of sight passes through the central hole. In this 
case $F_{\rm d} = F = j(\vec{r})\Delta V$.
Hence, for  
$R_{\rm d} > R_{\star}$ 
lines with different critical densities can be affected 
differently by the dust absorption because the inner radius of the line-emitting volume, 
$r_{\rm min}$, depends on $N_{\rm cr}$.

In Figs. \ref{profil_all} to \ref{t1000r1000} the results of the model calculations
are displayed for the linear $f(\theta$) law and various parameter combinations.
Figure \ref{profil_all} shows line profiles calculated for  
$D_0 =1.0$, 
$R_{\rm d} = 1\,R_{\star}$ 
different inclination angles $i$, and three values of the
optical depth 
$\tau_{\rm d}$ of 0.1, 1, and 1000. Note that for 
$R_{\rm d} = 1\,R_{\star}$ 
the profile shape does not depend on $D_0$. The profiles for 
$\tau_{\rm d} = 1000$ are identical to those shown by E87. 
A smaller $\tau_{\rm d}$ leads to a more symmetrical profile.
For large inclination angles
double-peaked profiles are produced. 
The $V/R$ ratio of the peak fluxes 
is always $\le 1$ and depends on $\tau_{\rm d}$. A small $\tau_{\rm d}$ leads to 
$V/R \approx 1$ for all inclination angles. For $\tau_{\rm d} \ga 1$ the $V/R$ ratio
deviates significantly from 1 except for large $i$. The peak separation is 
determined by $f(\theta = \pi/2)$ and depends also on the inclination $i$. Decreasing 
$f(\theta = \pi/2)$ and/or $i$ yields a smaller velocity difference of the line peaks.
There is also some  weak dependence of the peak separation on $\tau_{\rm d}$. 

In Figs. \ref{t1r1000} and \ref{t1000r1000} the influence of $D_0$ as a function of
inclination is shown for  
models with a dust ring. The inner radius of the ring is assumed to be 
$R_{\rm d} = 1000\,R_{\star}$. 
Two sets of models are shown for which the optical depth of the dust is $\tau_{\rm d}$ =  1 and 1000, 
respectively. 
Different $D_0$ values 
were adopted for a polar density at the base of the wind of $N_0 = 10^{11}$\,cm$^{-3}$
and the critical densities given in Table \ref{ncr} for [\FeII ], [\OI ], and [\NII ].  
This $N_0$ corresponds to an equatorial density of $\sim10^{12}$\,cm$^{-3}$.
For small $\tau_{\rm d}$ the profiles remain nearly symme\-tric with increasing $D_0$. 
Only a small decrease of the flux on the red wing can be seen. For 
intermediate inclination angles the split profiles show a trend of decreasing $V/R$ 
with increasing $D_0$. For large $\tau_{\rm d}$ the lines become more asymmetric for 
increasing $D_0$ because the emission for 
lines with a low critical density starts at a larger distance from the star.
Therefore less  emission can be seen through the central hole of the dust
ring. The configuration thus approaches that of Fig. \ref{profil_all}.  The receding 
part of the wind produces 
a weak bump on the red side of the profile  
if the  inclination angles is not too large. For high inclination the $V/R$ ratio decreases
with increasing $D_0$, passes through a minimum and then increases again. 
The models thus show that one line may show $V/R \approx 1$ 
and at the same time another line with a different $N_{\rm cr}$ can have  $V/R < 1$.

\begin{figure}
\resizebox{\hsize}{!}{\includegraphics[bbllx=45pt,bblly=48pt,bburx=375pt,bbury=770pt,clip=]{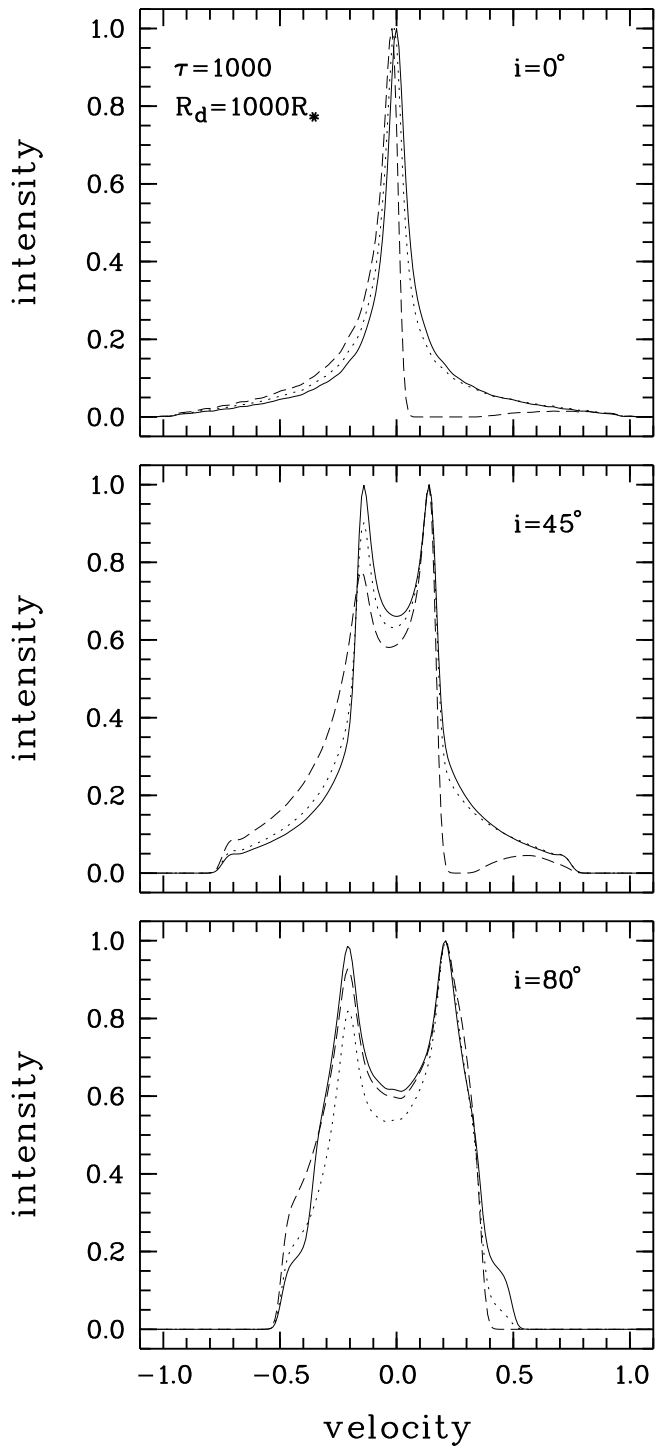}}
\caption[]{Line profiles for 
$D_0 = 5\,10^2$  (solid line), $3\,10^4$ (short dashed line), and $1\,10^6$ (long dashed line). 
For the dust ring an optical depth of $\tau_{\rm d} = 1$ and an inner radius 
of $R_{\rm d} = 1000\,\rstar\ $ was 
adopted.
}
\label{t1r1000}
\end{figure}

\begin{figure}
\resizebox{\hsize}{!}{\includegraphics[bbllx=45pt,bblly=48pt,bburx=375pt,bbury=770pt,clip=]{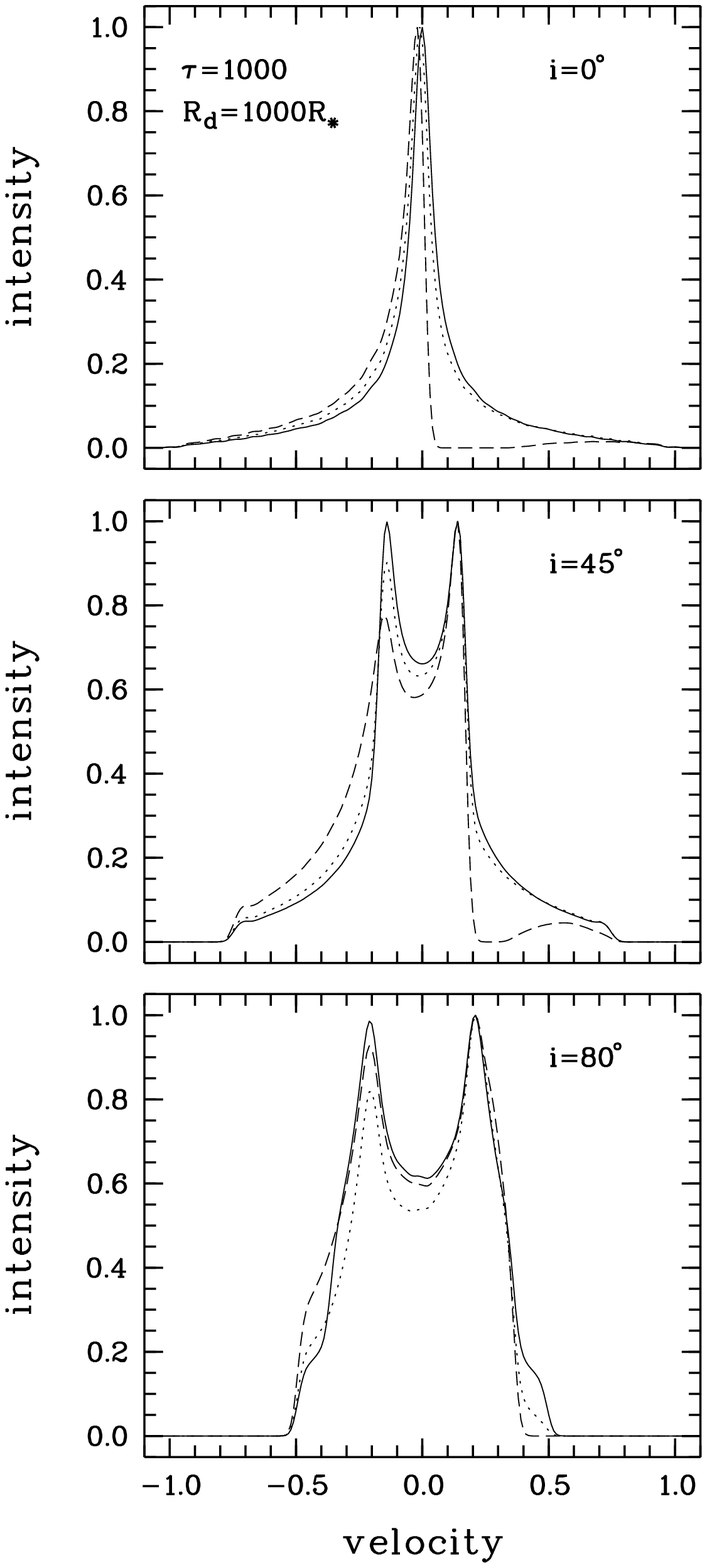}}
\caption[]{
Line profiles for 
$D_0 = 5\,10^2$  (solid line), $3\,10^4$ (short dashed line), and $1\,10^6$ (long dashed line). 
For the dust ring an optical depth of $\tau_{\rm d} = 1000$ and an inner radius 
of $R_{\rm d} = 1000\,\rstar\ $ was 
adopted.
}
\label{t1000r1000}
\end{figure}

Up to now a disk-like wind model with a {\em latitude-independent} ionization structure has been
considered.
However, 
lines with different ionization potentials may originate in different zones of the wind.
For example, the two-component stellar wind model by Zickgraf et al. (\cite{Zickgrafetal85})
suggests that the cool equatorial zone should 
give rise to neutral or low-ionisation lines of ions like [\OI ] and [\FeII ]. Lines with 
higher ionization potential like [\NII ] and [\SIII ] 
would 
originate in hotter yet tenuous regions at higher latitude towards the polar zone. 
A scenario like this is therefore characterized by a {\em latitude-dependent} ionization 
structure. It can be sketched by the hollow and filled cone model, respectively, 
of Appenzeller et al. (\cite{AJO84}). In the hollow cone geometry the emission is 
restricted to a volume within a certain angle from the equatorial plane. Correspondingly, 
the filled cone is the volume within a certain angle from the polar axis.
In the following the terms ``equatorial disk'' and ``polar cone'' will therefore be used
for the hollow and filled cone geometry, respectively.

Figs. \ref{t1konus} and \ref{t1000konus} show line profiles for the case of an opening
angle of 30\degr\ 
(measured from the equatorial plane) for the 
equatorial disk and  60\degr\ for the polar cone
(measured from the polar axis).
In the  case  
of  a small optical depth
of the dust ($\tau_{\rm d} = 1$) the resulting 
profiles for the 
polar cone are complex. 
The equatorial disk 
produces profiles 
similar to the 
latitude-independent ionization model 
discussed above (opening angle $90\degr$) . 
The FWZI of the equatorial disk line decreases with a decreasing opening angle of the disk.
With increasing $D_0$ or decreasing $R_{\rm d}$ the red peak of the polar cone line becomes
weaker.

\begin{figure}
\resizebox{\hsize}{!}{\includegraphics[bbllx=45pt,bblly=48pt,bburx=375pt,bbury=770pt,clip=]{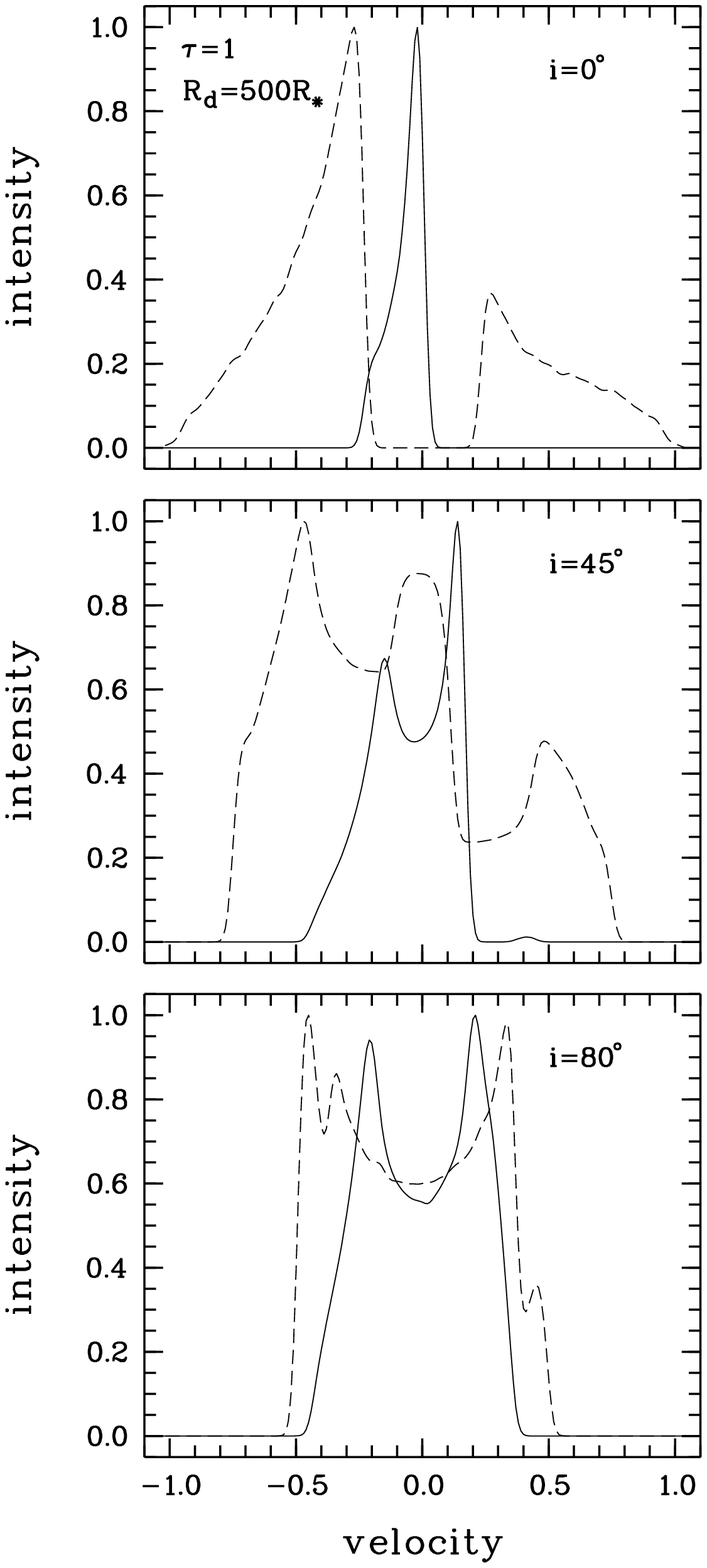}}
\caption[]{Line profiles for  
an equatorial disk and polar cone 
geometry with 
$\tau_{\rm d} = 1$, $D_0 = 3\,10^5$ and $R_{\rm d} = 500\,\rstar $. 
For the equatorial disk  
(solid line) an opening angle of $30\degr$ measured from the equatorial plane was adopted. The 
polar cone 
(dashed line) is the complementary volume with an opening angle of $60\degr$ measured 
from the polar axis.
}
\label{t1konus}
\end{figure}

\begin{figure}
\resizebox{\hsize}{!}{\includegraphics[bbllx=45pt,bblly=48pt,bburx=375pt,bbury=770pt,clip=]{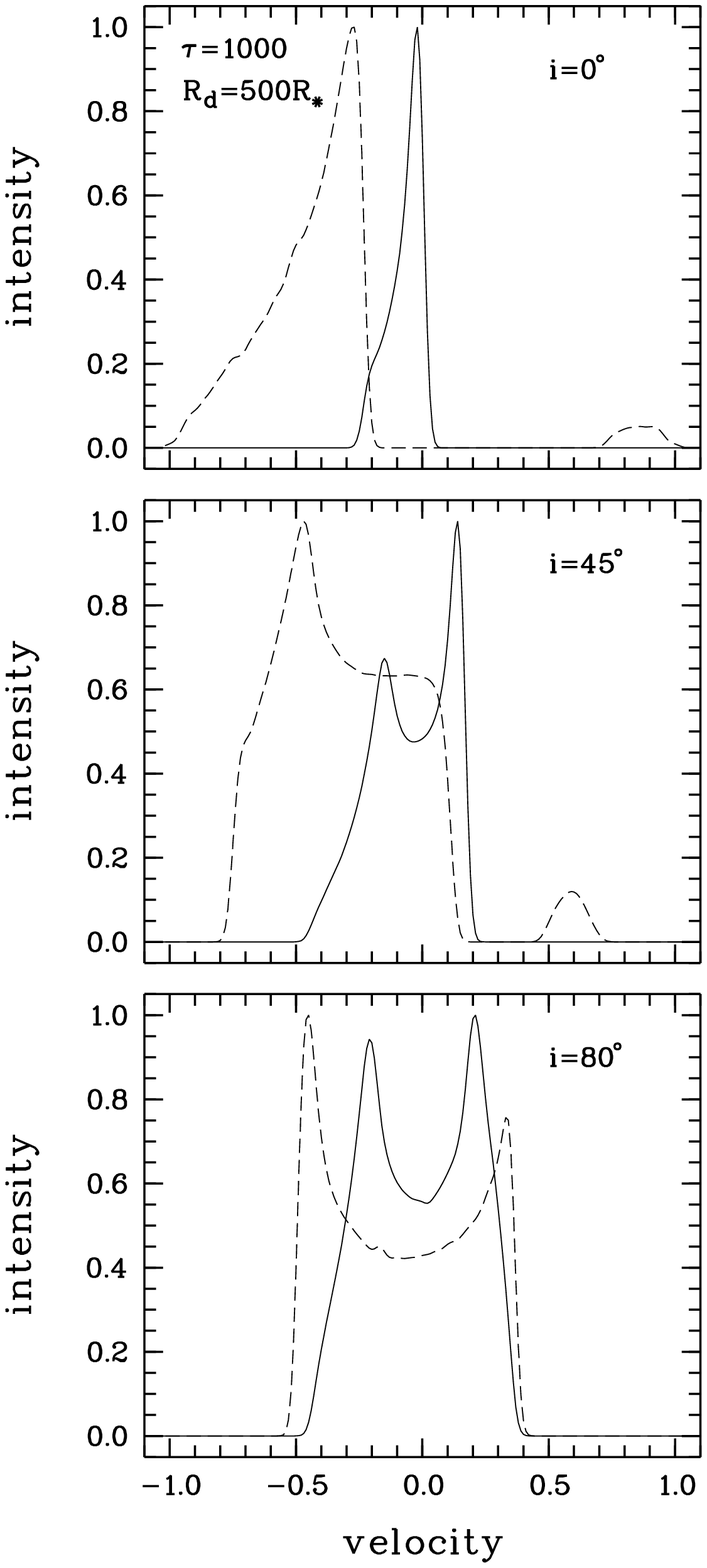}}
\caption[]{Line profiles for the same geometry as in Fig. \ref{t1konus} but for 
$\tau_{\rm d} = 1000$ and $D_0 = 1\,10^6$. 
The equatorial disk  and polar cone 
profiles are plotted as solid and dashed lines, respectively. 
}
\label{t1000konus}
\end{figure}

\subsection{Comparison with observations}
The comparison  of the observed profiles  in Fig. \ref{forbidden} and of the
models displayed in Figs. \ref{profil_all} to \ref{t1000konus} shows that the general
characteristics of 
the forbidden lines 
can qualitatively be reproduced. 
According to the model calculations split or asymmetric profiles are expected from the
latitude-dependent wind model with dust disks or rings of various optical depths.
The single and more or less symmetric emission peaks observed for a couple of stars can 
be produced with small inclination angles and small optical depths of the dust 
ring or disk. 
Furthermore, for a given inner radius of the dust ring  the different critical 
densities can lead to differences in the observed line profiles. Likewise, 
variations of the $V/R$ ratio from $<1$ to $>1$ are expected if the lines originate in 
different regions as e.g. in the equatorial disk  and polar cone 
configuration.

The [\OI ] lines of MWC\,17, MWC\,297, MWC\,349A, MWC\,645, MWC\,939, HD\,45677, and \cpdzw\
qualitatively resemble the profiles shown in Fig. \ref{profil_all} to \ref{t1000r1000}
for intermediate to large inclination angles. In most cases the wings appear symmetric 
(or nearly symmetric) as expected for a small optical depth of the dust ring.  
Exceptions are MWC\,645 and possibly Hen\,485 for which the line profiles resemble 
those for a large optical depth of the dust and intermediate inclination. 
In MWC\,645 the lines exhibit a strong asymmetry with pronounced blue wings and 
split peaks. In Hen\,485 the [\OI ] line and, to a lesser degree, the [\NII ] line 
also show  an asymmetric profile with a blue wing similar to MWC\,645. 
Line splitting is not discernible, though. 
The absence of splitting expected for an intermediate inclination could be due to 
turbulent broadening as discussed below.

The lines of MWC\,300 and HD\,87643 are similar to the profiles for a small inclination 
angle and small optical depth of the dust. A near pole-on viewing angle was suggested 
by Winkler \& Wolf (\cite{WinklerWolf89}) for MWC\,300 and by Oudmaijer et al. 
(\cite{Oudmaijeretal98}) for HD\,87643.

In Hen\,1191 and \cpdsi\ the [\NII ] lines are characterized by a broad  pedestal on top
of which a narrow peak is sitting.
This bears some resemblance with the profiles  calculated for small optical
depth of the dust and small to intermediate inclination. 

An interesting feature of the polar cone lines shown in Fig. \ref{t1konus} and 
\ref{t1000konus} is that they can have $V/R > 1$ if seen
under intermediate to large inclination angle. Neither 
the models with opening angles of $90\degr$ displayed in Figs. \ref{profil_all} to \ref{t1000r1000} nor the
equatorial disk lines in Figs. \ref{t1konus} and \ref{t1000konus} show this behaviour. 
This could explain the [\OI ] lines in MWC\,137, MWC\,342, and \cdvi , and the [\NII ] 
lines of MWC\,349A and \cpdsi . Likewise, the equatorial disk and polar cone model 
produces lines with different width, both FWHM and FWZI, which is not the case in the
latitude-independent ionization model.  

\begin{figure}
\resizebox{\hsize}{!}{\includegraphics[angle=-90,bbllx=52pt,bblly=62pt,bburx=314pt,bbury=318pt,clip=]{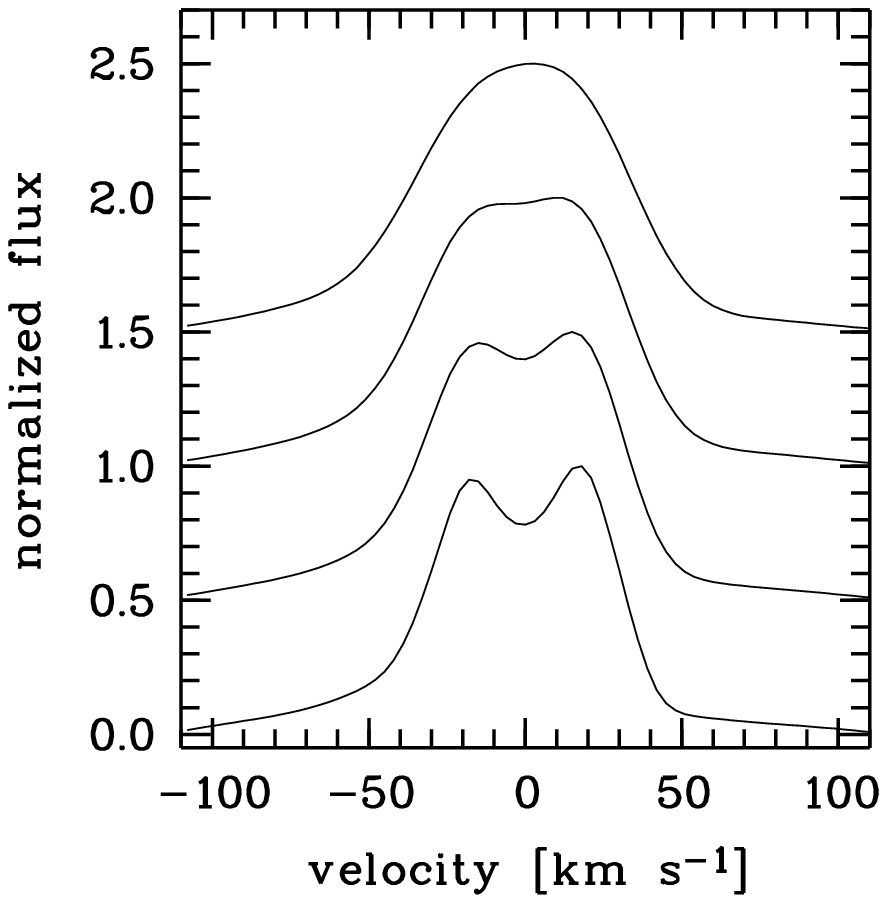}}
\caption[]{Line profiles for a wind model with $v_{\rm max} = 300$\,\kms\ at the pole, 
$v_{\rm eq} = 25$\,\kms\ at the equator, i.e. $f(\pi/2) = 0.08$, inclination $i =45\degr$, 
and macroturbulence velocities of $v_{\rm t} = $15, 20, 30, 35\,\kms , respectively 
(from bottom to top). $D_0$ and $R_{\rm d}$ are $3\,10^5$ and $1000\,\rstar $, respectively.
}
\label{turb}
\end{figure}

The complex profiles obtained for a equatorial disk and polar cone 
model with optically thin dust were not observed in the sample discussed here. However,
the forbidden lines of MWC\,17 and MWC\,349A show a strong resemblance with the 
profiles displayed in Fig. \ref{t1000konus}  for optically
thick dust  and inclination $i = 80\degr$.
In both stars the lines of [\NII ] and [\SIII ] are broader than [\OI ] and [\FeII ]. 
This is expected for the equatorial disk  and polar cone model
if the higher ionization lines originate in the polar zone (filled cone) and the lower
ionization lines in the equatorial region (hollow cone).
The profiles displayed in Fig. \ref{t1r1000} for optically thin dust 
also exhibit the double peak structure if the inclination is high. However,
in these models the line forming regions are not separate for lines with different 
ionization potential. Contrary to the observations the lines  
should thus show the same widths. Note that for MWC\,349A there is
observational evidence for the existence of a dust disk seen edge-on 
(White \& Becker \cite{WhiteBecker85}, Leinert \cite{Leinert86}, Hofmann et al.
\cite{Hofmannetal02}). The observations are thus
in favour of the equatorial disk and polar cone model with optically thick dust.

An interesting feature of this model in the case of large optical 
depth of the dust disk is the behaviour of the $V/R$ ratio.  
The polar cone line can have a flux ratio $V/R > 1$ 
and at the same time the corresponding equatorial disk line has $V/R < 1$. This could
explain  qualitatively the different appearance of the lines in 
MWC\,349A which shows simultaneously [\OI ], [\FeII ], and [\SIII ] lines with $V/R < 1$
and [\NII ] with $V/R > 1$.

The equatorial disk and polar cone model also  seems promising for MWC\,137, MWC\,342, 
and \cdvi . In these objects the [\OI ] exhibits a $V/R$ ratio $> 1$.  In  MWC\,137  
the line widths and the $V/R$ ratio suggest that [\OI ] is a polar cone 
line, whereas [\SIII ] and [\NII ] are equatorial disk lines.  

In Hen\,230 and MWC\,1055 the [\NII ] line shows a red wing indicating that this line
originates in a polar cone seen under an intermediate aspect
angle. In Hen\,230 the wing is also indicated in [\FeII ]. 

Though many line characteristics can thus be understood in terms of the latitude-dependent
wind  model there seems to be a problem in explaining simultaneously  
split and single-peaked forbidden lines as observed in several stars. In many of these 
cases insufficient spectral resolution or too low an $S/N$ ratio might be an explanation 
for the apparent differences of the profiles, e.g. in MWC\,297,
and MWC\,1055. In MWC\,17, MWC\,342, HD\,45677, and \cdvi , however, 
the differences seem to be real. 
A possible explanation could be the existence of  macroscopic 
turbulent motion of the order of $\sim10-30$\,\kms . It would broaden the local line 
profile and therefore smear out line 
splitting if the expansion velocities are small enough. As an example a hollow cone model 
with an opening angle of 30\degr\ and an inclination angle of 45\degr\ is shown in Fig. \ref{turb}. For the wind 
a polar velocity of $v_{\rm max} = 300$\,\kms , and an equatorial velocity of $v_{\rm eq} = 25$\,\kms , 
i.e. $f(\pi/2) = 0.08$, was adopted. The line broadening due to macro turbulence was assumed
to have a Gaussian shape
with a  FWHM given by the turbulent velocity $v_{\rm t}$.
Four values  were assumed for $v_{\rm t}$, i.e. 15, 20, 30, and 35\,\kms . 
The optical depth of the dust disk was $\tau_{\rm d} =1$. 
The figure shows that the split profile disappears for high $v_{\rm t}$.
If the lines originate in different regions, e.g. due to ionization effects, 
a location-dependent turbulence velocity  
could therefore lead to the observed
differences in the profiles. A high $v_{\rm t}$ could also be responsible for the 
sloping tops of the lines of \cpdzw . 
$V/R < 1$ and  [\NII ] with $V/R < 1$  (cf. Fig. \ref{forbidden}). 

\begin{figure}
\resizebox{\hsize}{!}{\includegraphics[angle=0,bbllx=45pt,bblly=505pt,bburx=370pt,bbury=770pt,clip=]{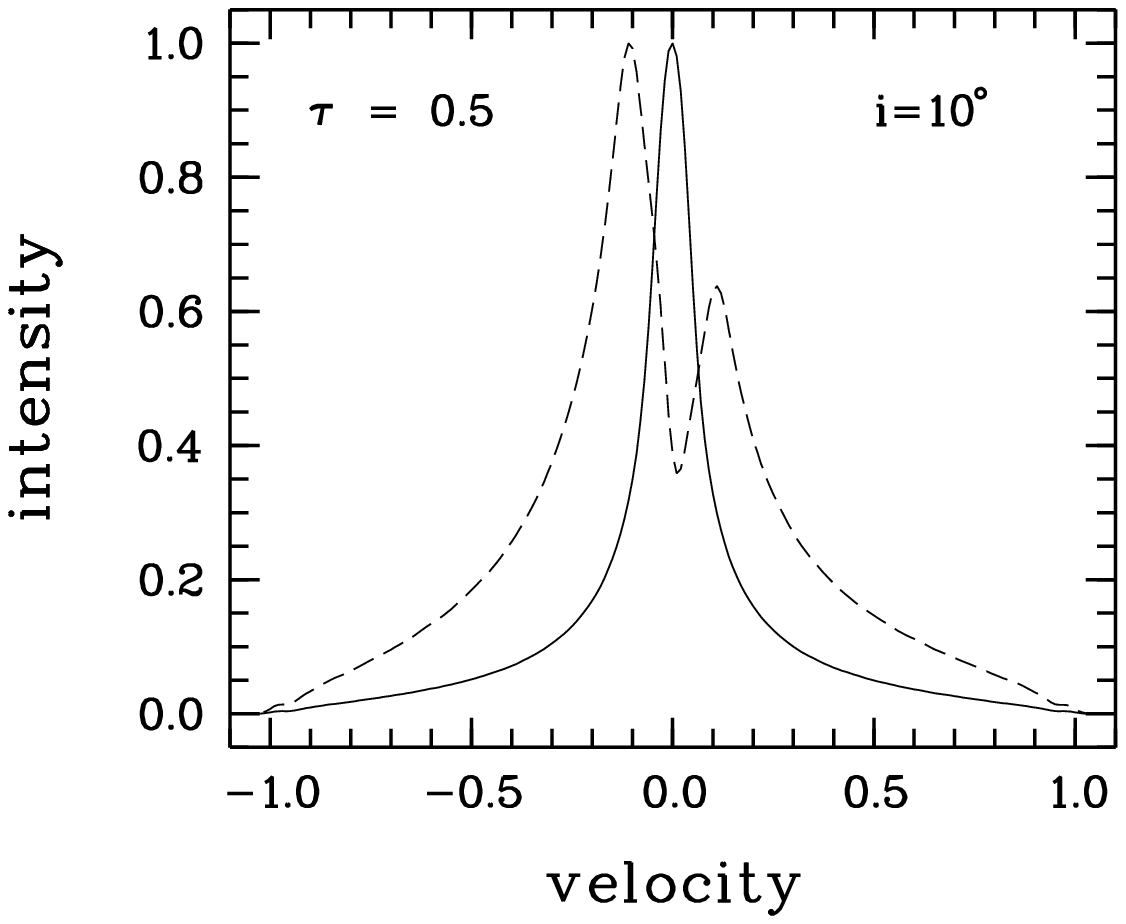}}
\caption[]{
Simultaneous appearance of split and single-peaked lines  for a polar cone with an 
opening angle of $80\degr$ (dashed line) 
and an opening angle of $90\degr$ (solid line) with $R_{\rm d} = 1000\,\rstar $,
$\tau_{\rm d} = 0.5$, $D_0 = 3\,10^{5}$ (dashed line) and $D_0 = 5\,10^{3}$ (solid line). 
The $D_0$ values correspond to [\OI ] and [\FeII ] for $N_0 =
10^{12}$\,cm$^{-3}$, respectively.
}
\label{singlesplit}
\end{figure}

Alternatively, the simultaneous appearance of split and single-peaked lines of 
the different ions can be 
explained with the equatorial disk and polar cone model. 
MWC\,342 and HD\,45677 show split [\OI ], but single-peaked [\FeII ] and [\NII ].
Surprisingly, such a combination of profiles can be produced in a nearly pole-on geometry. 
This is illustrated in Fig. \ref{singlesplit} for a model with an inclination angle 
of $i = 10\degr $. 
Here it is assumed that the split line originates in the polar cone with a large opening angle 
of about $\sim70\degr\ $ to $\sim 80\degr\ $. The line has a $V/R$ ratio of $>1$. 
The single-peaked line is produced with an opening
angle of $90\degr $, which means latitude-independent ionization. This confguration thus can
lead to split lines with $V/R > 1$ and simultaneously to single-peaked lines.

\section{Conclusions}
\label{sum}
High-resolution line profiles of selected permitted  and forbidden lines of \be -type
stars were discussed. A main result was the detection of double-peaked 
forbidden lines in a large fraction of the observed sample.
This strongly indicates that the line formation environment has a non-spherical
distribution, most likely in a disk-like configuration. 
Lines formed close to the star like \Ha\ and \HeI\ were compared with
forbidden lines formed in the tenuous outskirts. The comparison suggests that rotation 
could dominate in the inner zones but is overtaken by a radially expanding disk wind 
in the outer regions. 

Line profiles were calculated for the optically thin case.
Profiles similar to  the observed forbidden
lines are expected for a radially expanding latitude-dependent wind.  
This wind configuration can be considered a 
parametrization of the generally accepted two-component wind model for \be -type stars. 

For many objects in the sample the observed line profiles 
are consistent with  models assuming 
a latitude-independent ionization structure of the wind.
There are, however, two other groups of objects for which 
the models suggest (partly) separated line forming regions for the different ions. 
In one group the low ionization lines seem to 
originate in an equatorial disk. In the second group the neutral
line of [\OI ] appears to originate in a polar cone instead.  

It is clear, though, that due to simplifying assumptions of the 
model it cannot be expected to explain all details of the observed profiles. 
Rather, the discrepancies might suggest that the environments, although
in general having a disk-like structure, are apparently more complicated than 
assumed here. In particular, the ionization structure is taken into account only  
simplistically with the hollow and filled cone model because of the lack of information on
this parameter. For example, charge tranfer reactions  can have a strong influence on the
ionization balance of oxygen and nitrogen (e.g. Chamberlain \cite{Chamberlain56}, Butler \&
Dalgarno \cite{ButlerDalgarno79}). This has not been investigated for \be\ star disks so far.
Binarity could add further complexity. However, as long as an expanding wind dominates the 
circumstellar environment at large distances from the central object 
the latitude-dependent wind model could also be applicable to such a
subclass of  \be -type stars.
Thus even with the simplifying assumptions the results obtained here support the 
commonly adopted 
point of view that the forbidden emission lines of \be -type stars are formed in  
disk-like circumstellar environments.

\acknowledgements{I would like to thank the Deutsche For\-schungs\-gemeinschaft 
for granting travel funds (Zi\,420/7-1) and for financial support under grants Wo\,269/2-1 
and Wo\,269/2-2. 
I would also like to thank the referee, Dr. A. S. Miroshnichenko, for his critical  comments 
which helped to improve the paper.
This research has made use of the SIMBAD database, 
operated at CDS, Strasbourg, France.
}

\appendix
\section{Emission line parameters}
\label{parameter}
In Table \ref{para1} the line parameters of the observed emission lines 
are listed. Line peak intensities $I_{\rm line} = F_{\rm line}/F_{\rm cont}$ 
(for the stronger peak in case of split profiles)
and the ratio of blue and red
peak intensities, $V/R = (F_{\rm blue}-F_{\rm cont})/(F_{\rm red}-F_{\rm cont})$, 
are given in the first two columns. $W$ is the equivalent 
width in \AA , $\Delta v$ (in \kms ) is the 
peak separation for lines of type 3. 
FWZI and  FWHM  
(in \kms ) are the full width at zero intensity and full width at half 
maximum, respectively. Errors of  FWZI and  FWHM were found to be about
10-15\%. The meaning of ``FWHM'' is taken
literally by measuring the line widths at the 50\% flux level of the maximum peak 
flux independent of the line profile shape.   For type 3 profiles with 
$V/R < 0.5$ or type 1 P\,Cygni profiles this only measures 
the  width of the red emission peak which is then used as a rough estimate for the 
FWHM.  For \ha\ only a lower limit for  FWZI can be given due to the
small wavelength intervals covered by the spectra.

\begin{table}[h]
\tabcolsep=3pt
\caption[]{Line parameters. }
\begin{tabular}{lllllll}
\noalign{\smallskip}
\hline
\hline
\noalign{\smallskip}
\multicolumn{7}{c}{MWC\,17}\\
\noalign{\smallskip}	
\hline
\noalign{\smallskip}	
line     & $I_{\rm line}$ & $V/R$ & $\hspace{1em}W$ & FWZI  & FWHM & $\Delta v$ \\
\noalign{\smallskip}    
\hline
\noalign{\smallskip}    
\ha                    & 180.1 & 0.72 & $-680$  & $\ga$2050 & 178 & 90 \\
\HeI$\lambda$5876\AA\      & 5.7   & 0.68 & $-7.2$  & 200  & 97  & 54 \\
$[$\OI ]$\lambda$6300\AA\  & 25.4  & 0.91 & $-27.1$ & 103  & 51  & 20 \\
$[$\NII ]$\lambda$6583\AA\ & 3.4   & 0.92 &$-3.5$   & 156  & 70  & 16 \\
$[$\FeII ]$\lambda$7155\AA\ & 3.6   &  --  & $-2.5$  & 101  & 45  & -- \\
$[$\SIII ]$\lambda$6312\AA\ & 3.1   & 0.76 & $-3.5$  & 139  & 75  & 45 \\
\noalign{\smallskip}    
\hline
\noalign{\smallskip}    

\multicolumn{7}{c}{MWC\,84}\\
\noalign{\smallskip}    
\hline
\noalign{\smallskip}    
line     & $I_{\rm line}$ & $V/R$ & $\hspace{1em}W$ & FWZI  & FWHM & $\Delta v$ \\
\noalign{\smallskip}    
\hline
\noalign{\smallskip}    
\ha                    & 109	  &  --   &  $ -245$  & $\ga$1280 & 132 & -- \\
\HeI$\lambda$5876\AA\      &30.7	  &  --   &  $-52.5$  & 510  & 112 & -- \\
$[$\OI ]$\lambda$6300\AA\  &$\le1.05$ &  --   &  $\le0.1$ & --   & --  & -- \\
$[$\FeII ]$\lambda$7155\AA\ & 1.08  &  --  & $-0.19$  & 96  & (76:)  & -- \\
$[$\NII ]$\lambda$6583\AA\  &2.7	  &  1.00:  & $-2.5$   & 159  & 58  & 14: \\
$[$\SIII ]$\lambda$6312\AA\ &1.2	  &  --   &  $-0.30$  & 117  & 84  & -- \\
\noalign{\smallskip}    
\hline
\noalign{\smallskip}    

\multicolumn{7}{c}{MWC\,137}\\
\noalign{\smallskip}    
\hline
\noalign{\smallskip}    
line     & $I_{\rm line}$ & $V/R$ & $\hspace{1em}W$ & FWZI  & FWHM & $\Delta v$ \\
\noalign{\smallskip}    
\hline
\noalign{\smallskip}    
\ha (1987)                    & 86.9    &  --    & $ -395$ & $\ga$1400 & 197   & -- \\
\ha (2002)                   & 105.3     &  --    & $ -464$ & $\sim$1600 & 196   & -- \\
\HeI$\lambda$5876\AA\  (87)       & --  & --     & $+1.3$  & --  & --    &-- \\
\HeI$\lambda$5876\AA\  (02)       & 1.9 & --     & $-4.1$  & 796  & 270    &-- \\
$[$\OI ]$\lambda$6300\AA\    & 1.6	  & 1.20   & $-0.78$ & 141  & 58    & 29 \\
$[$\FeII ]$\lambda$7155\AA\  &  1.05   & --	   &$-0.09$  & 132:  & 77:   & -- \\
$[$\NII ]$\lambda$6583\AA\   & 1.45    & 0.96     &$-0.39$  & 95   & 34    & 11: \\
$[$\SIII ]$\lambda$6312\AA\  & 1.14    & --	   & $-0.10$ & 69   & 32    & -- \\
\noalign{\smallskip}    
\hline
\noalign{\smallskip}    

\multicolumn{7}{c}{MWC\,297}\\
\noalign{\smallskip}    
\hline
\noalign{\smallskip}    
line     & $I_{\rm line}$ & $V/R$ & $\hspace{1em}W$ & FWZI  & FWHM & $\Delta v$ \\
\noalign{\smallskip}    
\hline
\noalign{\smallskip}    
\ha                     & 98.2 &  --   & $ -203$ & $\ga$1650 & 173	& -- \\
\HeI$\lambda$5876\AA\        & 1.16 & --    & $-0.30$ & 299  & 139	& -- \\
$[$\OI ]$\lambda$6300\AA\    & 14.9 & 0.92  & $-8.0$  & 77	& 29	& 9 \\
$[$\NII ]$\lambda$6583\AA\   & 1.35 & --    & $-0.21$ & 82	& 22	& -- \\
\noalign{\smallskip}    
\hline
\noalign{\smallskip}    

\multicolumn{7}{c}{MWC\,300}\\
\noalign{\smallskip}    
\hline
\noalign{\smallskip}    
line     & $I_{\rm line}$ & $V/R$ & $\hspace{1em}W$ & FWZI  & FWHM & $\Delta v$ \\
\noalign{\smallskip}    
\hline
\noalign{\smallskip}    
\ha                     & 89.0   & 0.38    & $ -150$ & $\ga$1140  & 65   & 78 \\
\HeI$\lambda$5876\AA\        & 2.10   & --      & $-1.41$ & 201  & 59   & -- \\
$[$\OI ]$\lambda$6300\AA\    & 24.5	 & --	   & $-10.7$ & 75   & 21   & -- \\
$[$\NII ]$\lambda$6583\AA\   & 3.1   & --	   &$-1.33$  & 88   & 25   & -- \\
$[$\FeII ]$\lambda$7155\AA\  & 1.90   & --	   &$-0.61$  & 65   & 29   & -- \\
\noalign{\smallskip}    
\hline
\noalign{\smallskip}    

\multicolumn{7}{c}{MWC\,342}\\
\noalign{\smallskip}    
\hline
\noalign{\smallskip}    
line     & $I_{\rm line}$ & $V/R$ & $\hspace{1em}W$ & FWZI  & FWHM & $\Delta v$ \\
\noalign{\smallskip}    
\hline
\noalign{\smallskip}    
\ha (1987)              & 60.5   &  0.27  & $ -225$ & $\ga$1650 & 117   & 138 \\
\ha  (2000)             & 58.3   &  0.20  & $ -240$ & $\sim$1600 & 91    & 210 \\
\HeI$\lambda$5876\AA\        & 1.21   & --	  & $-1.20$ & 440  &281    & -- \\
$[$\OI ]$\lambda$6300\AA\    & 3.94   & 1.20   & $-3.90$ & 124  & 38    & 12 \\
$[$\NII ]$\lambda$6583\AA\   & 1.16   & --	  & $-0.14$ & 86   & 30    & -- \\
$[$\FeII ]$\lambda$7155\AA\  & 1.22   & --	  & $-0.15$ & 89   & 25    & -- \\
\noalign{\smallskip}    
\hline
\hline
\noalign{\smallskip}

\end{tabular}
\label{para1}
\end{table}

\addtocounter{table}{-1}
\begin{table}[tbp]
\tabcolsep=3pt
\caption[]{Line parameters, continued.
}
\begin{tabular}{lllllll}
\noalign{\smallskip}    
\hline
\hline
\noalign{\smallskip}    
\multicolumn{7}{c}{MWC\,349}\\
\noalign{\smallskip}    
\hline
\noalign{\smallskip}    
line     & $I_{\rm line}$ & $V/R$ & $\hspace{1em}W$ & FWZI  & FWHM & $\Delta v$ \\
\noalign{\smallskip}    
\hline
\noalign{\smallskip}    
\ha                     &  184  &0.62 & $ -189$  & $\ga$1230  & 152   & 72 \\
\HeI$\lambda$5876\AA\       & 11.4   &0.79 & $-22.7$  &257   &130    &86 \\
$[$\OI ]$\lambda$6300\AA\   & 3.94    &0.83 & $-3.74$  & 114  & 67    & 31 \\
$[$\NII ]$\lambda$6583\AA\  & 6.25    &1.09 & $-10.3$  & 175  & 119   & 69 \\
$[$\FeII ]$\lambda$7155\AA\ & 3.50    &0.90 & $-3.96$  & 123  & 66    & 32 \\
$[$\SIII ]$\lambda$6312\AA\ & 3.07    &0.91 & $-4.05$  & 173  & 122   & 90 \\
\noalign{\smallskip}    
\hline
\noalign{\smallskip}    

\multicolumn{7}{c}{MWC\,645}\\
\noalign{\smallskip}    
\hline
\noalign{\smallskip}    
line     & $I_{\rm line}$ & $V/R$ & $\hspace{1em}W$ & FWZI  & FWHM & $\Delta v$ \\
\noalign{\smallskip}    
\hline
\noalign{\smallskip}    
\ha                     & 478     & 0.38   &  $ -195$ & $\ga$1330 & 70  & 188 \\
$[$\OI ]$\lambda$6300\AA\   & 10.1     & 0.83   &  $-14.0$ & 279  & 74  & 19 \\
$[$\NII ]$\lambda$6583\AA\  & 1.24      & --     &  $-0.30$ & 105  & 69  & -- \\
$[$\FeII ]$\lambda$7155\AA\ & 5.55      & 0.82   &  $-8.3$  & 260  & 70  & 26 \\
\noalign{\smallskip}    
\hline
\noalign{\smallskip}    
\multicolumn{7}{c}{MWC\,939}\\
\noalign{\smallskip}    
\hline
\noalign{\smallskip}    
line     & $I_{\rm line}$ & $V/R$ & $\hspace{1em}W$ & FWZI  & FWHM & $\Delta v$ \\
\noalign{\smallskip}    
\hline
\noalign{\smallskip}    
\ha\ 1987                & 228    & 0.77    &  $ -391$ & $\ga$1260  & 174  & 96 \\
\ha\ 1988                & 191    & 0.87    &  $ -386$ & $\ga$1140  & 169  & 86 \\
\ha\ 2000                & 173    & 0.88    &  $ -270$ & 2700 & 160  & 83 \\
$[$\OI ]$\lambda$6300\AA\   &20.5	   & 0.89    &  $-15.8$ & 138  & 42   & 16 \\
$[$\NII ]$\lambda$6583\AA\  &2.57	   & 0.97    &  $-1.15$ & 56   & 33   & 11 \\
$[$\FeII ]$\lambda$7155\AA\ &5.13	   & 0.98    &  $-3.64$ & 148  & 32   & 10 \\
\FeII$\lambda$6456\AA\      &1.65	   & 1.68    &  $-1.43$ &153   & 78   & 36\\
\noalign{\smallskip}    
\hline
\noalign{\smallskip}    

\multicolumn{7}{c}{MWC\,1055}\\
\noalign{\smallskip}    
\hline
\noalign{\smallskip}    
line     & $I_{\rm line}$ & $V/R$ & $\hspace{1em}W$ & FWZI  & FWHM & $\Delta v$ \\
\noalign{\smallskip}    
\hline
\noalign{\smallskip}    
\ha\ (1987)                    &  68    & 0.04  & $ -181$ & $\ga1740$ & 101  &  (206)  \\
\ha\ (2000)                    &  62     & 0.03  & $ -113$ & 3300 & 113  & (330) \\
$[$\OI ]$\lambda$6300\AA\   & 2.54     & 0.86  & $-0.90$ & 54   & 28   & 7: \\
$[$\NII ]$\lambda$6583\AA\  & 1.07	   & --     &  $-0.03$ & 26:    & 13:   & -- \\
$[$\FeII ]$\lambda$7155\AA\ & 1.11     & --    & $-0.07$ & 47   & 26   & -- \\
\FeII$\lambda$6456\AA\      & 1.12     & --    & $-0.27$ & 218  & 99   & -- \\
\noalign{\smallskip}    
\hline
\noalign{\smallskip}    

\multicolumn{7}{c}{Hen\,230}\\
\noalign{\smallskip}    
\hline
\noalign{\smallskip}    
line     & $I_{\rm line}$ & $V/R$ & $\hspace{1em}W$ & FWZI  & FWHM & $\Delta v$ \\
\noalign{\smallskip}    
\hline
\noalign{\smallskip}    
\ha                     & 96.2    & --  &  $ -229$ & $\ga$1550  & 119  & -- \\
$[$\OI ]$\lambda$6300\AA\   & 2.26    & --  &  $-0.80$ & 88   & 25   & -- \\
$[$\NII ]$\lambda$6583\AA\  & 1.16    & --  &  $-0.05$ & 41   & 16   & -- \\
$[$\FeII ]$\lambda$7155\AA\ & 1.25    & --  &  $-0.09$ & 36   & 14   & -- \\
\FeII$\lambda$6456 \AA\     & 1.20    & --  &  $-0.33$ & 156  & 89   & -- \\
\noalign{\smallskip}    
\hline
\noalign{\smallskip}    

\multicolumn{7}{c}{Hen\,485}\\
\noalign{\smallskip}    
\hline
\noalign{\smallskip}    
line     & $I_{\rm line}$ & $V/R$ & $\hspace{1em}W$ & FWZI  & FWHM & $\Delta v$ \\
\noalign{\smallskip}    
\hline
\noalign{\smallskip}    
\ha\  1986                 & 45.6    & --      & $ -170$ & $\ga$1800  & 159  & -- \\
\ha\ 1988                  & 44.7    & 0.04:   & $ -156$ & $\ga$1830  & 133  & -- \\
\HeI$\lambda$5876\AA\          & 1.32    & --      & $-0.92$ & 374  & 127  & -- \\
$[$\OI ]$\lambda$6300 \AA\     & 1.34    & --      & $-0.34$ & 116  & 41   & -- \\
$[$\NII ]$\lambda$6583\AA\ (86)& 1.67    & --      & $-0.31$ & 53  & 22   & -- \\
$[$\NII ]$\lambda$6583\AA\  (88)& 2.86    & --      & $-0.92$ & 62  & 22   & -- \\
$[$\FeII ]$\lambda$7155\AA\    & 1.21    & --      & $-0.32$ & 153  & 45   & -- \\
\FeII$\lambda$4549 \AA\        & 2.44    & --      & $-2.12$ & 198  & 85   & -- \\
\noalign{\smallskip}    
\hline
\hline
\noalign{\smallskip}    

\end{tabular}
\label{para2}
\end{table}

\addtocounter{table}{-1}
\begin{table}[ht]
\tabcolsep=3pt
\caption[]{Line parameters, continued.
}
\begin{tabular}{lllllll}
\noalign{\smallskip}	
\hline
\hline
\noalign{\smallskip}	
\multicolumn{7}{c}{Hen\,1191}\\
\noalign{\smallskip}    
\hline
\noalign{\smallskip}    
line     & $I_{\rm line}$ & $V/R$ & $\hspace{1em}W$ & FWZI  & FWHM & $\Delta v$ \\
\noalign{\smallskip}    
\hline
\noalign{\smallskip}    
\ha                     &723   & 0.62  &  $ -821$ & $\ga$1280  & 93  & 69 \\
$[$\OI ]$\lambda$6300\AA\   &124    & --	 &  $-32.6$ & 90  & 18  & -- \\
$[$\NII ]$\lambda$6583\AA\  &2.36	 & --	 &  $-0.92$ & 93   & 24  & -- \\
$[$\FeII ]$\lambda$7155\AA\ &39.6	 & --	 &  $-12.7$ & 85   & 19  & -- \\
\FeII$\lambda$6456 \AA\     &22.6	 & --	 &  $-8.90$ & 136  & 17  & -- \\
\noalign{\smallskip}    
\hline
\noalign{\smallskip}    

\multicolumn{7}{c}{HD\,45677}\\
\noalign{\smallskip}    
\hline
\noalign{\smallskip}    
line     & $I_{\rm line}$ & $V/R$ & $\hspace{1em}W$ & FWZI  & FWHM & $\Delta v$ \\
\noalign{\smallskip}    
\hline
\noalign{\smallskip}    
\ha                     & 59.9   &0.70  &  $-194$  & $\ga$1140  & 172  & 66 \\
\ha , red peak        &   59.9    & 1.00 &   &    &    & 25  \\
$[$\OI ]$\lambda$6300\AA\   & 8.59   &1.00 &  $-5.10$ & 97   & 30   & 6 \\
$[$\NII ]$\lambda$6583\AA\  & 1.93   &--   &  $-0.48$ & 75   & 21   & -- \\
$[$\FeII ]$\lambda$7155\AA\ & 1.21   &--   &  $-0.27$ & 126  & 20   & -- \\
\FeII$\lambda$6456\AA\      & 1.34   &1.13 &  $-0.52$ & 249  & 67   & 29 \\
\noalign{\smallskip}    
\hline
\noalign{\smallskip}    

\multicolumn{7}{c}{HD\,87643}\\
\noalign{\smallskip}    
\hline
\noalign{\smallskip}    
line     & $I_{\rm line}$ & $V/R$ & $\hspace{1em}W$ & FWZI  & FWHM & $\Delta v$ \\
\noalign{\smallskip}    
\hline
\noalign{\smallskip}    
\ha\ ~1986                   & 55.0   &0.32 &  $ -274$ & $\ga$1500  & 146  & 215 \\
\ha\ ~1988                   & 92.0   &0.24 &  $ -323$ & $\ga$1520  & 108  & 184 \\
$[$\OI ]$\lambda$6300\AA\ (86)   & 1.92   &--   &  $-1.02$ & 180  & 37   & -- \\
$[$\OI ]$\lambda$6300\AA\ (88)   & 2.43   &--   &  $-1.23$ & 181  & 36   & -- \\
$[$\FeII ]$\lambda$7155\AA\    & 1.65   &--   &  $-0.60$ & 260  & 38   & -- \\
\FeII$\lambda$6456\AA\         & 2.06   &--   &  $-1.91$ & 357  & 69   & -- \\
\noalign{\smallskip}    
\hline
\noalign{\smallskip}    

\multicolumn{7}{c}{ CD\,$-24\degr$5721}\\
\noalign{\smallskip}    
\hline
\noalign{\smallskip}    
line     & $I_{\rm line}$ & $V/R$ & $\hspace{1em}W$ & FWZI  & FWHM & $\Delta v$ \\
\noalign{\smallskip}    
\hline
\noalign{\smallskip}    
\ha                      & 38.3  &0.65 &  $ -182$ & $\ga$1740  & 240 & 120 \\
\HeI$\lambda$5876\AA\        & --	 & --  &  $+0.89$ & --   & 310 & -- \\
$[$\OI ]$\lambda$6300\AA\    & 2.22  &1.04 &  $-1.15$ & 141  & 46  & 12 \\
$[$\NII ]$\lambda$6583\AA\   & 1.27  & 1.0  &  $-0.17$ & 69   & 26  & 6:  \\
$[$\FeII ]$\lambda$4287\AA\  & 1.48  &--  &  $-0.15$ & 49   & 24  & -- \\
\FeII$\lambda$4549 \AA\      & --	 &--   &  $+0.44$ & --   & 11   & -- \\
\noalign{\smallskip}    
\hline
\noalign{\smallskip}    

\multicolumn{7}{c}{CPD\,$-57\degr$2874}\\
\noalign{\smallskip}    
\hline
\noalign{\smallskip}    
line     & $I_{\rm line}$ & $V/R$ & $\hspace{1em}W$ & FWZI  & FWHM & $\Delta v$ \\
\noalign{\smallskip}    
\hline
\noalign{\smallskip}    
\ha                     &  23.2  &0.30 & $ -92.7$ & $\ga$1920  & 146 & 130 \\
$[$\OI ]$\lambda$6300\AA\   &  1.18  &1.00 & $-0.37$  & 194  & 96  & 31: \\
$[$\NII ]$\lambda$6583\AA\  &  1.19  &1.05 & $-0.29$  & 168  & 55  & 22 \\
$[$\FeII ]$\lambda$7155\AA\ &  1.13  &-- & $-0.41$  & 278  & 120 & -- \\
\FeII$\lambda$6456 \AA\     &  1.16  &0.88 & $-0.51$  & 322  & 175 & 99 \\
\noalign{\smallskip}    
\hline
\noalign{\smallskip}    

\multicolumn{7}{c}{ CPD\,$-52\degr$9243 }\\
\noalign{\smallskip}    
\hline
\noalign{\smallskip}    
line     & $I_{\rm line}$ & $V/R$ & $\hspace{1em}W$ & FWZI  & FWHM & $\Delta v$ \\
\noalign{\smallskip}    
\hline
\noalign{\smallskip}    
\ha                     & 15.8  &0.08 &   $ -54.9$ & $\ga$1460  & 174 & -- \\
$[$\OI ]$\lambda$6300\AA\   & 1.30  &0.72:&   $-0.40$  & 111  & 67  & 36: \\
$[$\FeII ]$\lambda$7155\AA\ & 1.21  &0.75:&   $-0.28$  & 119  & 65  & 39: \\
\FeII$\lambda$6456\AA\      & 2.09  &--   &   $-2.16$  & --   & 102 & --  \\
\noalign{\smallskip}    
\hline
\hline
\noalign{\smallskip}    

\end{tabular}
\end{table}

\section{Heliocentric radial velocities}
\label{helio}
In Table \ref{rad2} heliocentric radial velocities of the different lines are
listed. For double-peaked 
emission lines ``e$_{\rm b}$'' 
and ``e$_{\rm r}$'' denote the velocities of the blue and red emission components,
respectively. For single-peaked emission lines ``c'' 
denotes the velocities of the center of emission. 
In the case of double-peaked lines ``c'' is the velocity of the central absorption or 
of the central dip between the emission peaks.  For  pure absorption lines 
it is the central velocity of the absorption feature. Velocities of absorption 
components are additionally flagged by the letter ``a''. For type 1 profiles absorption velocities  are 
listed in the column ``e$_{\rm b}$/a''. The laboratory wavelengths used  for 
the forbidden lines are 6300.31\AA\ for  [\OI ], 7155.14\AA\ for [\FeII ], 
6583.37\AA\ for [\NII ],  and  6312.10\AA\ for [\SIII ]. 
Table \ref{natrium} lists the heliocentric radial velocities of the absorption components of the \NaI\,D
doublet. In some cases not all components were detectable in each of the doublet components.
This is often due to saturation effects in the strong line cores. 

\begin{table*}[h]
\caption[]{Heliocentric radial velocities of the emission lines. } 
\begin{tabular}{l|lll|lll|lll|lll}
\noalign{\smallskip}    
\hline
\hline
\noalign{\smallskip}    
  line                   & \multicolumn{3}{c|}{MWC\,17}&
                       \multicolumn{3}{c|}{MWC\,84}& 
                       \multicolumn{3}{c|}{MWC\,137}&
                       \multicolumn{3}{c}{MWC\,297}
                       \\
   & \multicolumn{1}{c}{e$_{\rm b}$/a} &\multicolumn{1}{c}{c} & \multicolumn{1}{c|}{e$_{\rm r}$}
   & \multicolumn{1}{c}{e$_{\rm b}$/a} &\multicolumn{1}{c}{c} & \multicolumn{1}{c|}{e$_{\rm r}$}
   & \multicolumn{1}{c}{e$_{\rm b}$/a} &\multicolumn{1}{c}{c} & \multicolumn{1}{c|}{e$_{\rm r}$}
   & \multicolumn{1}{c}{e$_{\rm b}$/a} &\multicolumn{1}{c}{c} & \multicolumn{1}{c}{e$_{\rm r}$}
   \\ 
\noalign{\smallskip}    
\hline
\noalign{\smallskip}    
\ha                    & $-101$ & $-59$& $-11$
                       &        &$-84$ &
                       &        &$+43$ & 
                       &        & $-5$ & 
                       \\
$[$\OI ]$\lambda$6300\AA  & $-59$&$-51$ & $-39$
                       &      &      & 
                       & $+25$& $+43$& $+54$
                       &$-15$ & $-13$& $-6$
                       \\
$[$\NII ]$\lambda$6583\AA  & $-53$ & $-47$ & $-37$
                       & $-49$ &$-42$  & $-35$
                       & $+38$ & $+45$ & $+49$
                       &       & $-5$  &
                       \\
$[$\FeII ]$\lambda$7155\AA &      &  $-46$ & 
                       &      &      & 
                       &      & $+48$&
                       &      &      &
                       \\
\HeI$\lambda$5876\AA       & $-78$ &  $-57$  & $-24$ 
                       & $-52$ &         & $+9$
                       &       & $+20$a   &
                       &       &  $+16$  &
                       \\
\HeI$\lambda$5876\AA  (2002) &       &         &     
                       &       &         &  
                       &       & $+36^f$   &
                       &       &         &
                       \\
\HeI$\lambda$6678\AA       & $-67$ &  $-53$  & $-20$ 
                       & $-59$ &         & 
                       &       &  $+29:$ & 
                       &       &         & 
                       \\
$[$\SIII ]$\lambda$6312\AA & $-70$ &  $-44$  & $-25$
                       &       &   $-45$ &
                       &       &  $+41:$ & 
                       &       &         &
                       \\
\noalign{\smallskip}    
\hline
\noalign{\smallskip}    
                       &
                       \multicolumn{3}{c|}{MWC\,300}&
                       \multicolumn{3}{c|}{MWC\,342}&
                       \multicolumn{3}{c|}{MWC\,349}& 
                       \multicolumn{3}{c}{MWC\,645}
                       \\
\noalign{\smallskip}    
\hline
\noalign{\smallskip}    
\ha                    &$-25$ & $-1$ & $+51$
                       & $-141$ & $-81$ & $-3$
                       & $-69$  &  $-24$& $+3$
                       & $-218$ & $-108$&$-30$
                       \\
$[$\OI ]$\lambda$6300\AA   &      & $+21$ &
                       & $-38$  & $-31$ & $-26$
                       & $-23$  & $-11$ & $+8$
                       & $-58$  &  $-54$ &$-39$
                       \\
$[$\NII ]$\lambda$6583\AA   &      & $+24$&
                        &      &$-31$ & 
                        & $-48$& $-4$ & $+21$
                        &      & $-47$&
                       \\
$[$\FeII ]$\lambda$7155\AA  &      &$+23$ & 
                        &      &$-29$ & 
                        & $-21$& $-6$ & $+11$
                        & $-65$& $-56$&$-39$
                       \\
\HeI$\lambda$5876\AA       & $-30^e$a&$+54$   & 
                       &         & $-38$  & 
                       & $-52$   &  $-20$ & $+34$
                       &         &        &
                       \\
\HeI$\lambda$6678\AA       &       &         & 
                       &  $-272$a & $-21$  & 
                       & $-56$ &  $-26$  & $+30$
                       &       &         &
                       \\
$[$\SIII ]$\lambda$6312\AA &      &      & 
                       &      &      & 
                       & $-55$     & $-10$     &$+35$
                       &      &      &
                       \\
\noalign{\smallskip}    
\hline
\noalign{\smallskip}    
                      &\multicolumn{3}{c|}{MWC\,939}&
                       \multicolumn{3}{c|}{MWC\,1055}&
                       \multicolumn{3}{c|}{Hen\,230}&
                       \multicolumn{3}{c}{Hen\,485}
                       \\
\noalign{\smallskip}    
\hline
\noalign{\smallskip}    
\ha (1986)             &      &      &
                       &      &      & 
                       &      &      &
                       &      & $+15$& 
                       \\
\ha\ (1987)             & $-43$  & $+7$   &$+53$
                       & $-283$ & $-221$ & $-77$
                       &        &        &  
                       &        &        &  
                       \\
\ha\ (1988)             & $-36$  &  $+6$  &$+47$
                       &        &        &  
                       &        &$+38$   & 
                       & $-245$a&$-3$    &
                       \\
\ha\ (2000)             & $-33^f$  &  $+5^f$  &$+50^f$
                       &  $-408$      &  $-315$      &  $-78$
                       &        &    & 
                       &  &     &
                       \\
$[$\OI ]$\lambda$6300\AA   & $+1$   & $+9$   & $+17$
                       &$-97:$  & $-94:$ & $-90$
                       &        & $+25$  & 
                       &        & $-10$ & 
                       \\
$[$\NII ]$\lambda$6583\AA  & $+7$     &  $+12$    & $+18$
                       &      & $-91^f$     & 
                       &      & $+29$&
                       &      & $-7$ &
                       \\
$[$\FeII ]$\lambda$7155\AA & $+6$     & $+10$  & $+16$
                       &      &   $-91^f$   & 
                       &      & $+27$&
                       &      &$-12$ &
                       \\
\FeII $\lambda$6456\AA     &      &      & 
                       &      & $-94^f$     & 
                       &      &$+19$ &
                       & $-28$& $-14$&$-6$
                       \\
\HeI$\lambda$5876\AA       &      &  $+24^f$    & 
                       &      &  $-112^f$    & 
                       &      &      &
                       &      & $-12$&
                       \\
\noalign{\smallskip}    
\hline
\noalign{\smallskip}    
                     & \multicolumn{3}{c|}{Hen\,1191}&
                       \multicolumn{3}{c|}{HD\,45677}&
                       \multicolumn{3}{c|}{HD\,87643}&
                       \multicolumn{3}{c}{CD$-24{\degr}5721$}
                       \\
\noalign{\smallskip}    
\hline
\noalign{\smallskip}    
\ha\ (1986)             &         &         & 
                       &         &         & 
                       & $-189$  & $-105$  &$+26$
                       & $-12$   & $+48$   &$+108$
                       \\
\ha\ (1988)             & $-50$  & $-17$    & $+19$
                       & $-17$  & $+6$     & $+49^a$
                       & $-170$ & $-108$   &$+14$
                       &        &          & 
                       \\
$[$\OI ]$\lambda$6300\AA  &        & $-8$  & 
                       &  $+15$ & $+18$ & $+21$
                       &        & $-9$  & 
                       & $+49$  & $+55$ & $+61$
                       \\
$[$\NII ]$\lambda$6583\AA  &      & $-7$ & 
                       &      & $+21$&
                       &      &      &
                       &  $+55$    &$+58$ &$+61$
                       \\
$[$\FeII ]$\lambda$7155\AA &      & $-7$ & 
                       &      & $+21$& 
                       &      & $-8$ &
                       &      &$+55^b$&
                       \\
\FeII $\lambda$6456\AA     &      & $-15$   & 
                       & $-2$ & $+16$   & $+27$
                       &      & $-21$  & 
                       &      & $+59$a$^c$ &
                       \\
\HeI$\lambda$5876\AA       &      &      & 
                       &      &      & 
                       &      & $+23$&
                       &      & $-14$a&
                       \\
\noalign{\smallskip}    
\hline
\noalign{\smallskip}    
                     & 
                       \multicolumn{3}{c|}{CPD$-57{\degr}2874$}&
                       \multicolumn{3}{c|}{CPD$-52{\degr}9243$}&
                       \multicolumn{6}{c}{}
                       \\
\noalign{\smallskip}    
\hline
\noalign{\smallskip}    
\ha                    & $-108$ & $-75$  & $+22$
                       & $-194$a&$-38$   & 
                       &\multicolumn{6}{c}{}
                       \\
$[$\OI ]$\lambda$6300\AA   &       & $+3$    &
                       &       & $-48^d$ &
                       &\multicolumn{6}{c}{}
                       \\
$[$\NII ]$\lambda$6583\AA  & $-3$ & $+8$ & $+19$
                       &      &      & 
                       &\multicolumn{6}{c}{}
                       \\
$[$\FeII ]$\lambda$7155& $-14$     & $+5$ & $+21$  
                       &      &$-50^d$& 
                       &\multicolumn{6}{c}{}
                       \\
\FeII $\lambda$6456\AA     & $-59$   & $-16$ & $+40$
                       &  $-238$a& $-52$ & 
                       &\multicolumn{6}{c}{}
                       \\
\HeI$\lambda$5876\AA       &$-148$a   &$+31$e   & $+133$a
                       &$-300$a   &    & 
                       &\multicolumn{6}{c}{}
                       \\
\noalign{\smallskip}    
\hline
\noalign{\smallskip}    
\multicolumn{4}{l}{$^a$two components: $+35$/$+60$\,\kms ;} &
\multicolumn{2}{l}{$^b$[\FeII ]$\lambda$4287\AA}&
\multicolumn{7}{l}{$^c$mean of \FeII $\lambda\lambda$4549,4556\AA ;}  \\
\multicolumn{4}{l}{$^d$centroid of asymmentric line} &
\multicolumn{9}{l}{$^e$additional narrow absorption component at $-100$\,\kms } \\
\multicolumn{9}{l}{$^f$observed with FOCES}\\
\end{tabular}
\label{rad2}
\end{table*}
\clearpage

\begin{table*}[ht]
\caption[]{Heliocentric radial velocities, $v_{\rm i}$, of the multiple absorption components, $i$, 
of the NaI\,D doublet in \kms .
}
\begin{tabular}{llllllll}
\noalign{\smallskip}    
\hline
\hline
\noalign{\smallskip}	
MWC\,17 & \multicolumn{1}{c}{$v_1$}  &\multicolumn{1}{c}{$v_2$}  & \multicolumn{1}{c}{$v_3$} & 
\multicolumn{1}{c}{$v_4$} & \multicolumn{1}{c}{$v_5$} & \multicolumn{1}{c}{$v_6$} & 
\multicolumn{1}{c}{$v_7$} \\ 
\noalign{\smallskip}
\hline
\noalign{\smallskip}
 \NaI\,D\,1 & $-48.7$ & $-41.5$ & $-24.5$ &  $-11.3$ & $-4.5$ & &  \\ 
 \NaI\,D\,2 & $-51.6$ & $-41.1$ & $-24.7$ & $-13.3$ &  $-1.7$ & & \\ 
\noalign{\smallskip}
\hline
\hline
\noalign{\smallskip}	
MWC\,84 &    &  &   & &  &  &  \\ 
\noalign{\smallskip}
\hline
\noalign{\smallskip}
 \NaI\,D\,1 &  & $-35.1$& $-6.5$& $+3.1$ & &  \\ 
 \NaI\,D\,2  &$-42.5$ &  $-32.3$&  $-7.2$& $+4.2$& &  \\ 
\noalign{\smallskip}    
\hline
\hline
\noalign{\smallskip}	
MWC\,137 &    &   &   &   &   &   &   \\ 
\noalign{\smallskip}
\hline
\noalign{\smallskip}
 \NaI\,D\,1 &  $+17.3$& $+25.9$& $+34.6$ & & &  \\ 
 \NaI\,D\,2 & $+18.$1& $+28.3$ & $+36.1$  & & &  \\ 
\noalign{\smallskip}	
\hline
\hline
\noalign{\smallskip}	
MWC\,297 &   &  &   & &  &  &  \\ 
\noalign{\smallskip}
\hline
\noalign{\smallskip}
 \NaI\,D\,1 &  $-44.8$   & $-26.6$ & $-17.0$  & $-9.7$&  &  &   \\  
 \NaI\,D\,2 &  $-44.2$  & $-26.5$ &   & $-7.7$&  &  &   \\ 
\noalign{\smallskip}	
\hline
\hline
\noalign{\smallskip}	
MWC\,300 &   &  &   & &  &  &  \\ 
\noalign{\smallskip}
\hline
\noalign{\smallskip}
 \NaI\,D\,1 &  $-23.8$  & $-6.7$ &        & $+6.2$ &  &  &  \\  
 \NaI\,D\,2  & $-22.5$  & $-8.9$ &  $-0.2$   & $+5.5$&  &  &  \\ 
\noalign{\smallskip}	
\hline
\hline
\noalign{\smallskip}	
MWC\,342 &   &  &   & &  &  &  \\ 
\noalign{\smallskip}
\hline
\noalign{\smallskip}
 \NaI\,D\,1  &  $-15.6$ & $-7.5$ &   & &  &  &    \\  
 \NaI\,D\,2  &  $-11.5$ & $-4.3$ &   & &  &  &     \\ 
\noalign{\smallskip}	
\hline
\hline
\noalign{\smallskip}	
MWC\,349 &   &  &   & &  &  &  \\ 
\noalign{\smallskip}
\hline
\noalign{\smallskip}
 \NaI\,D\,1  &    & $-7.3$ &$-0.3$	& &  &  &	 \\  
 \NaI\,D\,2  &  $-15.8$ & $-8.1$ &	$-0.7$& &  &  &	  \\ 
\noalign{\smallskip}	
\hline
\hline
\noalign{\smallskip}	
MWC\,939 &   &  &   & &  &  &  \\ 
\noalign{\smallskip}
\hline
\noalign{\smallskip}
 \NaI\,D\,1  &  $-8.9$  & $+11.5$ &   & &  &  &  \\ 
 \NaI\,D\,2  &  $-8.1$ & $+11.4$ &   & &  &  &  \\ 
\noalign{\smallskip}	
\hline
\hline
\noalign{\smallskip}	
MWC\,1055 &   &  &   & &  &  &  \\ 
\noalign{\smallskip}
\hline
\noalign{\smallskip}
 \NaI\,D\,1   &  $-209.7$  & $-188.7$ & $-43.5$  &$-14.4$ &  &  &  \\
 \NaI\,D\,2   &  $-207.9$  & $-191.1$  &$ -42.6$   & $-13.3$ &  &  &  \\
\noalign{\smallskip}	
\hline
\hline
\noalign{\smallskip}	
MWC\,485 &   &  &   & &  &  &  \\ 
\noalign{\smallskip}
\hline
\noalign{\smallskip}
 \NaI\,D\,1  &  $-73.8$  & $-66.3$  & $-45.2$   &$-29.7$  & $-15.8$  &  $-3.5$  &  $+6.4$ \\
 \NaI\,D\,2  &  $-74.3$  & $-67.1$ & $-45.4$   & $-29.3$& $-16.1$ &  $-4.0$  &  $+6.6$ \\ 
\noalign{\smallskip}	
\hline
\hline
\noalign{\smallskip}	
HD\,87643 &   &  &   & &  &  &  \\ 
\noalign{\smallskip}
\hline
\noalign{\smallskip}
 \NaI\,D\,1   &  $-56.0$   & $-20.0$ &  $-2.9$   & $+7.7$ & $+24.7$ &  &  \\ 
 \NaI\,D\,2   &   $-56.3$ &  $-19.7$ &   $-3.6$ & $+7.7$ & $+25.7$ &  &  \\ 
\noalign{\smallskip}	
\hline
\hline
\noalign{\smallskip}	
CPD$-$57\degr2874 &   &  &   & &  &  &  \\ 
\noalign{\smallskip}
\hline
\noalign{\smallskip}
 \NaI\,D\,1  & $-68.8$   & $-58.5$ & $-42.9$   &$ -18.0$ &  $-4.8$  &  $+9.6$ & $+34.3$ \\  
 \NaI\,D\,2  & $-68.4$  &  &   & $-18.6$ &   $-5.4$&   $+6.8$& $+34.2$ \\  
\noalign{\smallskip}	
\hline
\hline
\noalign{\smallskip}	
CPD$-$52\degr9243 &   &  &   & &  &  &  \\ 
\noalign{\smallskip}
\hline
\noalign{\smallskip}
 \NaI\,D\,1  &  $-176.6$ & $-38.7$ &  $$ & $ -6.7$ &  &  &  \\ 
 \NaI\,D\,2  & $-173.2$  &  $-39.3$  &  $-22.4$  &$-7.3$ &  &  &  \\ 
\noalign{\smallskip}
\hline
\hline
\noalign{\smallskip}
\end{tabular}
\label{natrium}
\end{table*}
\clearpage

\section{Remarks on individual objects}
\label{remarks}

\subsection{MWC\,17}
The nature of MWC\,17 is still unclear (classification uncl\be ), although 
some indications  for a post-AGB evolutionary status exist according to 
Leibowitz (\cite{Leibowitz77}), i.e. type cPN\be\ . 
So far this star was only studied using low to medium resolution spectra, 
e.g. recently by Jaschek \& Andrillat (\cite{JaschekAndrillat99}).

In the sample studied here \object{MWC\,17} it is the star with the second 
strongest \ha\ emission. 
It is nearly as strong as that of the cPNB[e] star Hen\,1191. 
The blue peak reaches 
an intensity of $\sim75$\% of the red peak (cf.  Sect \ref{remark_hen1191}
and Table \ref{para1}). 
The \ha\ profile of \object{MWC\,17} is twice as
broad as that of Hen\,1191. The shapes profile, however, 
are very similar. Unlike 
the cPNB[e] star Hen\,1191 most   
forbidden lines of MWC\,17  
show a double-peak structure. Only [\FeII ]$\lambda7155$\AA\   is a single-peaked 
emission line. Furthermore, the widths (FWHM) of the forbidden lines of 
MWC\,17 are a factor of 2-3 larger than those of Hen\,1191 and are among the broadest
lines in the sample.

Radial velocity measurements  were published 
by Swings \& Struve (\cite{SwingsStruve41}).  They measured $RV = -28$\,\kms\ 
for the Balmer lines for which the double peak was not resolved, and $-37$\,\kms\ for 
[\FeII ]. The latter value is in  reasonable agreement with the velocity  
of $-46$\,\kms\ measured here from the coud\'e spectrum.

\subsection{MWC\,84 (= CI\,Cam)}
In the studied sample MWC\,84 is exceptional. It is the known binary system 
CI Cam for which recently an X-ray-to-radio flare was observed 
(Frontera et al. \cite{fronteraetal98}, Orr et al. \cite{orretal98}). 
The X-ray properties of MWC\,84 suggest the presence 
of a compact companion. Frontera et al. discuss the possibility of a neutron 
star, black hole and white dwarf companion. 

Based on $K$-band spectra Clark et al. (\cite{Clarketal99}) suggest the 
classification sg\be . 
Miroshnichenko et al. (\cite{Miro02b}) discussed in detail high resolution 
spectra observed in 2002. They find a distance of less than 3\,kpc and a luminosity 
of $\log L/L_{\sun} \le 4.0$. The lack of significant [\OI ] emission 
(cf. Fig. \ref{prooi}), casts some doubt on the classification as \be\ star. 
This line seems to be always present in \be-type stars. 
Despite its similarities with \be\ stars MWC\,84 is therefore 
possibly not a typical member of this object class. 

\ha\ is a very strong single-peaked emission line, 
Likewise, \HeI$\lambda5876$\AA\ is a strong emission line with a  profile resembling 
closely that of \ha . Both lines show a bump on the red flank  
indicating a more complex structure (cf. also Miroshnichenko et al. \cite{Miro02b}).

\NaI\,D is also present in emission. 
The line of MgI$\lambda6318$\AA\ shows a split profile.  
There is some  indication for a split profile also for [\NII ]. 
The width of the weak line of [\FeII ]$\lambda$7155  is similar to that of [\SIII ], 
but appears narrower than [\NII ].

\subsection{MWC\,137}
\label{mwc137}
MWC\,137 is emdedded in the nebula S\,266. In the past, this star has usually 
been considered to be a Herbig Ae/Be star 
(e.g. Finkenzeller \& Mundt \cite{FinkenzellerMundt84}, Th\'e et al. \cite{Theetal94}). 
Recently,
Esteban \& Fernandez (\cite{EstebanFernandez98}) studied high resolution 
spectra (resolution $R \approx 20\,000$) and direct narrow band images
of MWC\,137 and S\,266. They measured a heliocentric radial velocity of 
$+18\pm2$\,\kms\ for the nebula. From this they derived a kinematical distance
of about 6\,kpc and concluded that MWC\,137 most likely is a \be\ supergiant.

The velocity of \ha\ of $+43$\,\kms\ and $+42$\,\kms\ measured from the Calar Alto 
coud\'e of 1987 and the FOCES spectrum of 2002, respectively, is
significantly higher than the nebular velocity quoted by Esteban \& Fernandez.
Unfortunately, they do not give the heliocentric radial velocity of the 
stellar component of \ha .
Measuring the wavelength of the \ha\ peak from their Fig. 2 yields 
$v_{\rm rad} \approx +20...+25$\,\kms , which is also significantly smaller 
than the velocity measured in the Calar Alto spectrum of 1987. More 
observations are needed to check whether these differences are due to  radial 
velocity variations.
Interestingly, the velocity measured from the \HeI $\lambda$5876\AA\ absorption 
line, $+20$\,\kms\ (s. Table \ref{rad2}), agrees remarkably well with the 
nebular velocity.  

The absorption features of the \NaI\,D doublet are each split into 
three components with heliocentric radial
velocities of $v_{\rm hel} =  +18, +27,$ and +36\,\kms , corresponding to 
velocities with respect to the local standard of rest (LSR) of 
$v_{\rm LSR} = +30,
+39,$ and $+48$\,\kms , respectively. Comparing these velocities with the
galactic rotation curve determined by Brand \& Blitz (\cite{brandblitz93})
reveals that for interstellar lines LSR velocities above 
$\sim30$\,\kms\ should not be found along the line of sight towards MWC\,137 
if the distance of 6\,kpc as estimated by Esteban \& Fernandez is correct. 
However, it is not clear whether all components are of interstellar 
origin. The distance estimate based on the radial velocity  should therefore be 
taken with caution. 

MWC\,137 appears to be spectroscopically variable. In contrast to the strong
and rather broad \HeI$\lambda5876$\AA\  emission line detected by 
Esteban \& Fernandez in the 1994 observations and in the FOCES spectrum of 
February 2002, the spectrum of 1987 shows a broad {\em absorption} feature. Note,
however, that in 1987 \HeI$\lambda6678$\AA\  appeared in emission (Fig. \ref{prohe66}). The heliocentric radial
velocity of \HeI$\lambda5876$\AA\  in 2002 was $+36$\,\kms .
The P\,Cyg profile reported  for \NaI\,D$_2$ by Esteban \& Fernandez
is not visible in the 1987 spectrum. Weak broad \NaI\ emission was visible 
in February 2002.

\subsection{MWC\,297}
MWC\,297 is generally classified as 
a
Herbig Be star  
(e.g. Sharpless \cite{Sharpless59}, Herbig \cite{Herbig60}, Finkenzeller \& Mundt
\cite{FinkenzellerMundt84}).
For an extensive line list  based on intermediate 
resolution spectra cf. Andrillat \& Jaschek (\cite{AndrillatJaschek98}). 
Drew et al. (\cite{Drewetal97}) carried out a detailed analysis of this star 
and concluded that it is B1.5Ve zero-age  main-sequence star with a rotational
velocity of $\sim350$\,\kms . Oudmaijer 
\& Drew (\cite{OudmaijerDrew99}) obtained spectropolarimetry  
but could not detect intrinsic polarization
from the observation of \ha . They suggest that the aspect angle might be 
close to pole-on. The spectroscopic observations presented
here show very narrow forbidden lines of [\OI ] and  [\NII ] with widths of 31\,\kms\
and 21\,\kms , respectively (Tab \ref{para1}). 
Likewise, \ha\ exhibits a single emission peak. These observations 
in fact seem to be consistent with a near pole-on viewing angle. However, the
[\OI ] line though being narrow shows a split profile suggesting a viewing angle
somewhere between intermediate 
and edge-on rather than pole-on. Likewise, the high rotational velocity 
found by Drew et al. (\cite{Drewetal97}) is inconsistent with a pole-on
viewing angle. The presence of a flat structure of the circumstellar matter 
around MWC\,297 seen at radio wavelength (Drew et
al. \cite{Drewetal97}) also suggests a more edge-on aspect angle.  
The question which viewing angle is correct thus still remains
controversial. However, the split [\OI ] profile argues for a
non-spherical distribution of the circumstellar envelope of MWC\,297.

\subsection{MWC\,300}
\label{mwc300}
MWC\,300 was considered for a long time to be a Herbig Be star
(Herbig \cite{Herbig60}, Finkenzeller \& Mundt \cite{FinkenzellerMundt84}).
Allen \& Swings (\cite{AS76}) listed this object as peculiar Be star with infrared excess.
Based on the analysis of high-resolution spectra Wolf \& Stahl (\cite{WolfStahl85}), 
suggested that this star is actually a B hypergiant of 
spectral type B1Ia$^+$. Here the classification as sg\be\ is adopted although the nature 
of MWC\,300 is still controversial. 

While \ha\ exhibits a double-peak type 3 profile a P\,Cygni profile of group 1 
is found for \HeI $\lambda$5876\AA . MWC\,300 is the only star in the observed sample
exhibiting a type 1 P\,Cygni profile of \HeI . This line shows an 
additional narrow blue shifted absorption component which is very likely also 
due to \HeI$\lambda$5876\AA . Both, [\OI ]$\lambda$6300\AA\  and 
[\NII ]$\lambda$6584\AA\ are single-peaked emission lines.
The emission components of \NaI~D are disturbed on the blue side by the
interstellar absorption components.

Winkler \& Wolf (\cite{WinklerWolf89}) analyzed the emission line spectrum of 
MWC\,300 using  ESO CASPEC spectra ($R \approx 20\,000$) observed in August 1984. 
The radial velocities measured for the forbidden lines of [\FeII] and [\OI ]
of $+22.5\pm0.7$\,\kms\ agree well with the velocities measured from the Calar Alto 
coud\'e spectra of $+23$\,\kms\ and $+21$\,\kms , respectively. Winkler
(\cite{Winkler86}) also lists detailed velocities for the Balmer line profiles 
based on the same spectroscopic material. He measured $-23$\,\kms\ and $+61$\,\kms\ 
for the two emissson peaks and $+2$\,\kms\ for the central absorption of \ha . The 
central and blue peak velocities are  in good agreement with the results listed in 
Table \ref{rad2}, while for the 
red emission peak a slight difference of 10\,\kms\ might be present.   

\subsection{MWC\,342 (=  V1972\,Cyg)}
The status of MWC\,342 is still poorly known leading to a classification of 
``uncl\be ''. Intermediate resolution spectra were described by Andrillat \& Jaschek 
(\cite{AndrillatJaschek99}). Miroshnichenko \& Corporon (\cite{MP99}) discussed 
high resolution spectra  and published radial velocities for the Balmer lines 
\ha\ to \hd . Comparison with the velocity of \ha\  
listed in Table \ref{rad2} shows good agreement with their measurement. 

Both, \ha\ and [\OI ] exhibit a split line profile. The velocity difference of the peaks
of [\OI ] is only 12\,\kms\ which is about a factor of 10 smaller than for \ha . 
[\FeII ]$\lambda7155$ and [\NII ]$\lambda6584$ show  single-peaked 
emission lines with FWHM of 25\,\kms\ and 30\,\kms , respectively, whereas  [\OI ] has 
a FWHM of 69\,\kms . 
The width of the \NaI\,D emission lines at continuum level (FWZI) is 250\,\kms . For  
\HeI\ a FWZI of 440 \,\kms\ was measured. The forbidden lines are significantly 
narrower at continuum level with FWZI of the order of 100\,\kms .

\subsection{MWC\,349A}
MWC\,349A has been studied extensively during the past decades. The observations have been
carried out mainly at radio
wavelengths and in the infrared, but there are still only a few spectroscopic 
studies in the optical wavelength region, e.g. Allen \& Swings (\cite{AS76}), 
Brugel \& Wallerstein (\cite{BrugelWallerstein79}), and
Hartmann et al. (\cite{Hartmannetal80}).   
For a recent investigation presenting a  line list  based on medium resolution 
data cf.  Andrillat et al. (\cite{Andrillatetal96}). 
Despite the observational efforts it s  still discussed controversially 
whether MWC\,349A  is a massive pre-main sequence or a post-main sequence object 
(type ``uncl\be '' in Table \ref{spectype}).

Based on velocity-resolved infrared spectroscopy Hamann \& Simon 
(\cite{HamannSimon86}) suggested a model  for MWC\,349A consisting of a disk and a 
bipolar outflow similar to the model for \be\ supergiants by Zickgraf et al. 
(\cite{Zickgrafetal85}).
Split line profiles observed  by Hamann \& Simon  (\cite{HamannSimon88}) 
in the wavelength region 7500-9300\,\AA\ with a 
resolution of 30\,\kms\  were consistent with this model. 
It is 
also supported by the speckle observations 
described by Leinert (\cite{Leinert86}) and Hofmann et al. (\cite{Hofmannetal02}). 
In both studies a flat disk-like structure oriented 
perpendicular to the radio lobe found by White \& Becker (\cite{WhiteBecker85}) 
could be resolved.

All lines observed in this work exhibit clear double-peaked profiles. The lines 
of the different ions show different line widths depending on excitation 
potential. With 130\,\kms\ the line  of \HeI$\lambda5876$\AA\ has the 
largest  width (FWHM). The lines of [\OI ]$\lambda6300$\AA\  and [\FeII ]$\lambda7155$\AA\  
are the narrowest with 67  and 66\,\kms\ , respectively. [\NII ]$\lambda6584$\AA\  and 
[\SIII ]$\lambda6584$ \AA\ 
are intermediate with 120\,\kms\ and 124\,\kms , respectively. 
MWC\,349A shows the strongest [\SIII ] line in the sample.
The line widths agree well with those given by 
Hamann \& Simon  (\cite{HamannSimon88}). 

\subsection{MWC\,645}
MWC\,645 is a poorly known object which is listed as uncl\be\ in Table \ref{spectype}. 
Swings \& Allen (\cite{SwingsAllen73}) studied the blue spectral region of MWC\,645
using 20\,\AA\,mm$^{-1}$ coud\'e spectra observed in 1971. They found that 
all strong emission lines of \FeII\ and [\FeII  ] were double with 
a radial velocity difference 
between the stronger red and weaker blue peak of 150\,\kms . The profiles of \hg\ and
\hd\ exhibited three components.
Swings \& Allen compared  the spectrum with that of $\eta$\,Car and found a strong 
resemblance even for the line profiles. 

Low-resolution spectra have been investigated by
Swings \& Andrillat (\cite{SwingsAndrillat81}). The peculiar line
profile of \ha\  shown in Fig. \ref{havel} was visible also in their
data. More recently, medium resolution spectra were studied by
Jaschek et al. (\cite{Jascheketal96}). They measured a heliocentric radial
velocity of $-76$\,\kms\ from the emission lines. This is in reasonable
agreement with the results listed in Table \ref{rad2} when the different spectral
resolutions are taken into account. 

The \ha\ profile  of MWC\,645 is very peculiar. It consists of a broad 
blue  and a narrow red
emission component. The other objects with \ha\ 
profiles of type 3 exhibit  blue and red components of similar witdhs.
A fit of Gaussian profiles to each component yields widths of 5.0\,\AA\ and 
1.3\,\AA\  (FWHM) for the blue and red component, respectively. 

The profiles of all other lines of MWC\,645  exhibit a 
characteristic asymmetry with a steep red flank and a wing on the blue side. 
This is not found in any other object of the sample studied here. 
In [\OI ] and  [\FeII ]  the central emission peaks  
are split by 19\,\kms\ and 26\,\kms , respectively. This is much less than reported by 
Swings  \& Allen (\cite{SwingsAllen73}) and indicates spectral variability. 
The profile of [\NII ] 
(observed with the lower resolution of 23\,000) 
is not split  although being well resolved with a FWHM of 64\,\kms . It 
exhibits a single  peak, however with an asymmetric line shape like the other 
metal lines. 

The line of \HeI$\lambda6678$ visible in the 1995 spectrum  of Jaschek 
et al. was absent in 1987 (cf. Fig. \ref{prohe66}). 
The identification of the emission line near $\lambda$6665\,\AA\ is uncertain. 
It could be due to a blend of \FeI\ lines around $\lambda$6667\AA . 

\subsection{MWC\,939}
MWC\,939 is a little studied object of unknown evolutionary status. 
Until now no high-resolution spectra have been published.
Based on medium resolution spectra Parthasarathy et al. (\cite{Partha00})
suggested 
a spectral type of B5. 

All emission lines observed exhibit double-peaked profiles with the red peak being
stronger than the blue peak. The velocity difference of the peaks of \ha\ is
$\sim90$\,\kms . This line shows some variability  with the $V/R$ ratio changing 
between 1987 and 1988. In 2000 the ratio was nearly the same as in 1988. 
The velocity of the central absorption did not vary.  

The metal lines have a much smaller line splitting than \ha\ of
about 10-15\,\kms . [\OI ] appears to be split slightly more than the lines of 
[\FeII ] and [\NII ]. The permitted line \FeII$\lambda$6456\AA\ is also split. 
However, contrary to the forbidden lines the  blue  peak is stronger than the red. 

\subsection{MWC\,1055}
Little is known about MWC\,1055. In Table \ref{spectype} it is
entered as uncl\be . The only spectroscopic study has been carried by
Allen \& Swings (\cite{AS76}) who classify the object as member of their group 1
comprising objects with an appearance similar to conventional Be stars. They found 
possible [\OI ] emission 
but \FeII\ was absent. 

The \ha\ profile is variable. In 1987 it exhibited a strong red and a very weak 
blue emission peak. The absorption component was weak
and did not reach below the continuum. 
In 2000 the absorption component was much broader 
and the blue emission peak was even weaker. The velocity of the blue 
absorption edge changed from  $-287$\,\kms\ in 1987 to  $-408$\,\kms\ in 2000.  
The velocity of the red emission peak remained constant.

The coud\'e spectrum shows clear [\OI ] emission with a possible double-peak 
structure. In the lower resolution FOCES spectrum  [\FeII ]$\lambda7155$\AA\ 
is present as weak single-peaked emission line. Likewise, numerous \FeII\ lines 
are weakly present in the FOCES spectrum.
\HeI$\lambda5876$\AA\ appears in absorption. The \NaI\,D
doublet shows a P\,Cyg profile  with red-shifted emission components and blue-shifted
possibly circumstellar absorption components in addition to interstellar absorption.  

\subsection{HD\,45677 (= FS CMa)}
\object{HD\,45677} is probably the best-studied \be\ star in the Milky Way. 
Nevertheless, its nature is still controversial (cf. Lamers et al. \cite{Lamersetal98}, 
Cidale et al. \cite{Cidaleetal01}). 

The 
first detailed spectroscopic investigation was carried out by 
Swings (\cite{Swings73a}). He detected split  \FeII\ lines. The photospheric 
absorption line spectrum was analyzed by Israelian et
al. (\cite{Israelianetal96}) who derived $T_{\rm eff} =22\,000$\,K and $\log 
g = 3.9$, consistent with the spectral type of B2V. Spectroscopic 
variablitity was studied by
Israelian \& Musaev (\cite{IsraelianMusaev97}). Short and long-term 
photometric and spectroscopic behaviour
was studied in detail by de Winter \& van den Ancker (\cite{deWinterAncker97}).

The  \ha\ line of \object{HD\,45677} observed in 1988 exhibits a complex profile similar 
to that
reported by 
de Winter \& van den Ancker (\cite{deWinterAncker97}). 
Their spectra were 
obtained 
in 1993 and 1994.
The line profile of 1988 shows a red emission component which is split 
into two subcomponents. 
The  peak separation   is  25\,\kms . Comparison 
with the high-resolution line profiles of de Winter \& van den Ancker 
(\cite{deWinterAncker97})
indicates some slight long-term variability of the line profile. 
The $V/R$ ratio of the subcomponents of the main  \ha\ peak changed from 1.0 in
March 1988 to  0.92 in October 1993 and 0.80 in January 1994, 
as measured from the plots
in de Winter \& van den Ancker (\cite{deWinterAncker97}).

Whereas [\OI ]$\lambda$6300\AA\ is  marginally split  by 
6\,\kms\ into two components, [\FeII ]$\lambda$7155\AA\ shows only a single 
emission peak although the FWHM and FWZI of the latter line are larger. Likewise,  
[\NII ] is a single-peaked emission line, however significantly narrower than [\OI ].
The permitted \FeII\ line at $\lambda$6456\AA\ exhibits two well separated peaks
with a velocity difference of 29\,\kms , similar to the subcomponents of \Ha . 
Swings (\cite{Swings73a}) measured a similar line splitting of 32\,\kms\ for the  
\FeII\ lines.

\subsection{HD\,87643}
HD\,87643 was originally classified  by Carlson \& Henize (\cite{CarlsonHenize79}) 
as P\,Cygni type or nova-like star. It is now generally listed as \be\ supergiant.

HD\,87643 is embeddded in a reflection nebula (Henize 
\cite{Henize62}) for which  Surdej et al. (\cite{Surdejetal81}) found an expansion
velocity of 
about 150\,\kms . 
On the other hand, a  much higher outflow 
velocity  of $\sim1400$\,\kms\ measured from the 
P\,Cygni profile of \ha\  was reported  by Carlson \& Henize (\cite{CarlsonHenize79}). 
From \hb\ de Freitas Pacheco et al. (\cite{deFreitasetal82}) obtained an outflow
velocity of 1200\,\kms .
A recent optical study of HD\,87643 was carried out by Oudmaijer et al.
(\cite{Oudmaijeretal98}). They found  blue edge velocities of the P\,Cygni profile of 
\ha\ of 1500-1800\,\kms\ in spectra observed in 1997 
with clear indication of variability on a time scale of 3 months. 

The  wavelength interval covered by the 1986/88 CES spectra is to small to include the 
broad P\,Cygni absorption component. It only covers the central part of the 
profile within a velocity range of $\pm650$\,\kms . The comparison of this part of the
line profile with the profiles displayed in Oudmaijer et al.
shows that the $V/R$ has changed significantly from $\sim0.3$ in 1986/88 to
0.55 in January 1997 and 0.8 in April 1997. 
From 1986 to 1988 the equivalent widths of \Ha\ and of [\OI ]$\lambda$6300\AA\ 
increased by about 20\%.    

The FWHM of the forbidden lines of [\OI ]$\lambda$6300\AA\ and 
[\FeII ]$\lambda$7155\AA\ are 37\,\kms\ and  43\,\kms , respectively. Oudmaijer et al.
(\cite{Oudmaijeretal98}) measured 40\,\kms\ for [\OI ] in 1997, which is in good
agreement with the CES spectra. The line of [\FeII ]$\lambda$7155\AA\ 
appears slightly asymmetric. It exhibits a faint blue wing which leads to a FWZI 
of 260\,\kms . The FWZI of  [\OI ] on the other hand is only 180\,\kms . 
Note, however, that  [\OI ]$\lambda$6300\AA\ is slightly disturbed on the blue side 
by  \FeI(62)$\lambda$6297\AA , which could mask a  blue wing as observed in the [\FeII ]
profile.  

The splitting of the emission peak of \ha\ is $\sim200$\,\kms\ with some 
indication for an increase by 30\,\kms\ from 1986 to 1988. This is mainly due to an
increased blue shift of the blue emission component. Oudmaijer et al. measured a line
splitting of 180\,\kms , again in good agreement with the CES spectra. The central 
dip of \ha\ is shifted to the blue relative to the forbidden lines by $\sim100$\,\kms . 

Oudmaijer et al. mention the possible presence of an absorption feature of 
\HeI$\lambda$5876\AA . In the CES spectrum of 1988 this line is clearly present in
absorption. Its heliocentric radial velocity is shifted to the red by 32\,\kms\
relative to the forbidden lines. The permitted  line of \FeII$\lambda$6456\AA\ on the
other hand is shifted to the blue by 12\,\kms\ relative to [\OI ] and [\FeII ].
Furthermore, this line and the \FeII\ lines around 4550\,\AA\  exhibit blue wings.

\subsection{Hen\,230 (= He\,2-17)}
Very little is known about this object of type uncl\be . It appeared in 
lists of planetary nebulae, but was recognized as ``misclassified'' PN by 
Acker et al. (\cite{Ackeretal87}). 

Swings (\cite{Swings73b}) described spectra of this object. He detected lines of 
\Ha , \hb, \hg, \hd, [\OI]$\lambda$6300\AA , [\SII]$\lambda$4068\AA , a possible blend
of [\FeII] and possible \HeI$\lambda$4471\AA . However, no indication of  [\NII] or 
[\OIII] lines was found. He concluded that Hen\,230 resembles more a peculiar Be star
than a planetary nebula. 

The CES spectra  show single-peaked profiles of all observed lines. With $\sim15$\,\kms\
the forbidden lines of [\FeII] and [\NII] are very narrow, but resolved. 
The permitted line of \FeII$\lambda$6456\AA\ 
is 6 times broader than the forbidden line [\FeII ]$\lambda$7155\AA . 
Furthermore, it is slightly blue shifted by $-6$\,\kms\ relative to the forbidden 
lines whereas the peak of \ha\ is red shifted by $+11$\,\kms .

\subsection{Hen\,485 (=Wray 15-642)}
\label{hen485}
First described by Allen \& Swings (\cite{AS76}) this stars was 
classified by Allen (\cite{Allen78}) using low resolution
spectra as Be!pec. He noted that it appeared to be surrounded by a low density 
emission nebula with high excitation. Apart from this not much is known about 
Hen\,485. Th\'e et al. (\cite{Theetal94}) list it among ``other Bep or \be\
stars with strong IR-excess and unknown spectral type''. It is classified
``uncl\be '' in Table \ref{spectype}.

\object{Hen\,485} exhibits a variable \ha\ line profile. 
In 1988 an absorption component was visible which was absent in 1986.
The heliocentric radial velocity of the blue edge in 1988 is $-361$\,\kms . Adopting 
the mean velocity of the forbidden lines of $-10$\,\kms\ as systemic velocity this 
leads to a wind expansion velocity of 371\,\kms . The forbidden lines are 
single-peaked whereas the permitted line \FeII$\lambda$6456\AA\ is split into two 
components by 29\,\kms .

\HeI$\lambda$5876\AA\ is a broad emission line with FWZI of 410\,\kms , centred on the
velocity of the forbidden lines but with a slightly asymmetric red wing. 
\NaI\,D shows emission components and possibly also a circumstellar absorption 
component. On the blue side of \NaI$\lambda$5889\AA\  an absorption feature is visible. 
If identified with this line the central radial velocity would be $-263$\,kms . 
The corresponding red doublet component would then be filled in  by the emission 
of the blue doublet component explaining its absence.

\subsection{Hen\,1191}
\label{remark_hen1191}
Le Bertre et al. (\cite{LeBertreetal89}) classified Hen\,1191 as a possible
proto-planetary nebula, i.e. as an object in a stage intermediate 
between the asymptotic giant branch and the PN stage. They detected a 
bipolar nebula which is even visible on the Digital Sky Survey. In Table \ref{spectype}
the star is therefore listed as cPN\be .

Hen\,1191 exhibits double-peaked  \ha\ and very narrow emission lines.  
The peak separation of \ha\ is 79\,\kms .
The average widths at half maximum is 19$\pm2$\,\kms\ for  
[\OI ], [\NII ] [\FeII ], and \FeII . Note that the lines are resolved. 
The forbidden lines of Hen\,1191 are the strongest of the sample.

Le Bertre et al. also observed a high-resolution \ha\ line profile using the same
instrumentation at ESO as for the observations presented here. The 
spectrum was obtained in 1988 just 10 days before the one shown in Fig.\ref{havel}. 
The radial velocities measured from the \ha\ profile by Le Bertre et al. 
seem to differ from the results given here in this paper in Tab \ref{rad2}. 
From the wavelengths they list in their Table 3 radial velocities of  
$-33$\,\kms\ and $-43$\,\kms\ can be calculated, which correspond to
a peak separation of 10\,\kms\ only. However, this contradicts 
the peak separation visible in the spectrum shown in their Fig. 6a, which 
is consistent with a  separation of $\approx 70$\,\kms , and hence  
is in good agreement with the line profile presented here in Fig. \ref{havel}.

\subsection{CD$-$24$\degr$5721}
The unknown evolutionary status leads to a classification of 
``uncl\be '' in Table \ref{spectype}.
Allen \& Swings (\cite{AS76}) detected P\,Cygni profiles in the Balmer lines 
from \ha\ to \hd . Furthermore, emission lines of [\SII ],  [\FeII ], and \FeII\ were
found. \HeI$\lambda$5876\AA\ appeared in emission while \HeI$\lambda$4471\AA\ was
present in absorption. 

The CES spectrum shows a double-peaked \Ha\ emission line with a peak separation of
120\,\kms . Likewise, [\OI ] is double-peak, however with a peak separation of only
12\,\kms .  [\NII ] and [\FeII ] are single-peaked emission lines. 
Unlike any other star in the sample the permitted lines of \FeII\ appear as
narrow absorption lines which are unshifted relative to the forbidden lines  (Fig.
\ref{profe45}). 
In this respect CD$-$24$\degr$5721 resembles strongly the shell-type classical Be stars.
Heliocentric radial velocities of the detected absorption lines are listed in Table 
\ref{cd24feii}.
The absorption line FWHM is 11\,\kms . Contrary to the observation by Allen \& Swings 
(\cite{AS76}) \HeI$\lambda$5876\AA\ is present as broad
absorption line with a FWHM of 310\,\kms , indicating spectroscopic variability. 
It is slightly blue shifted relative  to the forbidden lines.

\begin{table}
\caption[]{Metal absorption lines identified in the spectra of CD$-24\degr 5721$ 
with observed and laboratory wavelengths and
heliocentric radial velocities.
}

\begin{tabular}{llll}
\noalign{\smallskip}    
\hline
\hline
\noalign{\smallskip}    
line   &$\lambda_{\rm obs}$\,[\AA]   &$\lambda_{\rm lab}$\,[\AA]  &$v_{\rm rad}$\,[\kms ]\\
\noalign{\smallskip}
\hline
\noalign{\smallskip}
\CrII (31) & 4285.09 & 4284.210 & 62  \\
\FeII (27) & 4303.95 & 4303.166 & 55  \\
\FeII (28) & 4297.41 & 4296.567 & 52:  \\
\FeII (38) & 4550.36 & 4549.467 & 59  \\
\FeII (37) & 4556.79 & 4555.890 & 59  \\
\CrII (44) & 4559.54 & 4558.659 & 58  \\
\noalign{\smallskip}
\hline
\noalign{\smallskip}
\end{tabular}
\label{cd24feii}
\end{table}

\subsection{CPD$-$57$\degr$2874 (= Wray\,15-535, Hen\,394)}
The spectrum of CPD$-$57$\degr$2874 has been described by Carlson \& Henize 
(\cite{CarlsonHenize79}) as that of a fairly typical P\,Cygni star. They detected blue
shifted absorption components from \hg\ through  H8, but not in \hb . 
The average velocity of the Balmer emission lines is 28\,\kms , the average 
absorption velocity is -223\,\kms . Broad \HeI\
absorption lines at 4471\AA\ and 3964\AA\ were found. In 
\HeI$\lambda4471$\AA\ they suspected an emission component on the red side.
McGregor et al. (\cite{McGregoretal88}) classified this star as \be\ supergiant.

The CES observations show a double-peaked \ha\ line with a relatively broad 
red emission peak. The velocity of the red peak of +22\,\kms\ is in good agreement with 
the measurement of  Carlson \& Henize. The central absorption component has a 
velocity of $-75$\,\kms , 
i.e. it is less blue shifted than the absorption components of the higher Balmer
lines observed by  Carlson \& Henize. The helium line \HeI$\lambda$5876\AA\ clearly
shows the emission component suspected  by  Carlson \& Henize. It sits on the
red side of a broad  absorption feature. The blue absorption
component exhibits an absorption wing extending to a radial velocity of $-446$\,\kms .
The centre of the blue absorption component is at a velocity of  $-148$\,\kms .

[\OI ]$\lambda6300$\AA\ has a nearly flat-topped emission profile. 
[\FeII ]$\lambda7155$\AA\ 
exhibits a double-peaked
profile slightly broader than [\OI ]. [\NII ]$\lambda6583$\AA\ is narrower than
the two other forbidden lines, however it displays  a split line profile. Likewise, the
permitted line \FeII$\lambda6456$\AA\ has a double-peaked structure with the central
absorption blue shifted relative to the centres of the forbidden lines. The peak
separation is 99\,\kms . It is the broadest (FWHM) metal line observed for 
CPD$-$57$\degr$2874. 

\subsection{CPD$-$52$\degr$9243}
\label{cpd52}
Detailed studies of the \be\ supergiant CPD$-$52$\degr$9243 based on high-resolution 
spectra were carried out
by Swings (\cite{Swings81}) and Winkler \& Wolf (\cite{WinklerWolf89}). 
The latter detected Balmer lines exhibiting P\,Cygni profiles with a separation
between emission and absorption components of 145\,\kms . \HeI\ lines were present only
in absorption. Numerous lines of singly ionized and neutral metals were found, the
stronger lines exhibiting P\,Cygni profiles. Many of these lines displayed double-peaked 
emission components with peaks at heliocentric radial velocities of $-50$\,\kms\ and 
$-10$\,\kms . The only forbidden lines found were 
[\OI ]$\lambda\lambda$6300, 6364\AA . No forbidden singly ionized iron lines were 
found by these authors.

The CES spectra are only partly consistent with these findings. \ha\ has a P\,Cygni
profile similar to that shown by Winkler \& Wolf. The absorption component reaches 
below the continuum level. \FeII$\lambda$6456\AA\ also exhibits a P\,Cygni
profile. Its absorption component is more blue shifted by 44\,\kms\ than the 
P\,Cygni absorption of \ha\  . 

In addition to [\OI ]$\lambda$6300\AA\ the line [\FeII ]$\lambda$7155\AA\ is present.
This is the first detection of a forbidden line other than those of [\OI ]. 
The profiles of the forbidden lines are asymmetric. However, in contrast to the 
findings of Winkler \& Wolf no clear double-peak structure is discernible neither 
in the forbidden lines nor in the permitted line \FeII$\lambda$6456. This is surprising, 
since Swings (\cite{Swings81}) noted that a doubling of the red emission component exists 
for most of the strong emission lines.  

\HeI\ appears in absorption only. The centre of the absorption line of 
\HeI$\lambda$5876\AA\ shows the same velocity as the P\,Cygni absorption components,
i.e. it is blue shifted relative to the emission peaks. The radial velocity is  
$-300$\,\kms . The blue edge velocity is $-494$\,\kms . 

Whereas the \NaI\,D doublet in  the other stars shows very likely pure
interstellar absorption features, in CPD$-52\degr$9243 a circumstellar
absorption component is clearly visible. 
The \NaI\,D1 line
exhibits a rather broad 
absorption wing visible in the bluest feature of the multi-component 
absorption complex. The velocity of the blue edge is $-476$\,\kms , i.e. close to the
velocity of the blue edge of \HeI$\lambda$5876\AA . 
The \NaI\,D2 line
seems to show a similar profile. However, the extended absorption wing 
is strongly disturbed by the interstellar absorption lines of the blue doublet 
component.

\section{Observed spectra}
\label{plots}
\begin{figure*}[tbp]
\includegraphics[width=18cm,bbllx=75pt,bblly=55pt,bburx=585pt,bbury=685pt,clip=]{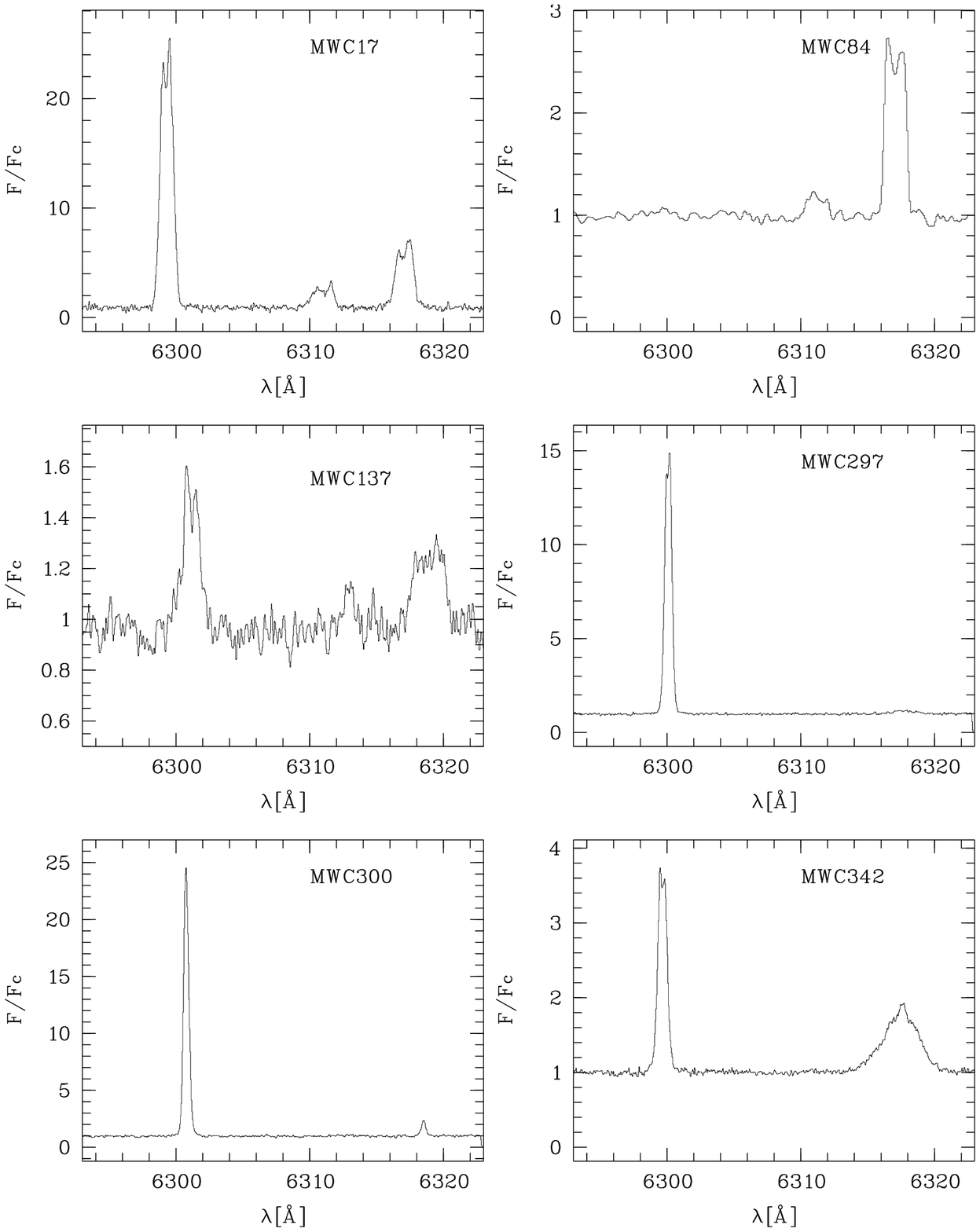}
\caption[]{Sections of the spectra around the lines of [\OI ]$\lambda$6300\AA , 
[\SIII ]$\lambda$6312\AA , and \MgII$\lambda$6318\AA . 
}
\label{prooi}
\end{figure*}

In this section the observed spectra are presented except \ha\ which is shown in 
Fig. \ref{havel}.  
The wavelength section around [\OI ]$\lambda$6300\AA\ shown in Fig. \ref{prooi} 
additionally contains the lines of [\SIII ]$\lambda$6312\AA\, and 
\MgI$\lambda$6318\AA . The profiles of [\NII ]$\lambda$6583\AA\ are displayed 
in Fig. \ref{pronii}. Note that the profiles of MWC\,137,  MWC\,342,
and MWC\,1055 were observed with FOCES.  MWC\,297 and MWC\,645 were observed with the 
lower coud\'e resolution of 23\,000. 
The spectrum in the wavelength section shown  contains additionally the line of 
\FeII$\lambda$6587\AA\ and in the case MWC\,84 of \CII$\lambda$6578\AA .
The wavelength section around the forbidden line of [\FeII ]$\lambda$7155\AA\ is
displayed in Fig. \ref{profe71}. The line of [\FeII ]$\lambda$7172\AA\ belonging
to the same multiplet 14F is also visible in most cases, however, heavily disturbed by
strong telluric absorption features which could not be well corrected. For two stars the
forbidden [\FeII ] lines at $\lambda\lambda$4276, 4287\AA\ were observed. They
are shown in Fig. \ref{profe42}.
The line profiles of the permitted \FeII\ line at $\lambda$6456 are displayed in
Fig. \ref{profe64}. In addition, sections of the spectra of three stars 
around $\lambda$4560\AA\ containing numerous \FeII\ lines are shown in 
Fig. \ref{profe45}.
The lines of \HeI$\lambda$5876\AA\ and of the \NaI\,D doublet are shown in
Fig. \ref{prohe58}. For four stars the wavelength region around \HeI$\lambda$6678\AA\ was
observed. They are shown in Fig. \ref{prohe66} 

 \newpage
\clearpage

\addtocounter{figure}{-1}
\begin{figure*}[tbp]
\includegraphics[width=18cm,bbllx=75pt,bblly=55pt,bburx=585pt,bbury=685pt,clip=]{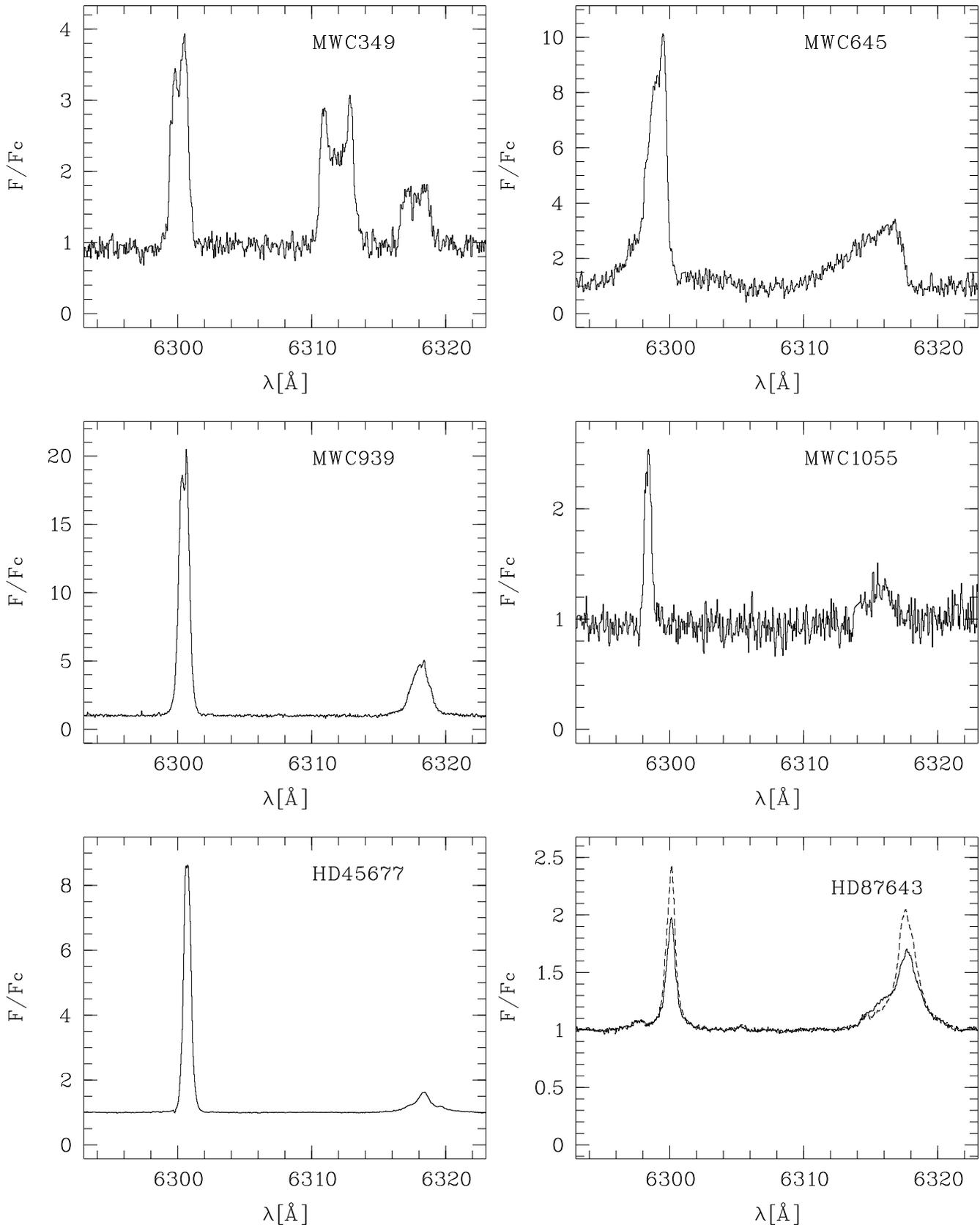}
\caption[]{[\OI ]$\lambda$6300\AA , continued. HD\,87643 was observed twice, in 1986 (solid line) 
and 1988 (dashed line).
}
\end{figure*}
\newpage
\clearpage

\addtocounter{figure}{-1}
\begin{figure*}[tbp]
\includegraphics[width=18cm,bbllx=75pt,bblly=55pt,bburx=585pt,bbury=685pt,clip=]{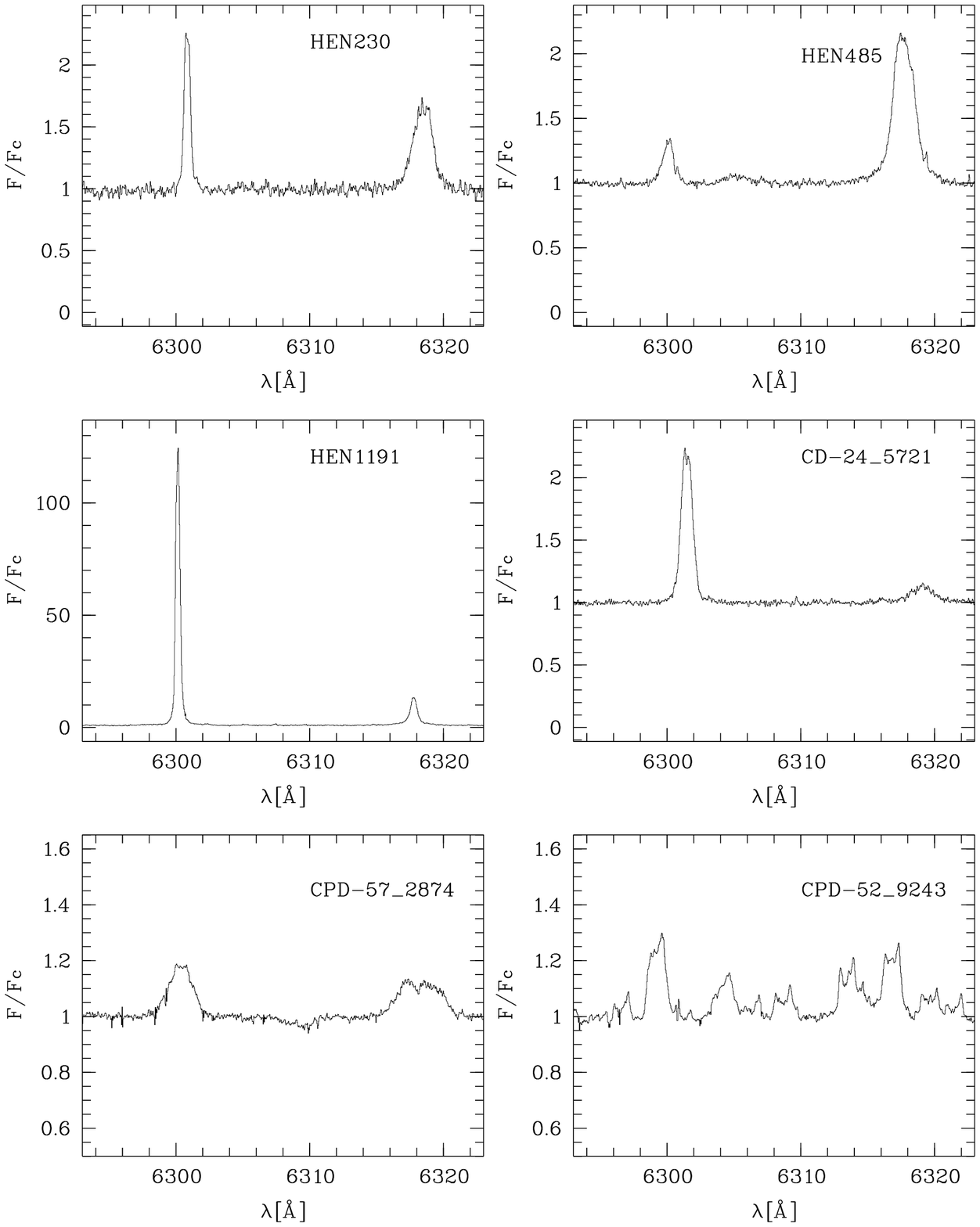}
\caption[]{[\OI ]$\lambda$6300\AA , continued.
}
\end{figure*}
\newpage

\begin{figure*}[tbp]
\includegraphics[width=18cm,bbllx=75pt,bblly=55pt,bburx=585pt,bbury=685pt,clip=]{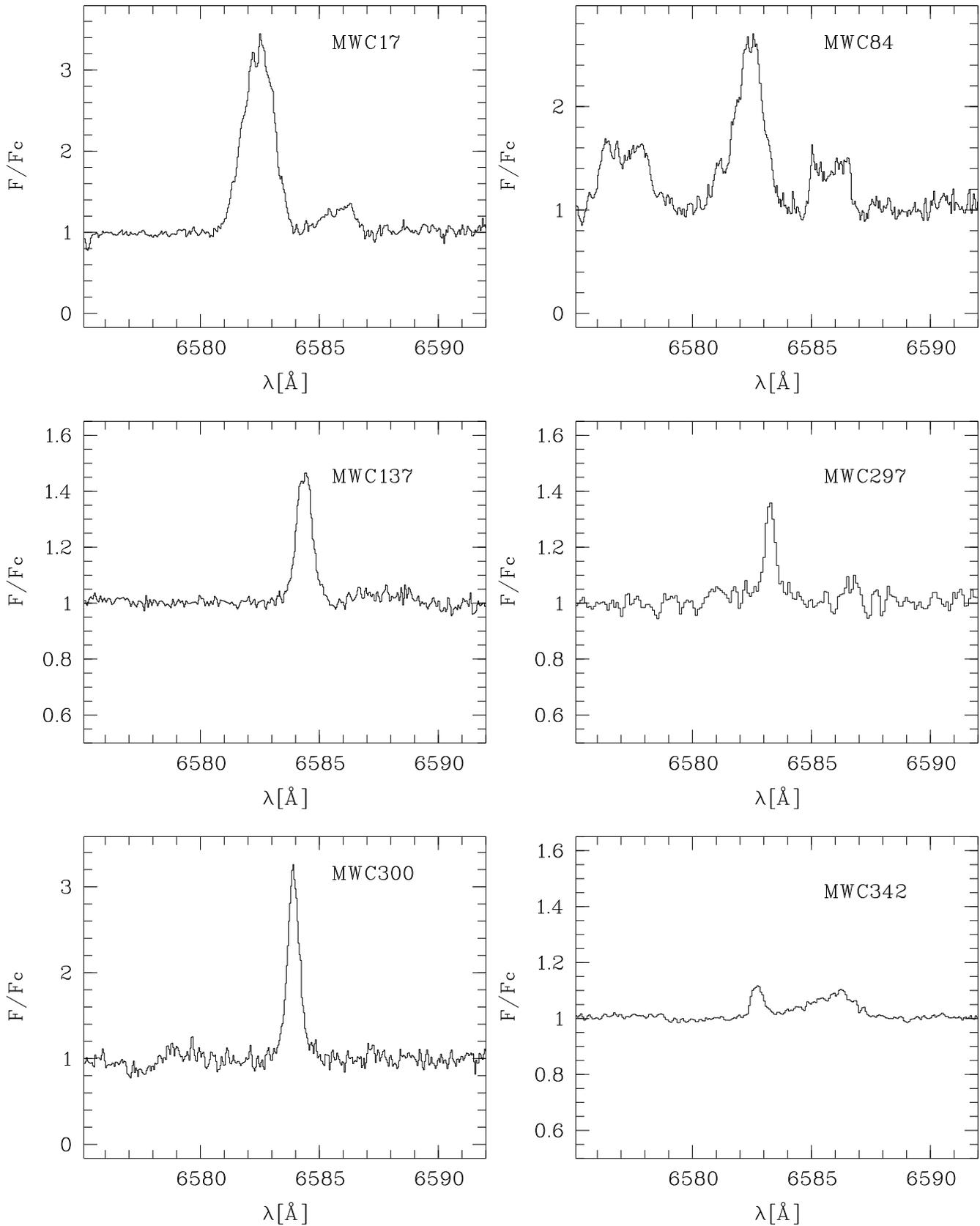}
\caption[]{Sections of the spectra around the line of [\NII ]$\lambda$6583\AA . Several 
stars also show emission of \FeII$\lambda$6587\AA  . MWC\,84 in addition exhibits
\CII$\lambda$6578\AA .
}
\label{pronii}
\end{figure*}
\newpage
\clearpage

\addtocounter{figure}{-1}
\begin{figure*}[tbp]
\includegraphics[width=18cm,bbllx=75pt,bblly=55pt,bburx=585pt,bbury=685pt,clip=]{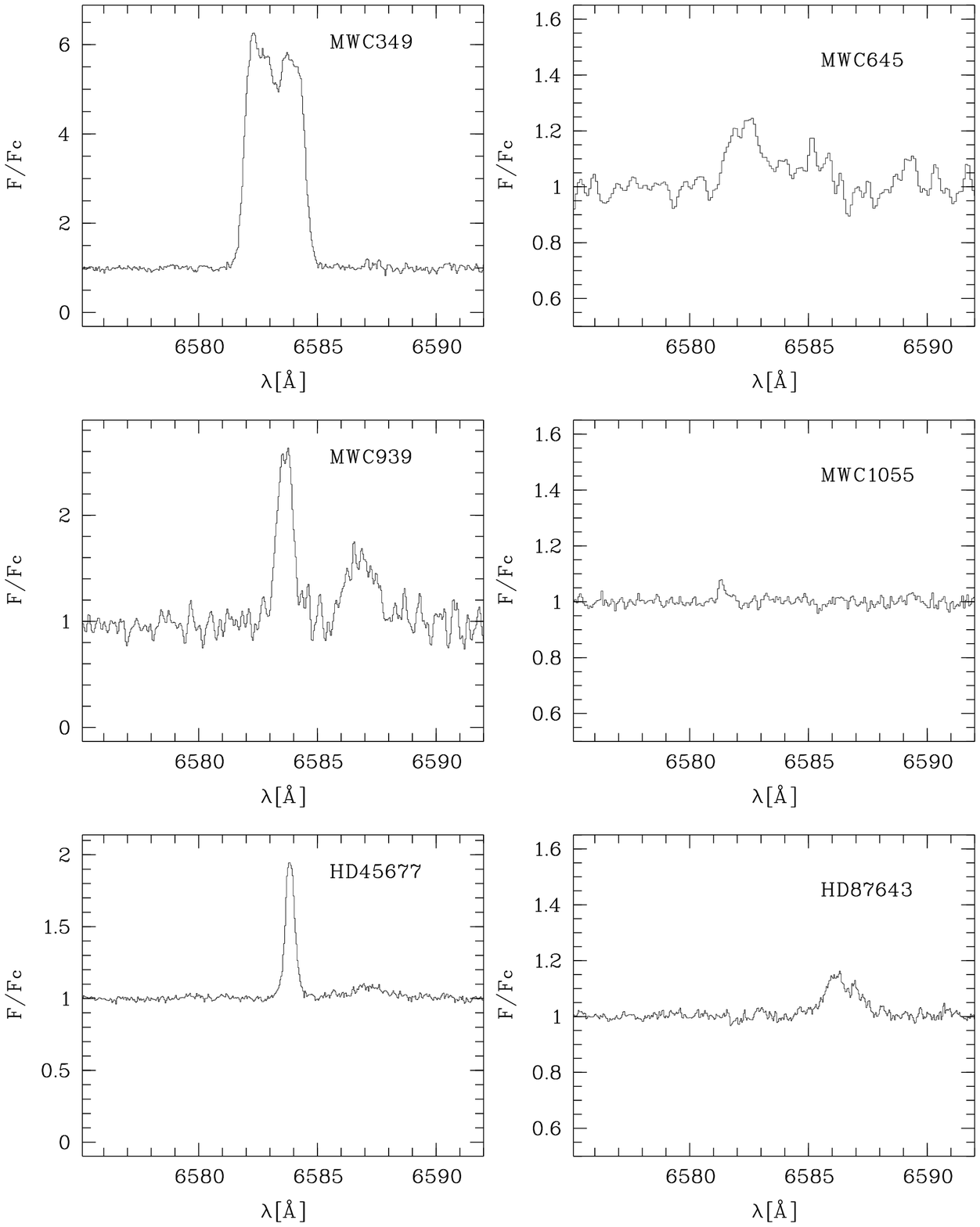}
\caption[]{[\NII ]$\lambda$6583\AA , continued. The line visible in  
HD\,87643 is \FeII$\lambda$6587\AA .
}
\end{figure*}
\clearpage
\newpage

\addtocounter{figure}{-1}
\begin{figure*}[tbp]
\includegraphics[width=18cm,bbllx=75pt,bblly=55pt,bburx=585pt,bbury=685pt,clip=]{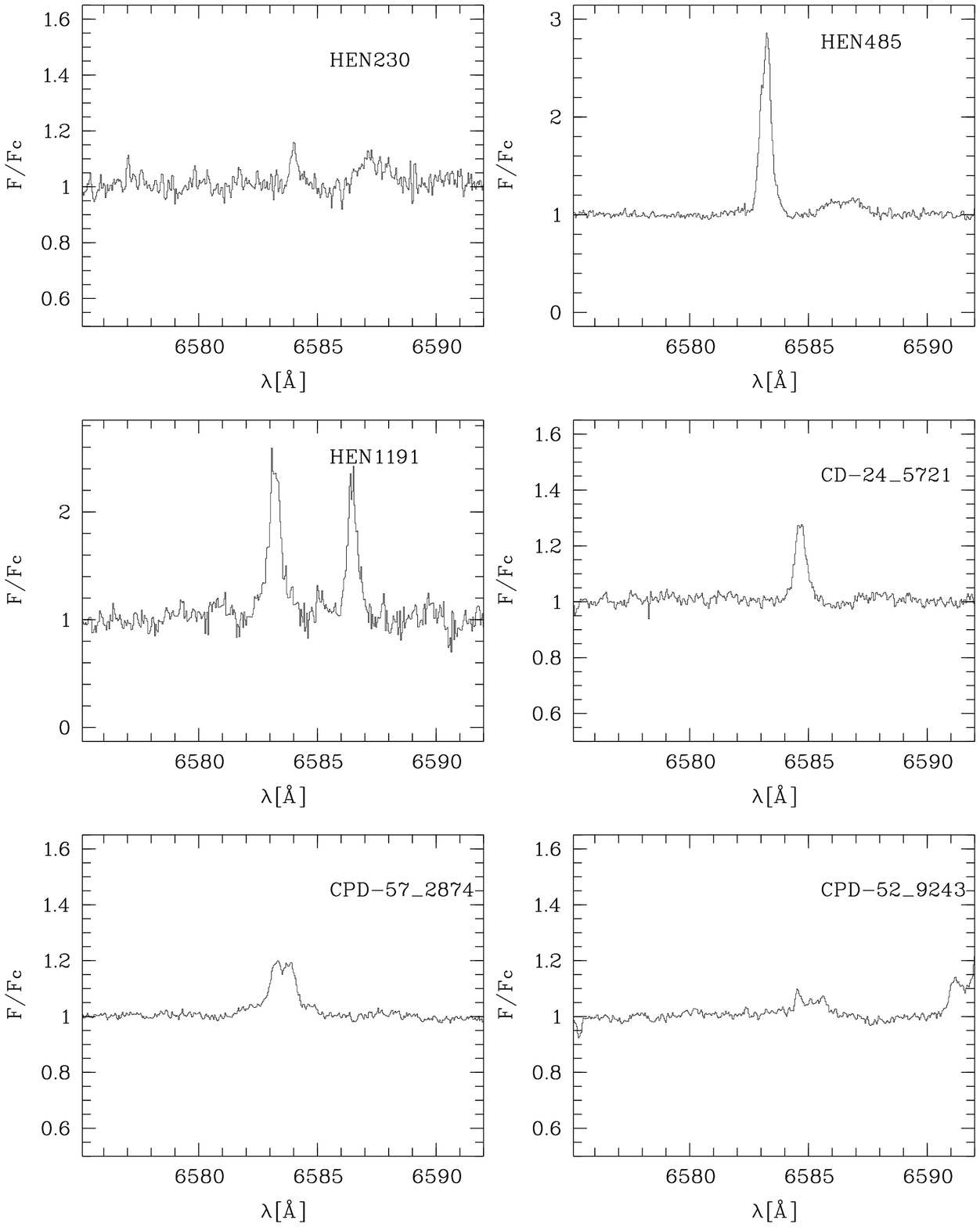}
\caption[]{[\NII ]$\lambda$6583\AA , continued. The line visible in CPD$-$52$\degr$9243 is 
\FeII$\lambda$6587\AA .
}
\end{figure*}
\newpage
\clearpage

\begin{figure*}[tbp]
\includegraphics[width=18cm,bbllx=75pt,bblly=55pt,bburx=585pt,bbury=685pt,clip=]{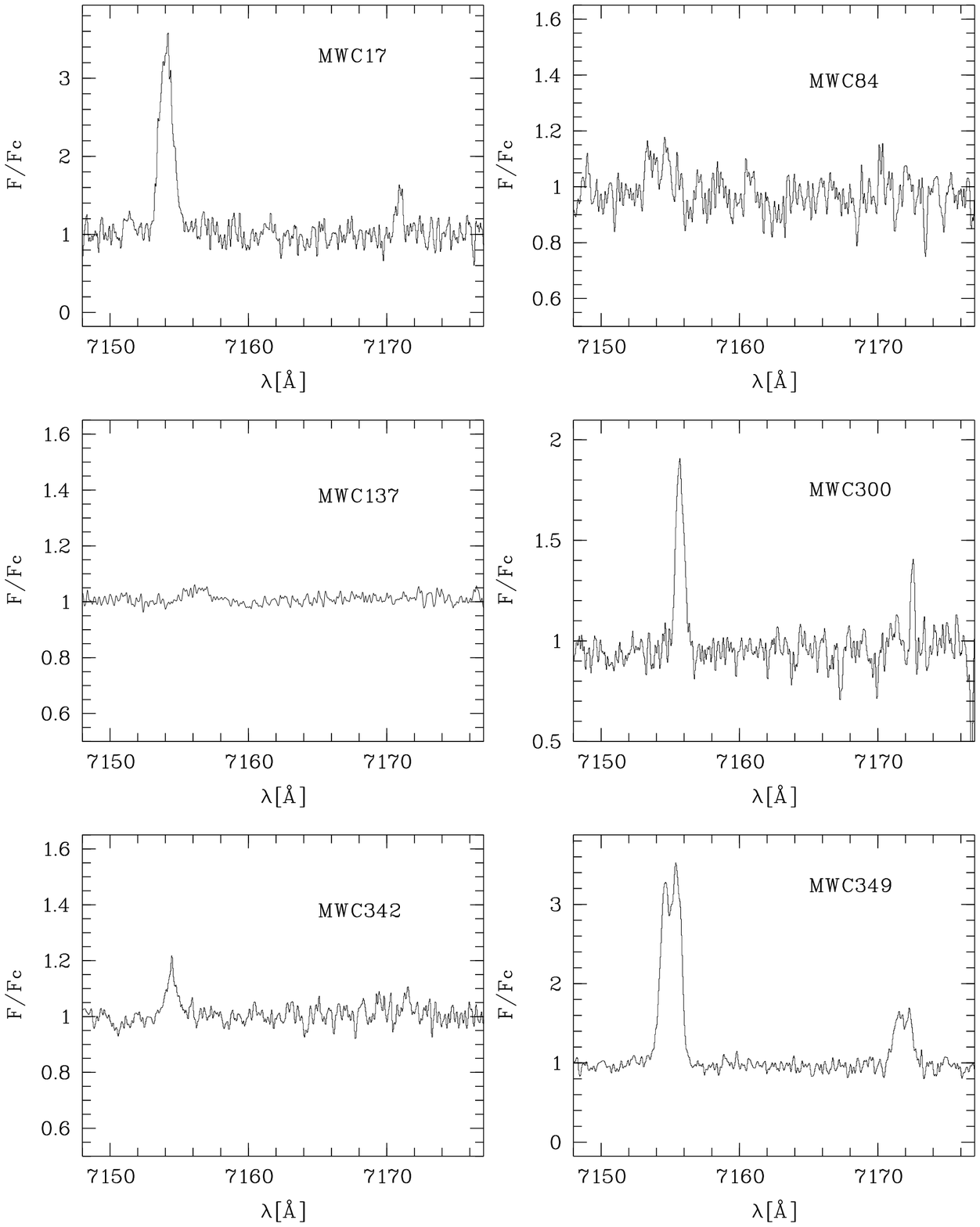}
\caption[]{Sections of the spectra around the line  [\FeII ]$\lambda$7155\AA . Also visible
is [\FeII ]$\lambda$7172\AA .
}
\label{profe71}
\end{figure*}
 \newpage
\clearpage

\addtocounter{figure}{-1}
\begin{figure*}[tbp]
\includegraphics[width=18cm,bbllx=75pt,bblly=55pt,bburx=585pt,bbury=685pt,clip=]{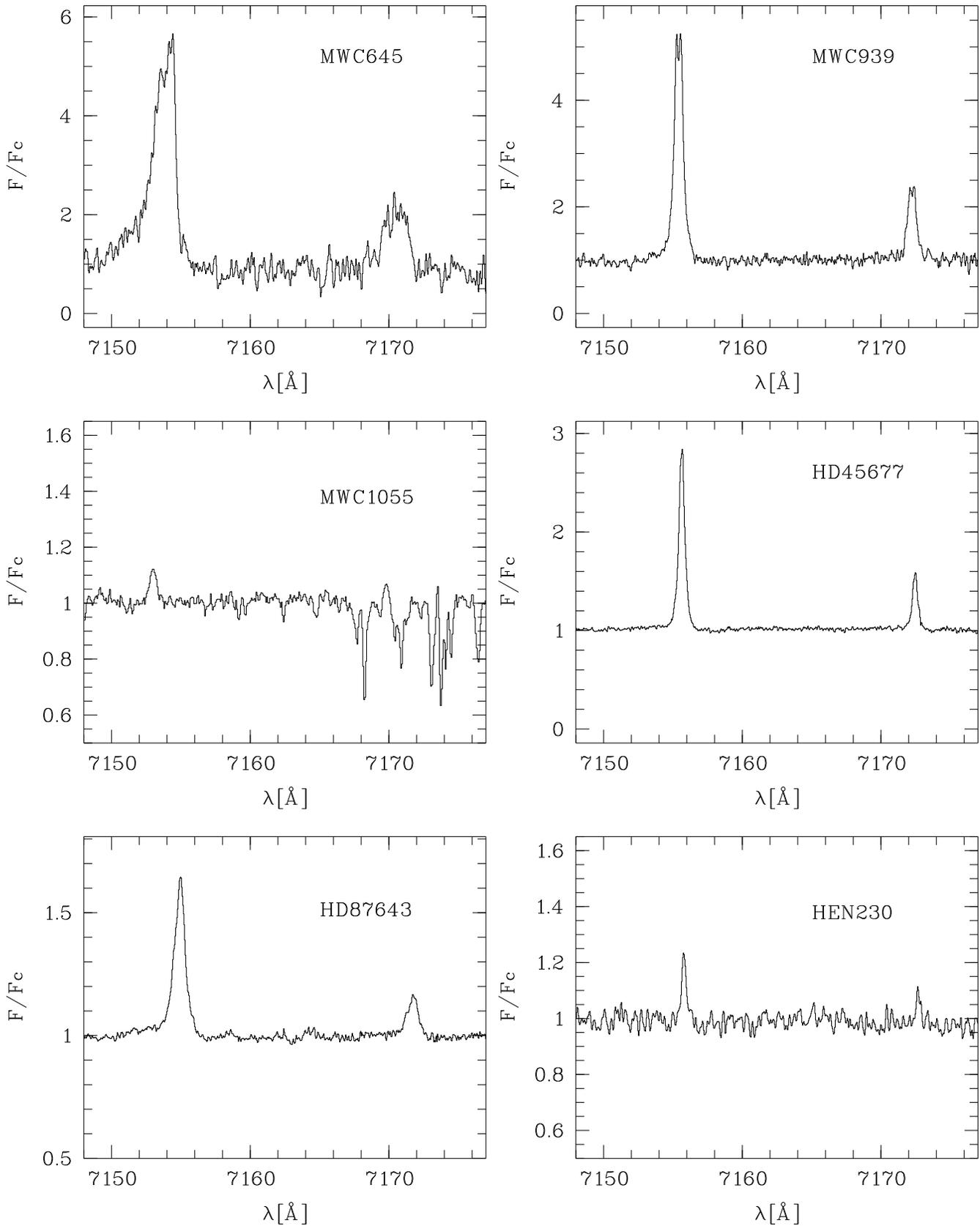}
\caption[]{[\FeII ]$\lambda$7155\AA , continued. For MWC\,1055 the telluric absorption lines
have not been corrected.
}
\end{figure*}
\newpage
\clearpage

\addtocounter{figure}{-1}
\begin{figure*}[tbp]
\includegraphics[width=18cm,bbllx=75pt,bblly=160pt,bburx=585pt,bbury=580pt,clip=]{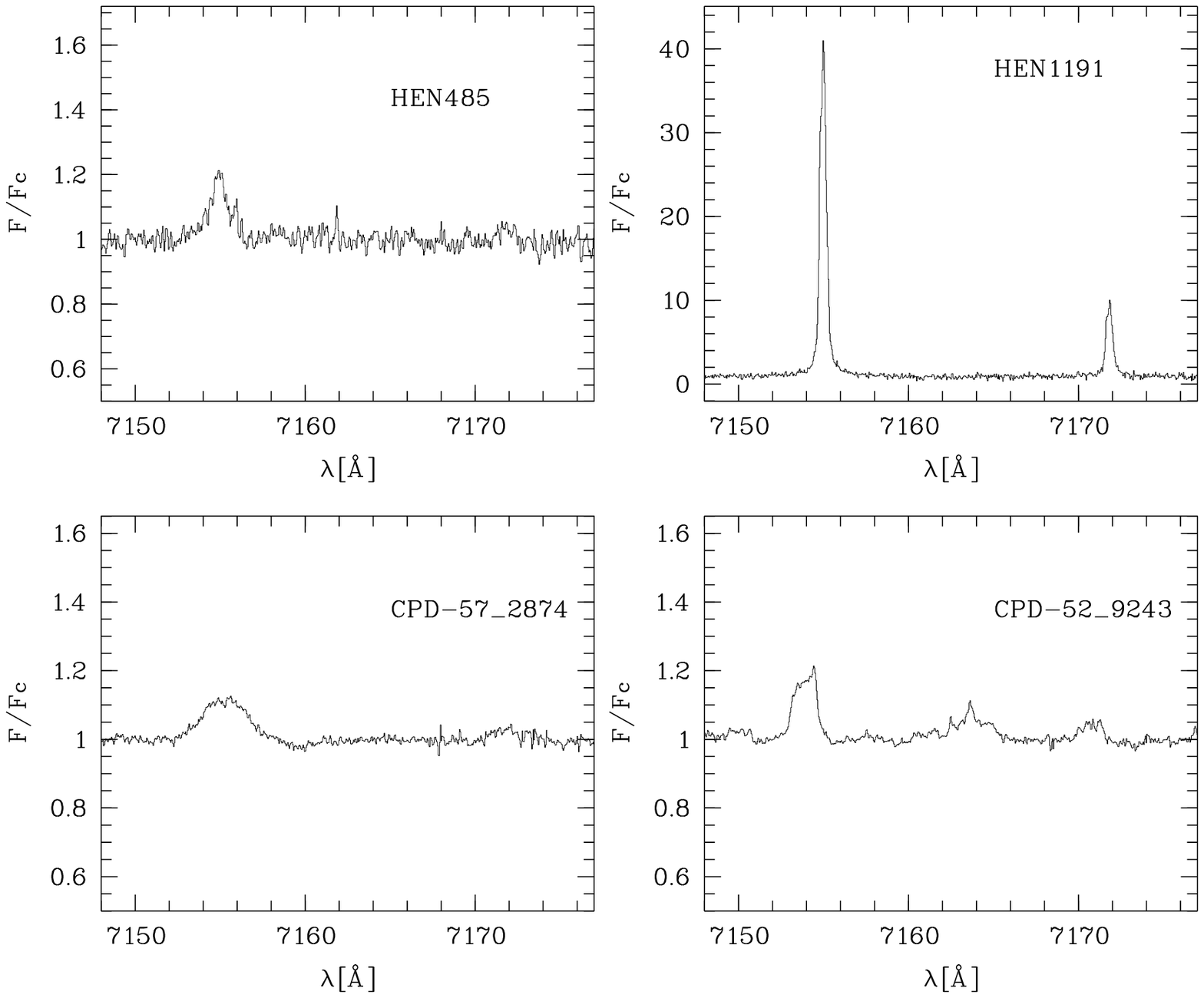}
\caption[]{[\FeII ]$\lambda$7155\AA , continued. 
}
\end{figure*}
\begin{figure*}[tbp]
\includegraphics[width=18cm,bbllx=75pt,bblly=260pt,bburx=585pt,bbury=475pt,clip=]{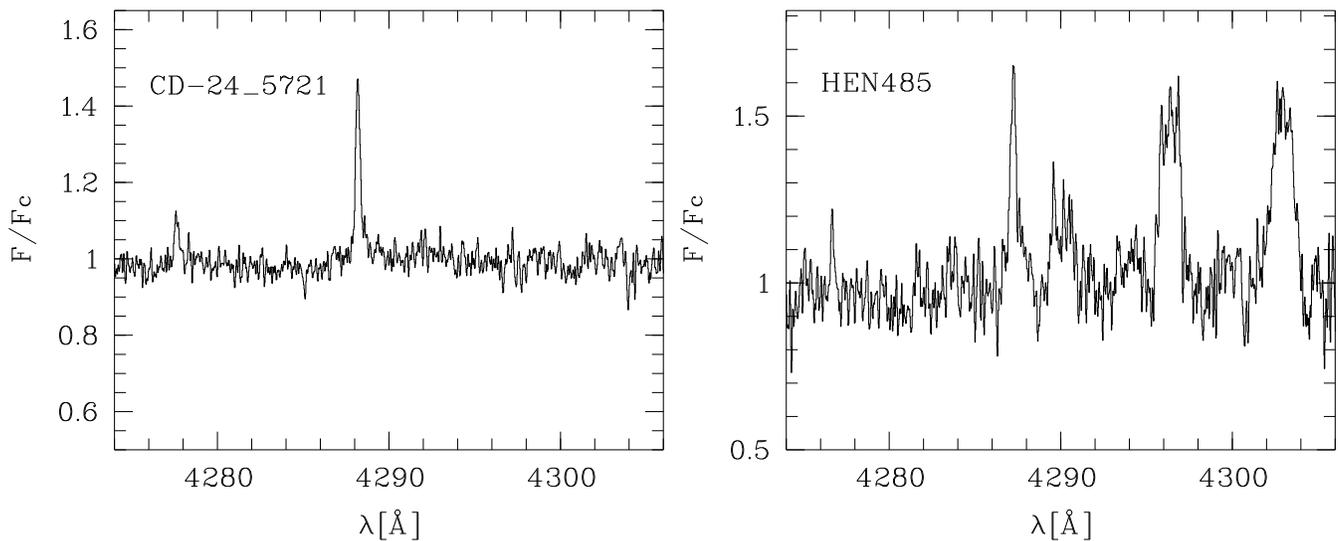}
\caption[]{Sections of  the spectra around [\FeII ]$\lambda$4287  observed for two stars.
}
\label{profe42}
\end{figure*}
\newpage
\clearpage

\begin{figure*}[tbp]
\includegraphics[width=18cm,bbllx=75pt,bblly=55pt,bburx=585pt,bbury=685pt,clip=]{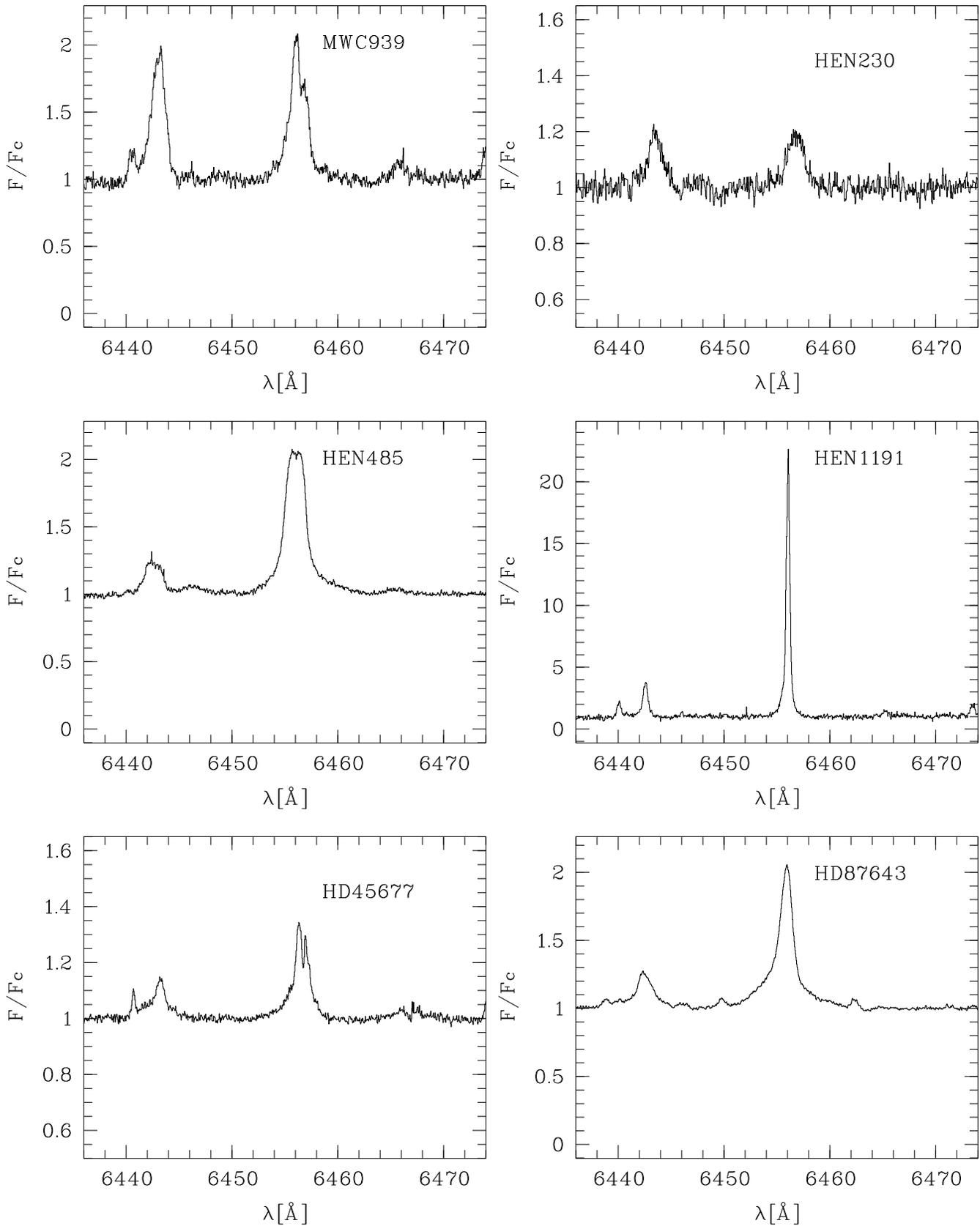}
\caption[]{Permitted \FeII\ line profiles. The figure shows sections of the 
spectra around the permitted line  of \FeII$\lambda$6456\AA .   
}
\label{profe64}
\end{figure*}
\newpage
\clearpage

\addtocounter{figure}{-1}
\begin{figure*}[tbp]
\includegraphics[width=18cm,bbllx=75pt,bblly=265pt,bburx=585pt,bbury=475pt,clip=]{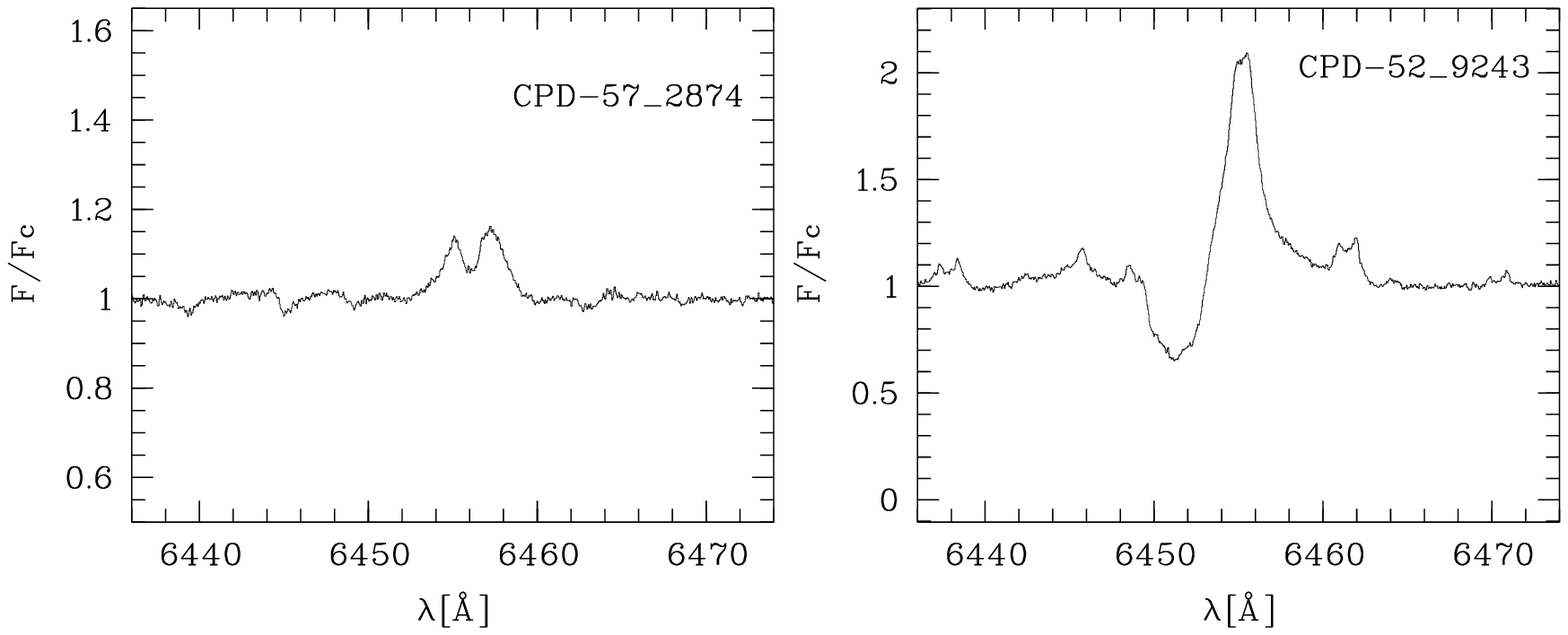}
\caption[]{\FeII$\lambda$6456\AA , continued. 
}
\end{figure*}

\begin{figure*}[tbp]
\includegraphics[width=18cm,bbllx=75pt,bblly=160pt,bburx=585pt,bbury=580pt,clip=]{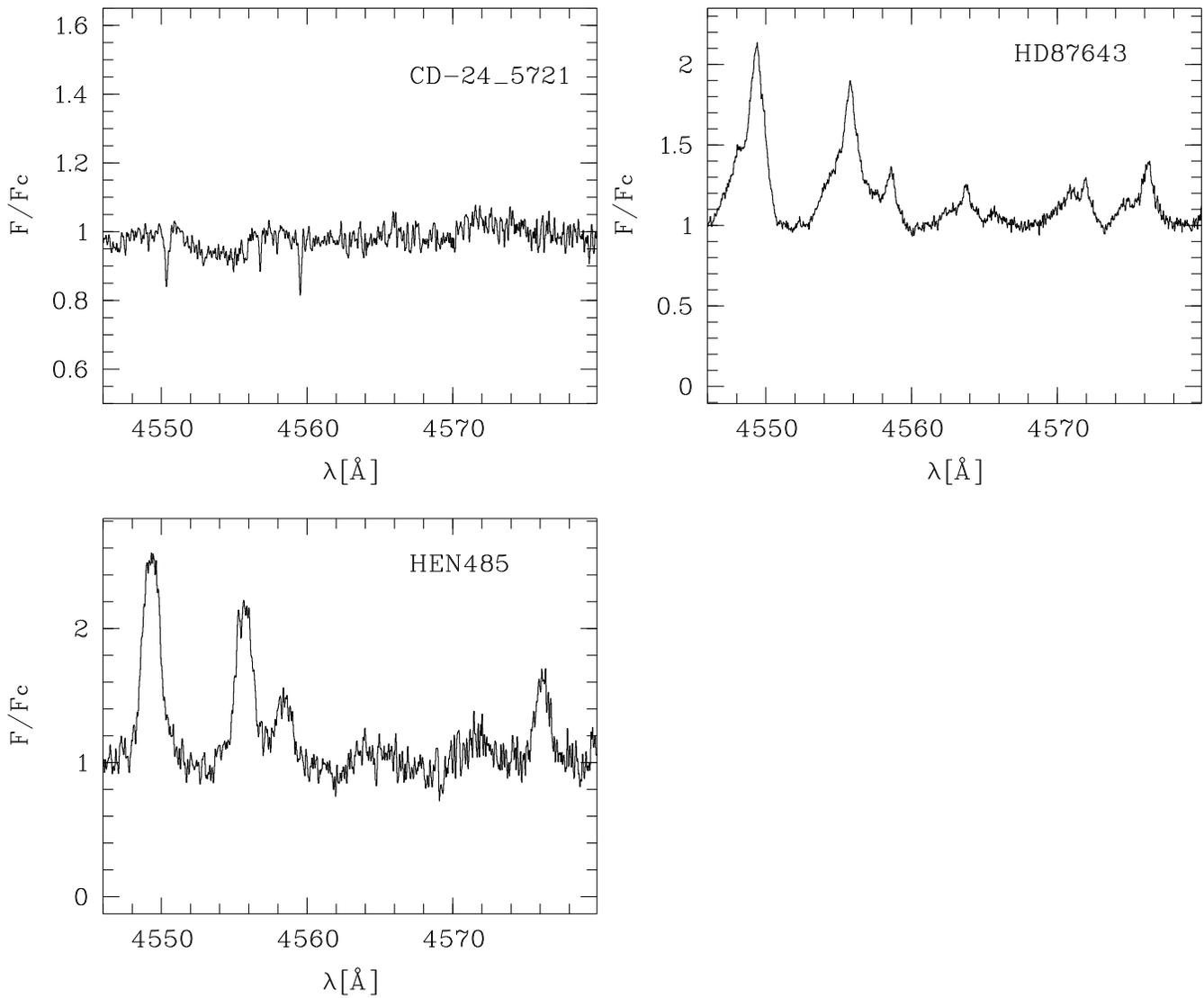}
\caption[]{Permitted \FeII\ lines around $\lambda$4560\,\AA\  observed for three stars.
}
\label{profe45}
\end{figure*}
\newpage
\clearpage

\begin{figure*}[tbp]
\includegraphics[width=18cm,bbllx=75pt,bblly=55pt,bburx=585pt,bbury=685pt,clip=]{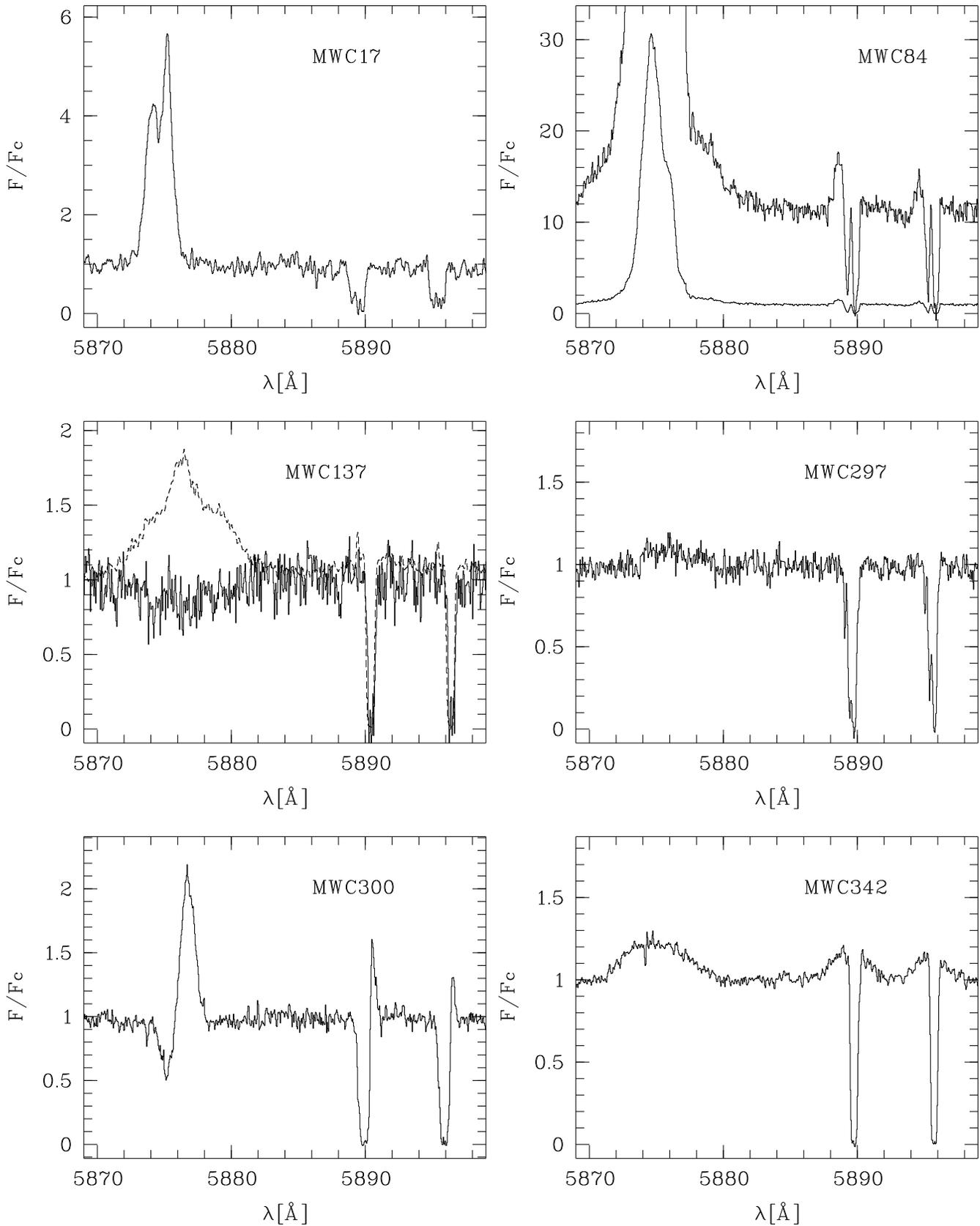}
\caption[]{Sections of the spectra around the lines of \HeI $\lambda$5876\AA\ and 
the \NaI\,D doublet. The upper spectrum of MWC\,84 has been overplotted with 
a stretch factor of 10 in the normalized flux.
}
\label{prohe58}
\end{figure*}
\newpage
\clearpage

\addtocounter{figure}{-1}
\begin{figure*}[tbp]
\includegraphics[width=18cm,bbllx=75pt,bblly=55pt,bburx=585pt,bbury=685pt,clip=]{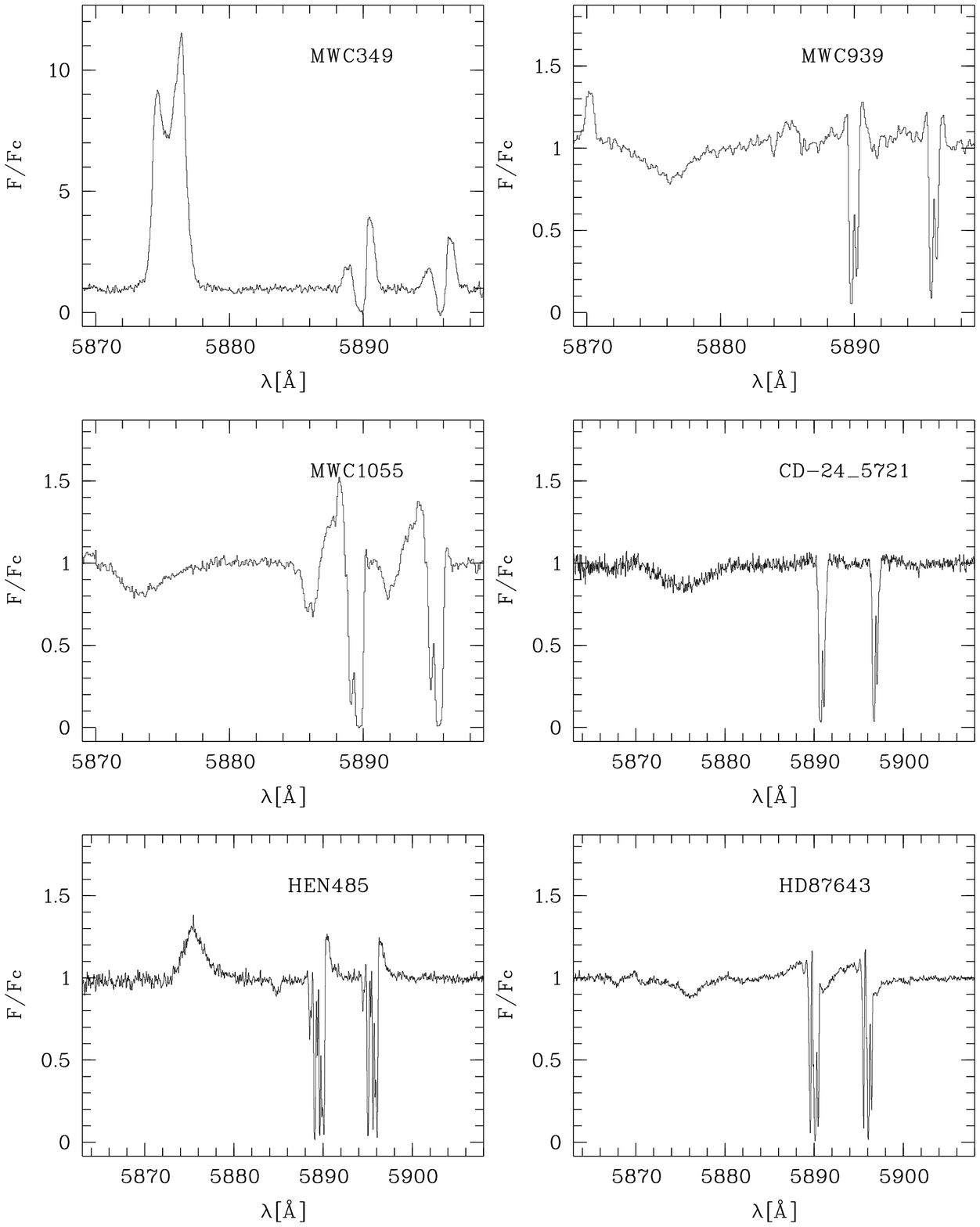}
\caption[]{\HeI $\lambda$5876\AA , continued.  
}
\end{figure*}
\newpage
\clearpage
 
\addtocounter{figure}{-1}
\begin{figure*}[tbp]
\includegraphics[width=18cm,bbllx=75pt,bblly=265pt,bburx=585pt,bbury=475pt,clip=]{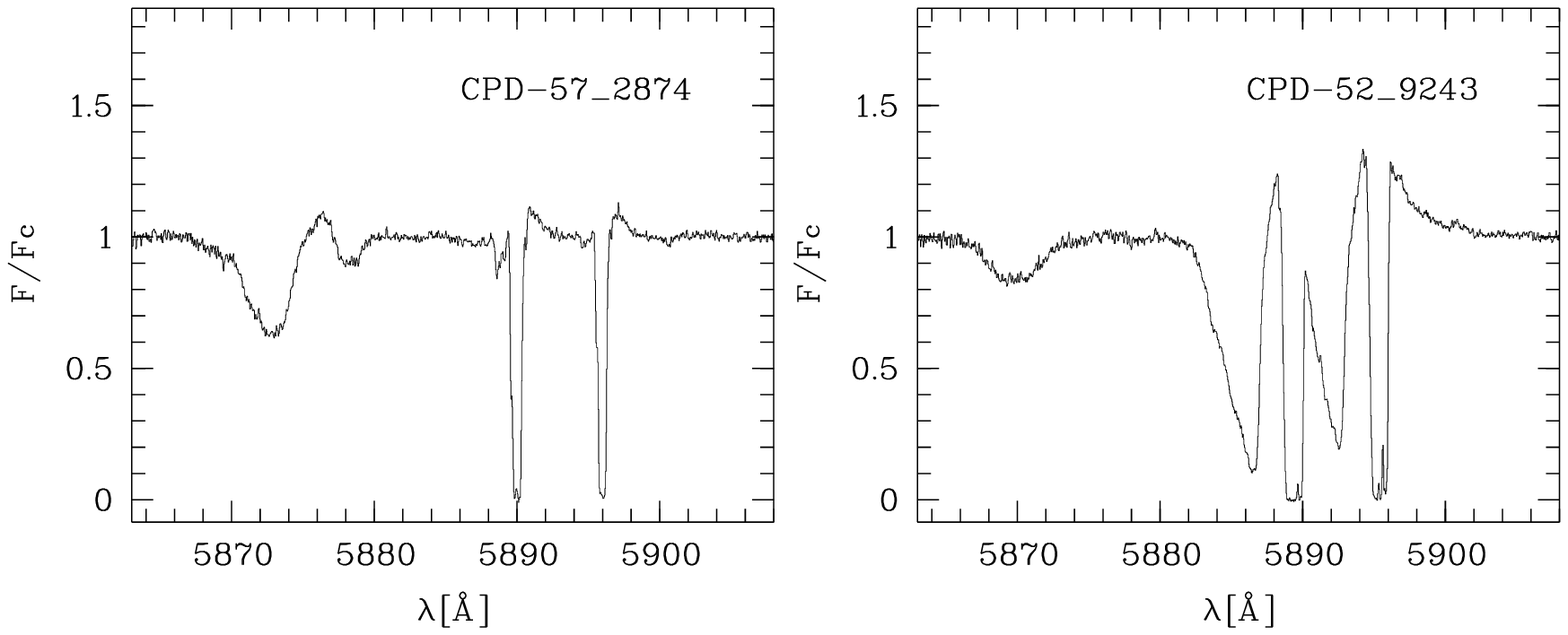}
\caption[]{\HeI $\lambda$5876\AA , continued. 
}
\end{figure*}

\begin{figure*}[tbp]
\includegraphics[width=18cm,bbllx=75pt,bblly=160pt,bburx=585pt,bbury=580pt,clip=]{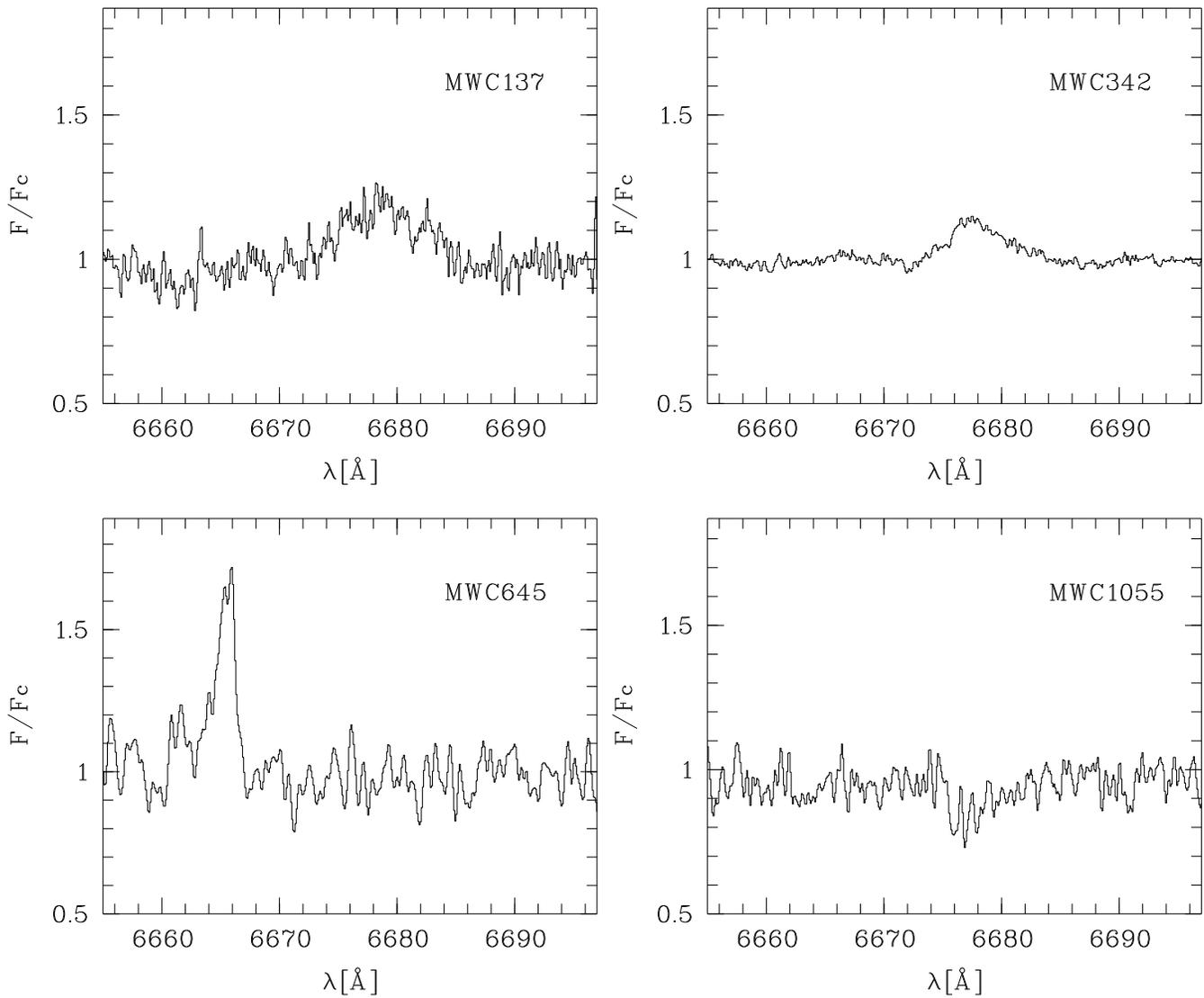}
\caption[]{Sections 
of the spectra around \HeI$\lambda$6678\AA\  observed for four stars. 
}
\label{prohe66}
\end{figure*}
\newpage
\clearpage

\end{document}